\documentstyle[makeidx, amsfonts, amsbsy, amssymb,
review, rev_body, graphicx]{report}

\begin{document}
\thispagestyle{myfirst}

\mytitle{Electron Phenomena in Layered Conductors}

\myauthor{O.V.~KIRICHENKO, Yu.A.~KOLESNICHENKO\\
and V.G.~PESCHANSKY} \vspace{-8pt}
\myadress{B.I. Verkin Institute for Low Temperature Physics and Engineering,\\
National Academy of Sciences of the Ukraine\footnotemark{} }
\vspace{-4pt} \footnotetext{\normalsize B.I. Verkin Institute for
Low Temperature Physics and Engineering, National Academy of
Sciences of Ukraine, 47 Lenin Ave, Kharkov, 310164, Ukraine.
E-mail: peschansky@ilt.kharkov.ua}

\myabstract{The quasi-two-dimensional nature of the charge
carriers energy spectrum in layered conductors leads to specific
effects in an external magnetic field. The magnetoresistance of
layered conductors in a wide range of strong magnetic fields
directed in the plane of the layers can increase proportionally to
a magnetic field value. The electromagnetic impedance and the
sound attenuation rate depend essentially on the polarization of
normal to the layers. Propagation of electromagnetic and acoustic
waves in these conductors involves virtually all charge carriers
in the transfer of acoustic pulses and electromagnetic field
spikes to the bulk of the conductor. The orbits of Fermi electrons
in a magnetic field are virtually indistinguishable, which allows
the inclusion of large number of conduction electrons in the
formation of peculiar oscillatory and resonant effects which are
absent in the case of ordinary metals. Investigation of these
effects introduce the possibilities for detailed study of the
dissipative processes in electron systems of layered conductors
and the charge carriers energy spectrum.

Point contact investigations of layered metals allow us to obtain
the information about electron and phonon spectra. The electron
focusing signal and the point contact spectrum are extremely
sensitive to the orientation of the magnetic field vector $\bs{H}$
in relation to the layers with a high electrical conductivity. The
values of $\bs{H}$ for which the electron focusing signal has
peaks can be used for determining velocities and extremal
diameters for the open Fermi surface. The dependence of the point
contact spectra on the magnitude and the relaxation of electrons
at various types of phonon excitations.} \thispagestyle{empty}
\setcounter{page}{3} \thispagestyle{empty}
\section*{\large\bf Electron Phenomena in Layered Conductors}
\vspace*{12pt}
\section*{{\LARGE\bf C}\Large\bf ONTENTS}
\contentsline {section}{\numberline {1.}Introduction}{3}
\contentsline {section}{\numberline {2.}Galvanomagnetic
Effects}{4}
\contentsline {section}{\numberline {3.}Propagation of
Electromagnetic Waves
                                        in Layered Conductors}{19}
\contentsline {subsection}{\numberline {3.1}Normal Skin
Effect}{24} \contentsline {subsection}{\numberline {3.2}Anomalous
Skin Effect}{28} \contentsline {subsection}{\numberline
{3.3}Weakly Damping
                                            Reuter--Sondheimer Waves}{30}
\contentsline {section}{\numberline {4.}Acoustic Transparency of
                                        Layered Conductors}{36}
\contentsline {subsection}{\numberline {4.1}The Rate of Sound
Attenuation}{38} \contentsline {subsection}{\numberline
{4.2}Fermi-Liquid Effects}{50}
\contentsline {section}{\numberline {5.}Point-Contact Spectroscopy
of Layered
                                       Conductors}{55}
\contentsline {subsection}{\numberline {5.1}Point-Contact
Investigation
                                            of Electron Energy Spectrum}{55}
\contentsline {subsection}{\numberline {5.2}Resistance of
Point-Contact
                                            between Layered Conductors}{74}
\contentsline {subsection}{\numberline {5.3}Point-Contact
Spectroscopy of
                                            Electron--Phonon Interaction}{85}
\contentsline {section}{{Conclusion}}{93} \contentsline
{section}{}{} \contentsline {section}{{\it References}}{95}
\contentsline {section}{}{} \contentsline {section}{{\it
Index}}{99}
\section{INTRODUCTION}
The search for new superconducting materials has focused
attention on conductors of organic origin, which possess
layered of thread structure with a pronounced anisotropy in
the normal (non-superconducting) state. Many of them have
the metal-type electrical conductivity, i.e. their
resistance increases with increasing temperature and are
known as artificial or synthetic metals. However, the
electron properties of low-dimensional conductors differ
essentially from those of ordinary metals, and the
utilization of layered and thread conductors in different
spheres of the modern electronics appears to be more
effective than metals and semiconductors. In this
connection it is useful to make a theoretical analysis of
the electron processes proceeding in low-dimensional
conductors.

It is of interest to investigate to what extent layered
conductors are also the preferred objects for the
investigations of the Kapitza's effect.

A considerable part of organic superconductors are layered
structures, and their conductivity along the layers is
significantly higher than along the normal $\bs{n}$ to the
layers. Many layered conductors, in particular halides of
tetraseleniumtetracen and salts of tetrahiafulvalene,
exhibit the metal type conductivity even across the layers.
Thus there are grounds for making use of the concept of
quasiparticles carrying a charge $e$, analogous to
conduction electrons in metals, in order to describe
electron processes in such conductors. Evidently, sharp
anisotropy of the electrical conductivity is connected with
the anisotropy of the charge carriers velocity
$\bs{v}=\partial\varepsilon(\bs{p})/\partial p$ on the
Fermi surface $\varepsilon(\bs{p})=\varepsilon_{\rm F}$,
i.e. their energy
%1
\be
\varepsilon(\bs{p})=\sum_{n=0}^\infty
\varepsilon_n(p_x,p_y)\cos\left\{\frac{anp_z}{\hbar}\right\}
\ee
is weakly dependent on the quasi-momentum projection
$p_z=\bs{pn}$. The Fermi surface of such conductors
represents a weakly corrugated cylinder and, probably, some
more closed cavities referring to the small groups of
charge carriers.

Here $a$ is the separation between the layers, $\hbar$ is
the Planck constant, the maximum values on the Fermi
surface of the function $\varepsilon_1(p_x, p_y)$ is
$\eta\varepsilon_{\rm F}\ll\varepsilon_{\rm F}$, and of the
functions $\varepsilon_n(p_x,p_y)$ with $n\ge2$ being equal
to $A_n$, is still smaller, i.e. $A_{n+1}\ll A_n$. Such
form is characteristic of the charge carriers dispersion
low in the strong coupling approximation when the overlap
of the wave functions of electrons belonging to different
bands is negligible.

Below we consider electron phenomena in layered conductors
with the metal-type conductivity for the most general
assumptions on the form of the quasi-two-dimensional
electron energy spectrum.

%2
\section{GALVANOMAGNETIC EFFECTS}
In contrast to the case of a metal, in the layered
conductors placed in a magnetic field $\bs{H}$ both the
absence of a response to the action of the field and the
intensification of galvanomagnetic effects, characteristic
of metals, are possible.

In 1928, P.L. Kapitza observed a wonderful phenomena -- the
linear growth with a magnetic field of the resistance of
metals at liquid air and liquid carbon oxide temperature
ranges  [1]. For this purpose Kapitza created powerful magnets, in
 which the magnetic field attains 30--50 tesla. At that
time in Leiden experimental investigations at lower
temperatures were carried out, which raised the
effectiveness of less strong magnetic fields by increasing
the charge carriers mean free path. At liquid hydrogen
temperatures instead of the linear increase of the magnetic
field, the more complicated dependence of the resistance of
single bismuth crystals was observed by Shubnikov and de
Haas [2], and at helium temperatures the oscillatory
dependence of the resistance on the inverse magnetic field
-- the Shubnikov--de Haas effect -- was clearly
demonstrated [3]. The oscillatory dependence is a common
effect for metals, which is connected with the presence of
the singularities of the density of states of the charge
carriers while their energy spectrum is quantized by a
magnetic field [4]. The amplitude of the quantum
oscillations in metals is substantially less than the
amplitude of the oscillations observed in bismuth, which is
caused, by the sharp anisotropy of the Fermi surface of
bismuth-type semimetals.

The Shubnikov--de Haas effect is also clearly manifested in
the tetrathiafulvalene salts and halides of
tetraseleniumtetracen, which have a pronounced layered
structure [5--14]. A considerable increase in the
amplitude of the quantum oscillations of the
magnetoresistance in the layered conductors arises from the
quasi-two-dimensial character of their electron energy
spectrum. In metals, the Shubnikov--de Haas effect is
formed only by the small fractions of charge carriers on
the Fermi surface. These are electrons near the Fermi
surface cross-sections whose areas cut by the plane $p_{\rm
H}=\bs{pH}/H=\mbox{const}$ are close to the extreme magnitude
[15--16], or conduction electrons near the self-intersecting
orbit [17].  In contrast to metals, in the quasi-two-dimensional
conductors almost all charge carriers with the Fermi energy
contribute to the quantum oscillatory effects because at
$\eta\ll 1$ all closed cross-sections of the Fermi surface cut by
the plane $p_{\rm H}=\mbox{const}$ are almost indistinguishable.

It is also of interest to find out to what extent the
layered conductors are preferable objects for the
investigations of the Kapitza's effect as well.

The linear growth of the resistance observed by Kapitza was
not in agreement with the main principles of the electron
theory of metals, because in accordance with the Onsager
principle of symmetry of kinetic coefficients [18] the
resistance of a conductor must be an even function of a
magnetic field. The first attempt to explain the results of
the Kapitza's experiments was made only 30 years later. The
agreement with the Onsager principle of the linear increase
with $\bs{H}$ of the resistance is related to the
complicated form of the dependence of the charge carriers
energy $\varepsilon$ upon their quasimomentum $\bs{p}$. The
fundamental characteristic of the electron energy spectrum
-- the Fermi surface -- is open for almost all metals, and
in a magnetic field the orbits of electrons with the Fermi
energy $\varepsilon_{\rm F}$ can pass through a large number of
cells in the momentum space. The period of revolution of
conduction electrons on such strongly elongated orbits
$T=2\pi/\Omega$ may be greater than its free path time $\tau$ at
however strong magnetic field. As a result averaging of the
resistance of a polycrystal wire over all possible orientations
of crystallites and, consequently, over electron orbits leads to
the linear dependence of the resistance on the magnitude of a
strong magnetic field ($\Omega_0\tau\gg1$, where $\Omega_0$ is
the maximum rotational frequency of the Fermi electron in a
magnetic field) [19, 20]. If the thickness of a polycrystal
specimen of a metal f with an open Fermi surface significantly
exceeds the crystallite dimension, then in a strong magnetic
field the resistance $\rho$ is proportional to $H^{4/3}$ [21,
22], i.e.  the $H$-dependence of $\rho$ is close to the linear
one.  If there are saddle points on the Fermi surface the
frequency of revolution takes on all values within the interval
between zero and $\Omega_0$ in a single crystal. This is the case
when self-intersecting orbits, on which an electron cannot make a
total revolution, may take place.  However, a number of electrons
near the self-intersecting orbit, for which the period $T$ is
greater than the free path time, is proportional to
$\exp(-\Omega_0\tau)$, because the period as a function of
$p_{\rm H}=\bs{pH}/H$ diverges logarithmically on the
self-intersecting orbit.  As a result, in a very narrow range of
magnetic fields the complicated dependence of $\rho$ upon $H$
changes for the saturation or the quadratical growth with
increasing $H$.

In the quasi-two-dimensional conductors the period of
revolution of charge carriers in a magnetic field is weakly
dependent on the momentum projection $p_{\rm H}$, and there are
grounds to expect that in such conductors a number of
electrons near the self-intersecting orbit, for which $T>\tau$,
is essentially greater than in an ordinary metal. These
electrons, make a major contribution to conductivity in a
significantly wider range of magnetic fields ($\Omega\tau\gg1$),
and averaging of the frequencies of revolution will give another
result in comparison with the case of a metal. In order to find
out the connection between the current density
%2.1
\be
j_i=\frac{2}{(2\pi\hbar)^3}\int ev_if(\bs{p})\,\mbox{d}^3p
=\sigma_{ik}E_k
\ee
and electric field $\bs{E}$ it is necessary to solve the kinetic
equation for the charge carriers distribution function $f(\bs{p})$
%2.2
\be
\left(e\bs{E}+\frac{[\bs{v}\bs{H}e]}{c}\right)
\frac{\partial f}{\partial p} =W_{\rm col}\{f\}.
\ee

At small electric field the deviation of the
distribution function
%2.3
\be
f(\bs{p})=f_0(\varepsilon)-eE_i\psi_i(\bs{p})
\frac{\partial f_0(\varepsilon)}{\partial\varepsilon}
\ee
from the equilibrium Fermi function $f_0(\varepsilon)$ is small
and the kinetic equation (2.2) can be linearized in small
perturbation of the conduction electrons system. In this
approximation the collision integral $W_{\rm col}$ represents a
linear integral operator applying to $\psi_i$, which is the
function to be found. At low temperatures when conduction
electrons are scattered mainly by impurities and crystal defects,
we may take, with sufficient accuracy, the collision integral to
be the operator of multiplication by the collision
frequency $1/\tau$ of the unequilibrium correction to the Fermi
function $f_0(\varepsilon)$, i.e. the solution of the kinetic
equation appears to be the proper function of the integral
operator of collisions. When taking into account the other
mechanisms of dissipation, the solution of the kinetic
equation should be found in the basis of proper functions
of the collision integral operator, but a correct solution
of this complicated mathematical problem only enables us to
improve unessential numerical factors of the order of unity
and does not touch the functional behaviour of the physical
characteristics of a conductor, i.e. their dependence on
external parameters. In order to find out the dependence of
the resistance and the Hall field in the layered conductor
on the magnitude and orientation of a magnetic field, the
$\tau$-approximation for the collision integral is used. In
this approximation the kinetic equation takes the simple
enough form
%2.4
\be
\frac{e}{c}[\bs{v}\times\bs{H}]\frac{\partial\psi_i}{\partial\bs{p}}
+\frac{\psi_i}{\tau}=v_i,
\ee
and its solution
%2.5
\be
\psi_i(t,p_{\rm H},\varepsilon)=\int^{t}_{-\infty}
v_i(t',p_{\rm H},\varepsilon)\exp\left\{\frac{t'-t}{\tau}\right\}\,
\mbox{d}t'
\ee
allows us to determine the components of the electrical conductivity
tensor
%2.6
\ba
\sigma_{ik}&=&-\frac{2e^3 H}{c(2\pi\hbar)^3}
\int\mbox{d}\varepsilon\,\frac{\partial f_0}{\partial\varepsilon}\,
\int\mbox{d}p_{\rm H}\,
\int_0^T\mbox{d}t\,v_i(t)\nonumber\\
&\times&
\int^{t}_{-\infty}v_k(t')\exp\left\{\frac{t'-t}{\tau}\right\}\, \mbox{d}t'
\nonumber\\
&=&\langle e^2v_i\psi_k\rangle .
\ea
As the
variables in the momentum space we have chosen the integrals of motion
$\varepsilon$, $p_{\rm H}$ and the time $t$ of electron motion in a magnetic
field $\bs{H}=(0,H\sin\theta,H\cos\theta)$ according to the equations
%2.7
\ba
\frac{\partial p_x}{\partial t}&=&
        \frac{eH}{c}(v_y\cos\theta-v_z\sin\theta);\nonumber\\
\frac{\partial p_y}{\partial t}&=&
        -\frac{eH}{c}v_x\cos\theta;\nonumber\\
\frac{\partial p_y}{\partial t}&=&
        \frac{eH}{c}v_x\sin\theta.
\ea

\begin{figure}[h]
\centering
\includegraphics[width=10cm]{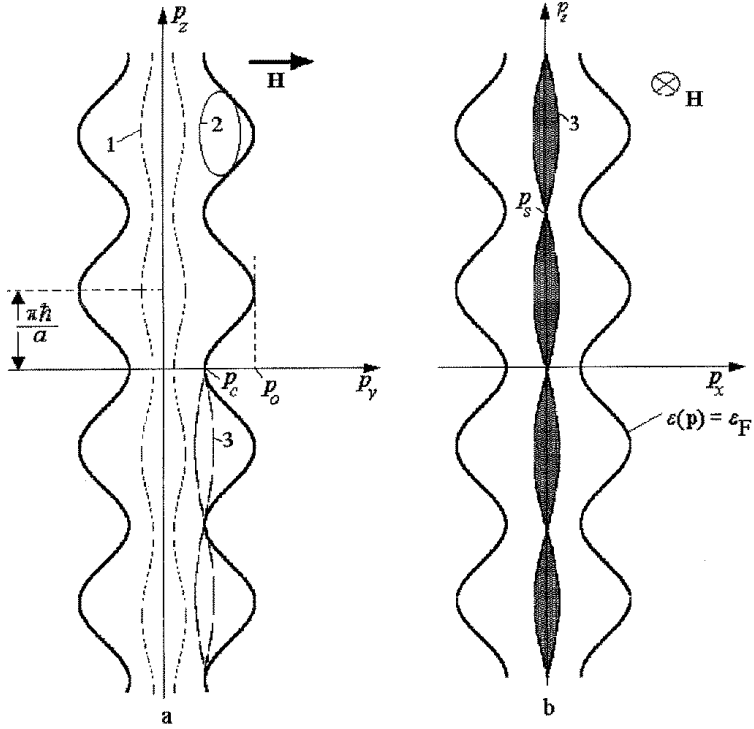}
\caption{Different types of electron trajectories in the momentum
space in a magnetic field applied parallel to the layers: {\it 1},
an open trajectory; {\it 2}, a closed orbit; {\it 3}, the
self-intersecting orbit containing the saddle points $\bs{p}_{\rm
s}$. The cross-section $p_y=p_{\rm c}$ separates the region of
open cross-sections from the closed ones.}
\end{figure}

\begin{figure}
\centering
\includegraphics[width=10cm]{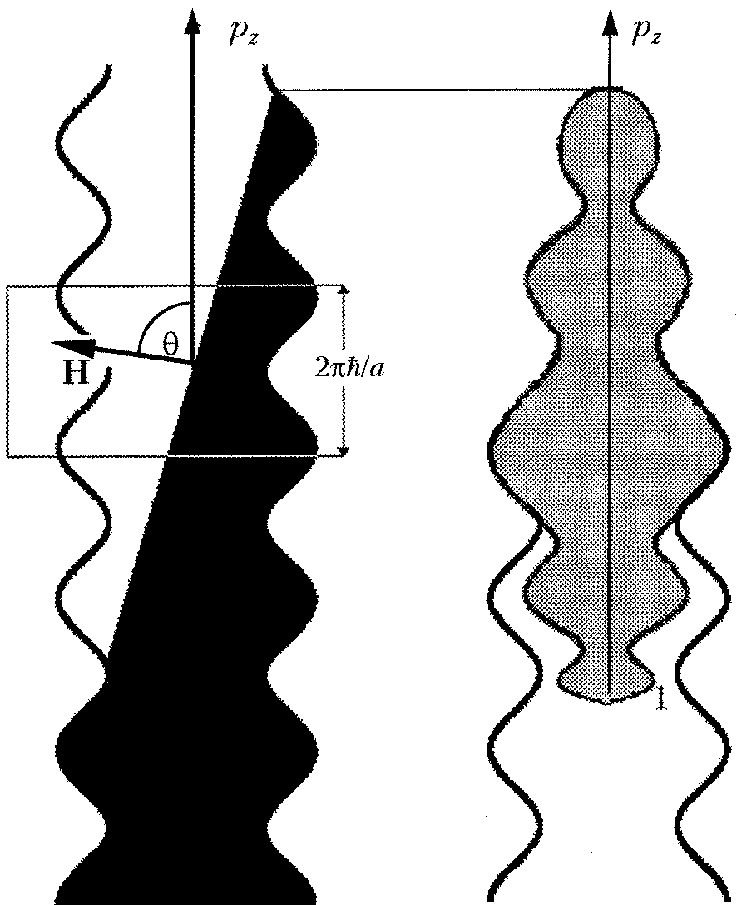}
\caption{Electron trajectories on the Fermi surface in an oblique
magnetic field ($\theta$ is the angle of deviation of the vector
$\bs{H}$ from the layers-plane). An electron orbit ({\it 1})
passes through a few number of cells in the momentum space.}
\end{figure}

Suppose the Fermi surface of the layered conductor with the
quasi-two-dimensional energy spectrum to be only one weakly
corrugated cylinder with the direction of ``openness'' aligned
with the $p_z$-axis \linebreak[10000] (Fig.~1). If $\theta$
differs from $\pi/2$, all cross-sections of the weakly corrugated
cylinder cut by the plane $\bs{pH}=\mbox{const}$ are closed and
almost indistinguishable for $\eta\ll 1$ (Fig.~2).  The angle
$\theta$ approaching $\pi/2$, closed electron orbit becomes
strongly elongated, and at $\theta=\pi/2$ transforms into two open
orbits (Fig.~1). The period of motion changes discontinually and
on the open cross-section cut by the plane $p_{\rm
H}=p_y=\mbox{const}$, takes the form
%2.8
\be T=\frac{2\pi m^*c}{eH}
=\frac{c}{eH}\int\limits_0^{2\pi\hbar/a}\frac{\mbox{d}p_z}{v_x} .
\ee

When
approaching the limiting cross-section $p_y=p_{\rm c}$, which separates the
region of open cross-sections from the small layer of closed ones, the
period of motion $T(p_y)$ increases without limit. This results from the
fact that the cross-section $p_y=p_{\rm c}$ contains the saddle points
$\bs{p}_{\rm c}=(0,p_{\rm c}, 2\pi\hbar n/a)$, where open orbits touch one another,
and electrons make a long stay near the points of self-intersection because
their velocity in the plane orthogonal to $\bs{H}$ is negligible
(Fig.~l{\it b}).

At $\theta=\pi/2$ the mean value of the velocity $v_x$ differs from zero
%2.9
\be
{\bar v}_x=T^{-1}\int\limits_{0}^{T}\mbox{d}t\,v_x(t)
=\frac{2\pi\hbar ec}{aHT} =\frac{h}{am^*},
\ee
and all directions of drift the charge carriers cover the whole
$xy$-plane. In this case the components of the electrical conductivity
tensor $\sigma_{xx}$ and $\sigma_{yy}$ coincide in order of magnitude with
the conductivity along the layers in the absence of a magnetic field.
However, at $\eta\ll 1$ the components of the tensor $\sigma_{ij}$ with one
or two $z$ indices are incredible small and in a strong magnetic field they
also decrease with increasing $H$ because ${\bar v}_z=0$.

Being determined up to terms proportional to $\eta$, the period
of motion along an orbit sufficiently distant from the
self-intersecting one, is inversely proportional to $v_x(0)$ (we
have taken the central cross-section $p_z=0$ as the origin of
the variable $t$). When an electron approaches the
self-intersecting orbit $p_y=p_{\rm c}$, its velocity along the $x$-axis
decreases and allowance for small corrections in $\eta$ becomes
necessary. The period of motion of electrons along the
orbit on which $p_{\rm H}$ is close to $p_{\rm c}$ is very large, since
electrons stay for a long time near the saddle point
$\bs{p}_{\rm c}=(0,p_{\rm c},0)$, where $v_x=v_z=0$ (see Fig.~l{\it b}). In
the vicinity of the cross-section $p_{\rm H}=p_{\rm c}$ the electron
velocity projection $v_x$ is a complicated function of $t$, but far from
this cross-section the small corrections in $\eta$ depending on $t$ can be
omitted in the expression for $v_x(t)$.  The period of charge carriers
motion has the form
%2.10
\be
T(p_{\rm H})=\frac{2\pi\hbar c}{aeHv_y(0)}=\frac{2\pi v_{\rm
F}}{\Omega_0v_x(0)}
\ee
and at $v_x\ll v_{\rm F}=\varepsilon_{\rm F}a/\hbar$ may become comparable
with the free path time. As a result the charge carriers with the small
value of the velocity projection $v_x$ give the major contribution into
$\sigma_{iz}$ and $\sigma_{zi}$.  Regardless of the small corrections
depending on $t$, $p_z$ is the linear function of the time of motion in a
magnetic field, and the velocity projection
%2.11
\be
v_z=-\sum\limits_{n=1}^{\infty}
        \frac{an}{\hbar}\varepsilon_n(p_x,p_y)\sin(n\Omega t)
\ee
is determined mainly by the first term in Equation (2.11).
It is not difficult to calculate the conductivity component
$\sigma_{zz}$ in this case:
%2.12
\be
\sigma_{zz}=\frac{2e^2\tau}{(2\pi\hbar)^3)}
\int2\pi m^*\,\mbox{d}p_{\rm H}\,
\sum\limits_{n=1}^{\infty}\frac{\{\varepsilon_n(p_x,p_y)an/\hbar^2\}}{1+(n\Omega\tau)^2}.
\ee

In the vicinity of the cross-section $p_{\rm H}=p_{\rm c}$ the numerators in
Equation (2.12) should be improved by changing them for $|v_z^n|^2$.
Since for these charge carriers $p_z$ is a complicated
function of $t$, the Fourier transforms of the velocities
$v_z^n$ need not be decreasing with $n$, and the contribution in
the asymptotic value of $\sigma_{zz}$ from electrons belonging to the
vicinity of the saddle point is not limited only by the
first harmonic in the Fourier expansion of the function $v_z(t)$.
At the limiting cross-section $p_y=p_{\rm c}$ the maximum value of the
velocity of electron motion along the $x$-axis is equal in
the order of magnitude to $\eta^{1/2}v_{\rm F}$ and $t$-dependence of $v_x$
is of no importance only at $v_x(0)\gg\eta^{1/2}v_{\rm F}$.

Let $\sigma_{zz}$ be represented in the form
$\sigma_{zz}=\sigma_{zz}^{(1)}+\sigma_{zz}^{(2)}+\sigma_{zz}^{(3)}$, where
$\sigma_{zz}^{(1)}$ describes the contribution of the charge carriers whose
velocity $v_z$ is given by Equation~(2.11), the second term is the
contribution to $\sigma_{zz}$ from conduction electrons on the open orbits
for which $v_x\le\eta^{1/2}v_{\rm F}$, and the last term
%2.13)
\be
\sigma_{zz}^{(3)}=\frac{4e^2\tau}{(2\pi\hbar)^3}
\int\limits_{p_{\rm c}}^{p_0}\mbox{d}p_{\rm H}\,2\pi m^*(p_{\rm H})
\sum\limits_{m=1}^{\infty}\frac{|v_z^n|^2}{1+(n\Omega\tau)^2}
\ee
is connected with the charge carriers on the closed orbits.

Here $p_0$ is the maximum value of $p_{\rm H}$ in the Fermi surface
reference point $\bs{p}_0=(0,p_0,\pi h/a)$ (Fig.~l{\it a}).

At $\gamma_0=1/\Omega_0\tau>\eta^{1/2}$ the integration in $p_{\rm H}$ over
a small interval where $\eta^{1/2}v_{\rm F}<v_x\ll v_{\rm F}$ leads to the
following result
%2.14
\be
\sigma_{zz}^{(1)}=\sigma_0\eta^2\gamma_0,
\ee
where $\sigma_0$ is of the order of the conductivity along the
layers in the absence of a magnetic field.

Near the reference point of the Fermi surface the charge
carriers cyclotron mass $m^*$ is proportional to $\eta^{-1/2}$ and
increases when approaching to the cross-section $p_{\rm H}=p_{\rm c}$, on
which it becomes infinite. At $\eta^{1/2}\ll\gamma_0$ conduction electrons
on the closed cross-sections have no time to make a total revolution along
the orbit and give the following contribution into $\sigma_{zz}$:
%2.15)
\be
\sigma_{zz}^{(3)}=\frac{4e^2\tau}{(2\pi\hbar)^3}
\int\limits_{p_{\rm c}}^{p_0}2\pi m^*(p_{\rm H})\,\mbox{d}p_{\rm H}\,
{\bar v}_z^2=\sigma_0\eta^{5/2}.
\ee

Here and below we shall omit unessential factors of the
order of unity in formulas for $\sigma_{zz}$.

Thus, at $\eta^{1/2}\ll\gamma_0\ll1$ a small fraction of conduction
electrons on the open orbits moving slowly along the $x$-axis gives the main
contribution to $\sigma_{zz}$.

A number of conduction electrons for which $T>\tau$ decreases
with increasing magnetic field, and the contribution to $\sigma_{zz}$
of the charge carriers on the orbits close to the cross-section
$p_{\rm H}=p_{\rm c}$ becomes essential. Near the self-intersecting
orbit the charge carriers velocity projection $v_x$ is small,
i.e. their energy is weakly dependent on $p_x$, and we can
easily find the $p_{\rm H}$-dependence of the period of motion from
the expansion of the energy in the powers of $p_x$. We retain
only the first two terms in Equation (1.1), viz.
%2.16
\be
\varepsilon(\bs{p})=\varepsilon(0,p_y)+\frac{p_x^2}{2m_1}
+\varepsilon_1(0,p_y)\cos\left(\frac{ap_z}{\hbar}\right).
\ee

Using Equation (2.7) we obtain
%2.17
\be
T(p_{\rm H})=
\frac{\hbar c}{aeH}\left\{\frac{m_1}{\varepsilon_1(0,p_y)}\right\}^{1/2}
\int\limits_{0}^{\pi}\mbox{d}\alpha\,(\xi^2+\sin^2\alpha)^{1/2},
\ee
where
$$
\xi^2=\frac{\varepsilon_0(0,p_{\rm c})-\varepsilon(0,p_y)
+\varepsilon_1(0,p_{\rm c})-\varepsilon_1(0,p_y)}{2\varepsilon_1(0,p_y)}.
$$

At $\xi\ll1$ the period of motion of charge carriers
%2.18)
\be
T(p_{\rm H})\cong\Omega_0^{-1}\eta^{-1/2}\ln\left(\frac{1}{\xi}\right)
\ee
diverges logarithmically, and the contribution into $\sigma_{zz}$ of
electrons belonging to the small vicinity (of the order of
$\Delta p_{\rm H}=p_{\rm c}-p_{\rm H}\cong p_{\rm c}\eta$) of the limiting
cross-section has the form
%2.19
\be
\sigma_{zz}^{(2)}=\sigma_0\eta^{5/2}\gamma_0^2
\int\limits_{\infty}^{1}\mbox{d}u\,
\frac{u^3\exp\{-u\}}{u^2\gamma_0^2+\eta}
\cong \sigma_0\frac{\gamma_0^2\eta^{5/2}}{\gamma_0^2+\eta}.
\ee

In ordinary metals the period of motion of charge carriers
is greater or comparable with the free path time only in a
small region (about $\exp[-\Omega_0\tau]$) of the Fermi surface
self-intersecting cross-section. In contrast to the case of a
metal, in the quasi-two-dimensional conductors the
condition $T\ge \tau$ is satisfied in a wider range of electron
orbits, where $\xi$ can be of the order of unity.

At $\eta^{1/2}\ll\gamma_0$ the contribution into $\sigma_{zz}$ of charge
carriers in the vicinity of the self-intersecting cross-section of the order
of $p_0\eta$ is very incredibly small and
$\sigma_{zz}\cong\sigma_{zz}^{(1)}$, whereas in the opposite case these
charge carriers contribute to conductivity on a level with other conduction
electrons. It is easy to determine that at $\eta>\gamma_0^2$ the
conductivity $\sigma_{zz}^{(1)}$ also decreases proportionally to
$\gamma_0^2$ as a magnetic field increases. As a result, in the range of
strong magnetic fields when $\gamma_0\le\eta^{1/2}$, we have
%2.20
\be
\sigma_{zz}\cong\sigma_0\eta^{3/2}\gamma_0^2.
\ee

In the layered conductors the Hall field also behave
differently to metals. In order to demonstrate the results
obtained more visibly, we consider the galvanomagnetic
phenomena using the simple model of the charge carriers
dispersion law
%2.21)
\be
\varepsilon(\bs{p})=\frac{p_x^2+p_y^2}{2m}
-\eta\frac{v_{\rm F}\hbar}{a}\cos\left(\frac{ap_z}{\hbar}\right).
\ee
This is the approximation at which charge carriers are
assumed to be almost free in the layers-plane. Since the
main contribution to the electrical conductivity across the
layers is given by electrons with small $v_x$-velocity and the
dependence of $\varepsilon_n(p_x,p_y)$ on $p_y$ is nonessential at
$\theta=\pi/2$, the analysis of the galvanomagnetic effects given below
applies.

Making use of the equation of charge carriers motion in a
magnetic field (2.7) at $\theta=\pi/2$ and of the dispersion law
(2.21), we have
%2.22
\be
\psi_z=\gamma(\psi_x-v_x\tau),
\ee
where $\gamma=mc/eH\tau$. From this it follows that
%2.23)
\be
\sigma_{zz}=\gamma\sigma_{zx};\quad
\sigma_{xz}=\gamma(\sigma_{xx}-\sigma_0);\quad
\sigma_{zx}=-\sigma_{xz},
\ee
and the matrix $\sigma_{ij}$ has the form
%2.24
\be
\sigma_{ij}=\left(
\begin{array}{ccc}
\sigma_0-\gamma^{-2}\sigma_{zz}&0&-\gamma^{-1}\sigma_{zz}\\
0&\sigma_0&0\\
\gamma^{-1}\sigma_{zz}&0&\sigma_{zz}
\end{array}
\right).
\ee
For the inverse tensor of the resistance we have
%2.25)
\vspace{6pt}
\be
\rho_{ij}=\left(
\begin{array}{ccc}
\sigma_0^{-1}&0&-(\gamma\sigma_0)^{-1}\\
0&\sigma_0^{-1}&0\\
(\gamma\sigma_0)^{-1}&0&\sigma_{zz}^{-1}-\sigma_0^{-1}\gamma^{-2}
\end{array}
\right).
\vspace{6pt}
\ee

It can be seen that the resistance along the normal to the
layers $\rho_{zz}$ to a good accuracy is equal to $1/\sigma_{zz}$ and grows
linearly with a magnetic field at $\eta^{1/2}\ll\gamma\ll1$. The Hall field
\vspace{5pt}
%2.26)
\be
\bs{E}_{\rm Hall}=R[\bs{j}\times\bs{H}]
\vspace{5pt}
\ee
is also proportional to $H$, and the Hall constant $R$ is
inversely proportional to the volume within the Fermi
surface. In metals, the Hall constant is of the same form
only in the absence of the Fermi surface open
cross-sections.

{\looseness=3
The absence of the magnetoresistance $\Delta\rho=\rho(H)-\rho(0)$
for the
current directed along the layers is connected with the quadratical
dispersion of the charge carriers in the $xy$-plane. For
more complicated dependence of the charge carriers energy
on $p_x$ and $p_y$ the resistance grows with increasing $H$ and
tends to a finite value in a high magnetic field, as
demonstrated in metals. However, the opposite occur in case
of a metal, and the quasi-two-dimensional conductors $\Delta\rho$ is
very small and disappears at $\eta=0$. This follows from the fact
that the only projection onto the normal to the layers of a
magnetic field, which disappears at $\theta=\pi/2$, fundamentally
affects the charge carriers dynamics.

}

For $(\pi/2)\le\eta$ the conductivity along the layers is similar to that
in the absence of a magnetic field, and the Equations
(2.14) and (2.20) are valid for $\sigma_{zz}$ until $\gamma_0\ge\eta$. But
at $(\pi/2-\theta)\gg\eta$ there are no self-intersecting orbits, and in
actually attained strong magnetic fields the in-plane resistance tends to a
finite value in a large range of angles $\theta$ of deviation of the field
from the layers-plane.

Using Equation (2.21) for the cargo carriers dispersion
law, we have for the conductivity tensor

%2.27)
\be
\sigma_{ij}=\left(
\begin{array}{ccc}
\displaystyle\frac{\sigma_{xy}\gamma_0}{\cos\theta}
&\sigma_{xy}
&\displaystyle-\frac{\sigma_{zz}\gamma_0\sin\theta}{\gamma_0^2+\cos^2\theta}
\\[8pt]
-\sigma_{xy}
&\displaystyle\sigma_0-\frac{\sigma_{xy}\cos\theta}{\gamma_0}
&\displaystyle\frac{\sigma_{zz}\sin\theta\cos\theta}{\gamma_0^2+\cos^2\theta}
\\[8pt]
\displaystyle\frac{\sigma_{zz}\gamma_0\sin\theta}{\gamma_0^2+\cos^2\theta}
&\displaystyle\frac{\sigma_{zz}\sin\theta\cos\theta}{\gamma_0^2+\cos^2\theta}
&\sigma_{zz}
\end{array}
\right);
\ee
and for the resistance tensor
%2.28)
\be
\rho_{ij}=\left(
\begin{array}{ccc}
\sigma_0^{-1}
&\displaystyle\frac{H\cos\theta}{Nec}
&\displaystyle-\frac{H\sin\theta}{Nec}
\\[8pt]
\displaystyle-\frac{H\cos\theta}{Nec}
&\sigma_0^{-1}
&0
\\[8pt]
\displaystyle\frac{H\sin\theta}{Nec}
&0
&\displaystyle
\sigma_{zz}^{-1}-\frac{\sin^2\theta}{\sigma_0(\gamma^2+\cos^2\theta)}
\end{array}
\right);
\ee
where
$$
\sigma_{xy}=\frac{\gamma_0\cos\theta}{(\gamma_0^2+\cos^2\theta)^2}
[\sigma_0(\gamma_0^2+\cos^2\theta)-\sigma_{zz}\sin\theta\cos^2\theta],
$$
$N$, is the charge carriers density.

The matrix $\rho_{ij}$ given above is valid at any value of a
magnetic field (including weak fields), the Hall constant
being equal to $1/Nec$ for an arbitrary orientation with
respect to the layers of both the magnetic field and
electric current.

For an arbitrary charge carriers dispersion law the
magnetoresistance for the current, coplanar with the
layers, differs from zero, as it occurs at $\theta=\pi/2$, and its
magnitude depends on the angle of deviation of a magnetic
field from the layers-plane. The magnetoresistance
increases with increasing $\theta$ and becomes comparable with
that in the absence of a magnetic field. On the contrary,
the resistance of the layered conductor along the ``hard''
direction, i.e. along the normal to the layers, is very
sensitive to the orientation of a magnetic field, and for
small $\eta$ its asymptotic value may change essentially at
some values of $\theta$. This follows from the fact that the
velocity of the charge carriers drift along the normal to
the layers ${\bar v}_z(p_{\rm H},\theta)$ disappears not only at
$\theta=\pi/2$, but also at an infinite number of the values
$\theta=\theta_{\rm c}$.  On the central cross-section of the Fermi surface
cut by the plane $p_{\rm H}=0$ the charge carriers velocity averaged over
the period disappears, since
%2.29)
\be
{\bar v}_z(0,\theta)=\sum\limits_{n=1}^{\infty}\frac{an}{\hbar T}
\int\limits_{0}^{T}\mbox{d}t\,\varepsilon_n(t,0)
\sin\left(\frac{anp_y(t,0)\tan\theta}{h}\right)=0.
\ee
When $\theta$ differs from zero essentially there always is such
a value $\theta=\theta_{\rm c}$, and not a single one, at which
%2.30
\be
\sum\limits_{n=1}^{\infty}\frac{an}{\hbar T}
\int\limits_{0}^{T}\mbox{d}t\,\varepsilon_n(t,0)
\cos\left(\frac{anp_y(t,0)\tan\theta_{\rm c}}{h}\right)=0.
\ee
and near the central cross-section the expansion of ${\bar v}_z$
starts from cube terms in $p_{\rm H}$.  Since for $\eta\ll1$ the charge
carriers velocity is weakly dependent on $p_{\rm H}$, at
$\theta=\theta_{\rm c}$ the expansion of the conductivity tensor components
$\sigma_{iz}$ and $\sigma_{zj}$ in a power series in the small parameters
$\eta$ and $\gamma_0/\cos\theta$ start with terms of higher order than at
$\theta\neq\theta_{\rm c}$. It is easy to make sure that the expansion in a
power series in $\eta$ however small of the components $\sigma_{iz}$ and
$\sigma_{zj}$ start with terms of the second or higher order.  This results
from the fact that for $\cos\theta\gg\eta$ not only the velocity, but also
the momentum projection $p_i(t,p_{\rm H}=p_i(t)+\Delta p+i(t,p_{\rm H})$ are
weakly dependent of $p_{\rm H}$.

When calculating the asymptotic value of $\sigma_{zz}$
%2.31
\ba
\sigma_{zz}(\eta, H)&=&\frac{2e^2H}{c(2\pi\hbar)^3}
\int\limits_{0}^{2\pi\hbar\sin\theta/a}\mbox{d}p_{\rm H}\,
\Biggl[
1-\exp\left(-\frac{T}{\tau}\right)^{-1}
\nonumber\\
&\times&
\int\limits_{0}^{T}\mbox{d}t\,
\int\limits_{t-T}^{t}\mbox{d}t'\,
\sum\limits_{n,m}\varepsilon_n(t,p_{\rm H})\varepsilon_m(t',p_{\rm H})
\nonumber\\
&\times&
\sin\left(
an\frac{p_{\rm H}/\sin\theta-p_y(t,p_{\rm H})\cot\theta}{\hbar}
\right)
\nonumber\\
&\times&
\sin\left(
am\frac{p_{\rm H}/\sin\theta-p_y(t',p_{\rm H})\cot\theta}{\hbar}
\right)
\Biggr]\exp\left(\frac{t'-t}{\tau}\right)
\nonumber\\
\ea
we may omit $\Delta p_i$ in Equation (2.33), if we confine ourselves
to quadratical in $\eta$ terms.  This gives
%2.32
\ba
\sigma_{zz}&=&\sum\limits_{n=1}^{\infty}\int\limits_{0}^{T}\mbox{d}t\,
\int\limits_{t}^{-\infty}\mbox{d}t'\,\left(\frac{an}{\hbar}\right)^2
\varepsilon_n(t)\varepsilon_n(t')\exp\left(\frac{t-t'}{\tau}\right)
\nonumber\\
&\times&\frac{e^3H\cos\theta}{ac(2\pi\hbar)^2}
\cos\left\{\frac{an}{\hbar}(p_y(t)-p_y(t'))\tan\theta\right\};
\ea
where all functions under the integral sign depend on $t$
and $t'$.

As a result, for small $\eta/\cos\theta$ and
$\gamma_0\ll\cos\theta$ the component $\sigma_{zz}$ takes the
form
%2.33)
\ba
\sigma_{zz}&=&\frac{e^2\tau am^*\cos\theta}{8\pi^3\hbar^4}
\sum\limits_{n=1}^{\infty}n^2|I_n(\theta)|^2
\nonumber\\
&+& \sigma_0\eta^2
\left\{\eta^2 f_1(\theta)
+\left(\frac{\gamma_0}{\cos\theta}\right)^2f_2(\theta)\right\};
\ea
where
%2.34
\be
I_n(\theta)=T^{-1}\int\limits_{0}^{T}\mbox{d}t\,\varepsilon_n(t)
\exp\left\{\frac{ian}{\hbar}p_y(t)\tan\theta\right\},
\ee
$f_1(\theta)$ and $f_2(\theta)$ are about unit and depend on the concrete
form of the charge carriers dispersion law. Reference to them is essential
only for those $\theta=\theta_{\rm c}$, at which in the sum over $n$ the
main term $I_1(\theta)$ disappears.

For $\tan\theta\gg 1$ the expression under the integral sign in Equation
(2.34) is a rapidly oscillating function, and $I_n(\theta)$ can be
calculated easily by means of the stationary phase method.
If there are only two points of stationary phase, where $v_x$
disappears, the asymptotic value of $I_n$ takes the form
%2.35
\ba
I_n(\theta)&=&T^{-1}\varepsilon_n(t_1)
\left|\frac{2\pi\hbar c}{anv'_y(t_1)\tan\theta}\right|^{1/2}
\nonumber\\
&\times&
\cos\left\{\frac{anD_p\tan\theta}{2h}-\frac{\pi}{4}\right\}.
\ea
Here $D_p$ is the diameter of the Fermi surface along the
$p_y$-axis, the prime denotes differentiation with respect to
$t$ in the stationary phase point, where $v_x(t_1)=0$.

As it follows from Equation (2.35), zeros of the function
$I_1(\theta)$, repeat periodically with the period
%2.36)
\be
\Delta(\tan\theta)=\frac{2\pi\hbar}{aD_p}.
\ee
When the current is directed along the normal to the
layers, the resistance of the layered conductor is
determined mainly by the component $\sigma_{zz}$, i.e.
$\rho_{zz}\cong\sigma_{zz}^{-1}$, and the Fermi surface diameter can be
found by measuring the period of the angular oscillations of the
magnetoresistance.  Changing the orientation of a magnetic field in the
$xy$-plane enables the anisotropy of the Fermi surface diameters to be
determined. The possibility of studying the diameters anisotropy in the
layered conductors is due to the presence of strongly elongated orbits,
which pass through a great number of cells in the momentum space.

If the terms in the sum over $n$ in Equation (2.32)
decrease rapidly with $n$ (so that $I_n(\theta)$ with $n\ge 2$ is less than
$v_{\rm F}n\gamma_0/\sin\theta$), then at $\theta=\theta_{\rm c}$
and $\eta<\gamma_0/\cos\theta\ll 1$ the resistance along the normal to the
layers grows quadratically with $H$ and tends to a finite value about
$\sigma_0^{-1}\eta^{-4}$ only in the range of stronger magnetic fields when
$\gamma_0\ll\eta\cos\theta$.

The $\tau$-approximation used for the collision integral is
applicable to the analysis of the galvanomagnetic phenomena
in the layered conductors with the quasi-two-dimensional
electron energy spectrum of the tetrathiafulvalene salts
type, because it does not contradict various experimental
results. The measured angular dependence of the
magnetoresistance [5, 6, 11] verify convincingly the
existence of the orientation effect -- the essential
alteration of the asymptotic behaviour of the
magnetoresistance along the normal to the layers for
certain orientations of a magnetic field about the layers.

In the layered high-temperature conductors on the basis of
oxi\-cup\-ra\-tes the free path lengths are not great and
realization of the case of a strong magnetic field ($\gamma\ll 1$) is
faced with difficulties. In a weak magnetic field ($\gamma\gg 1$) the
role of different mechanisms of the electron relaxation in
the magnetoresistance is more essential than their dynamics
in a magnetic field. In order to interpret the measured
anomalies of the magnetoresistance of bismuth
high-temperature superconductors (the nonmonotonical
temperature dependence of the magnetoresistance, the
negative magnetoresistance along the normal to the layers),
the more correct account of the collision integral is
necessary.

\begin{figure}
\centering
\includegraphics[width=10cm]{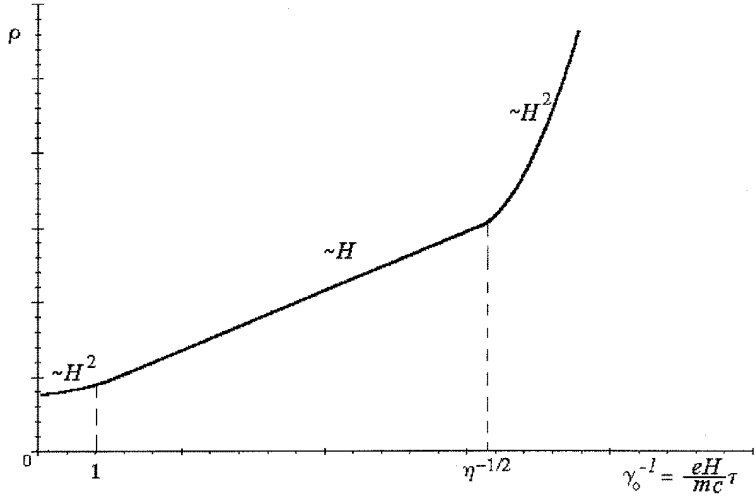}
\caption{The dependence of the transverse resistance
($\bs{H}\bot\bs{j}$) along the normal to the layers on a magnetic
field.}
\end{figure}

For $\eta^{1/2}<\gamma_0\ll1$ the free path time is not contained in the
asymptotic expression for the magnetoresistance, and the
problem of the collision integral does not arise. When it
is possible to obtain more perfect single crystals of the
high-temperature metal-oxide conductors, for which the
condition of the strong magnetic field can be satisfied,
then at $\bs{H}\bot\bs{j}$ the linear growth with $H$ of their
magnetoresistance will be expected (Fig.~3).

%3.
\section{PROPAGATION OF ELECTROMAGNETIC WAVES IN \newline
LAYERED CONDUCTORS}
The depth of penetration of an electromagnetic wave into
the layered conductor with the quasi-two-dimensional
electron energy spectrum depends essentially on the
polarization of the incident wave. A linearly polarized
wave whose electric field is aligned with the normal to the
layers can penetrate {\it a priori} to a greater depth than
a wave with the electric vector coplanar with the layers.
Under the normal skin-effect conditions, when the skin
layer depth is much more than the charge carriers free path
length $l$, the attenuation depth $\delta_\|$ for the electric field
$E_z(\bs{r})$ is $1/\eta$ times greater than the skin depth $\delta_\bot$
for the electric field along the layers
%3.1
\be
\delta_\bot=\delta_\|\eta.
\ee
In the case of the anomalous skin effect, when the skin
layer depth $\delta_\|$ is much less than $l$, the relation between
$\delta_\bot$ and $\delta_\|$ has the form
%3.2)
\be
\delta_\bot=\delta_\|\eta^{3/2}.
\ee
In a magnetic field, especially under the anomalous
skin-effect conditions, the relations between $\delta_\bot$ and $\delta_\|$
are more variable.

We consider the propagation of electromagnetic waves in the
half space $x\ge0$ occupied by the layered conductor in a
magnetic field $\bs{H}=(H\sin\phi,H\cos\phi\sin\theta,H\cos\phi\cos\theta)$,
where $\phi$ is the angle of deviation of a magnetic field from the sample
surface $x_{\rm s}=0$.

The complete set of equations representing the problem
consists of the Maxwell equations
%3.3 3.4)
\ba
(\nabla\times\bs{H})&=&-i\omega\bs{E}+\frac{4\pi\bs{j}}{c};\\
(\nabla\times\bs{E})&=&i\omega\bs{B};\nonumber\\
\bs{B}&=&\bs{H}+4\pi\bs{M};
\ea
and the kinetic equation for the charge carriers
distribution function
$f(\bs{p},x,t)=f_0(\varepsilon)-\psi(\bs{p},x)\exp\{-i\omega t\} \partial
f_0(\varepsilon)/\partial\varepsilon$:
%3.5
\be
v_x\frac{\partial\psi}{\partial x}
+\frac{e}{c}(\bs{v}\times\bs{H})\frac{\partial\psi}{\partial\bs{p}}
+\left(\frac{1}{\tau}-i\omega\right)\psi
=e\bs{vE}(x);
\ee
the solution of which allows us to determine the relation
between the current density and the electric field of the
wave.

Here $\bs{M}$ is the magnetization of a conductor. Usually the
magnetic susceptibility $\chi_{ij}=\partial M_i/\partial B_j$ of nonmagnetic
conductors is negligible, and below we shall not distinguish between the
magnetic field and the magnetic induction $\bs{B}$, except for some special
cases, when the de Haas--van Alfven effect is most pronounced and results in
the appearance of diamagnetic domains [24, 25]. The perturbation of the
charge carriers system caused by the electromagnetic wave is supposed to be
sufficiently weak, and here we confine our consideration to a linear
approximation in a weak electric field. The Maxwell equations in this
approximation become linear and we can also suppose the electromagnetic wave
to be monochromatic with the frequency $\omega$, without loss of the
generality.  This follows from the fact that the solution of the problem in
the case of an arbitrary time-dependence of the fields presents the
superposition of the solutions for different harmonics. For this reason the
time differentiation in the Maxwell equations (3.3) and (3.4) is equivalent
to multiplication by $(-i\omega t)$. Below $t$ will indicate the time of
the charge motion in a magnetic field according to Equations (2.9).

The kinetic equation (3.5) should be supplemented with the
boundary condition allowing for the scattering of charge
carriers by the sample surface $x_{\rm s}=0$
%3.6
\ba
\psi(\bs{p}_+,0)&=&q(\bs{p}_-)\psi(\bs{p}_-,0)\nonumber\\
&\times&
\int\mbox{d}^3p\,W(\bs{p},\bs{p}_+)\{1-\Theta[v_x(\bs{p})]\}\psi(\bs{p},0);
\ea
where the sample surface specularity parameter $q(\bs{p})$ is the
probability for a conduction electron incident onto the
surface $x_{\rm s}=0$ to be reflected specularly with momentum $\bs{p}$. The
specularity parameter is related to the scattering
indicatrix $W(\bs{p},\bs{p}_+)$ through the expression
%3.7
\be
q(\bs{p}_-)=1-\int\mbox{d}^3p\,W(\bs{p},\bs{p}_+)\{1-\Theta[v_x(\bs{p})]\};
\ee
$\Theta(\zeta)$ is the Heaviside function and the momenta $\bs{p}_-$ and
$\bs{p}_+$ (of incident and scattered electrons, respectively) are related
by the specular reflection condition, which conserves energy and momentum
projection on the sample surface.

The integral term in the boundary condition (3.6) ensures
no current through the surface. However, in the range of
high frequencies $\omega$ the solution to the kinetic equation
depends weakly on this functional of the scattering
indicatrix and, without reference to it, has the form
%3.8)
\ba
\psi(t_{\rm H},p_{\rm H},x)&=&\int\limits_{\lambda}^{t_{\rm H}}\mbox{d}t\,
e\bs{v}(t,p_{\rm H})\bs{E}(x(t,p_{\rm H})-x(\lambda,p_{\rm H}))
\nonumber\\
&\times&
\exp\{\nu(t-t_{\rm H})\}
+\frac{q(\lambda,p_{\rm H})}{1-q(\lambda,p_{\rm H})\exp\{\nu(2\lambda-T)\}}
\nonumber\\
&\times&
\int\limits_{\lambda}^{T-\lambda}\mbox{d}t\,
e\bs{v}(t,p_{\rm H})\bs{E}(x(t,p_{\rm H})-x(\lambda,p_{\rm H}))
\nonumber\\
&\times&
\exp\{\nu(t-t_{\rm H}+2\lambda-T)\};
\ea
where $\nu=-i\omega+1/\tau$, and $\lambda$ is the nearest to $t_{\rm H}$
root of the equation
%3.9)
\be
x(t,p_{\rm H})-x(\lambda,p_{\rm H})=
\int\limits_{\lambda}^{t}v_x(t',p_{\rm H})\,\mbox{d}t'=x;
\ee

For electrons that do not collide with the specimen
boundary, i.e. at $\{x(t_{\rm H},p_{\rm H})-x_{\rm min}\}<x$, one should put
$\lambda=-\infty$.

After several collisions with the surface $x_{\rm s}=0$ in an oblique
magnetic field electrons either move into the bulk of the
conductor or tend to approach the surface. The relative
number of electrons is not large and they do not contribute
markedly to an alternating current at $\phi\cong 1$. The contribution
of the remaining electrons, naturally, depends on the
nature of their interaction with the surface, but the state
of the surface influences unessential numerical factors of
the order of unity in the expression for the surface
impedance.

Following Reuter and Sondheimer [26], we continue the
electric field and the current density in an even manner to
the region of negative values of $x$, and apply a Fourier
transformation, viz.
%3.10
\ba
\bs{E}(k)&=&2\int\limits_{0}^{\infty}\mbox{d}x\,\bs{E}(x)\cos kx;
\nonumber\\
\bs{j}(k)&=&2\int\limits_{0}^{\infty}\mbox{d}x\,\bs{j}(x)\cos kx.
\ea
As a result, the Maxwell equations after the exclusion of
the magnetic field of the wave takes the form
%3.11)
\be
\left\{k^2-\frac{\omega^2}{c^2}\right\}E_\alpha(k)
-\frac{4\pi i\omega j_\alpha(k)}{c^2}=-2E_\alpha'(0);\quad
\alpha=(y,z),
\ee
where the prime denotes differentiation with respect to
$x$.

The solution of the kinetic equation (3.8) allows us to
find the relation between the Fourier transforms of the
electric field and the current density.
%3.12)
\be
j_i(k)=\sigma_{ij}(k)E_j(k)+
\int\mbox{d}k'\,Q_{ij}(k,k')E_j(k');
\ee
where
%3.13)
\ba
\sigma_{ij}(k)&=&\frac{2e^3H}{c(2\pi\hbar)^3}
\int\mbox{d}p_{\rm H}
\int\limits_{0}^{T}\mbox{d}t\,v_i(t,p_{\rm H})
\int\limits_{-\infty}^{t}\mbox{d}t'\,v_j(t',p_{\rm H})
\nonumber\\
&\times&
\exp\{\nu(t'-t)\}\cos k\{x(t',p_{\rm H})-x(t,p_{\rm H})\}
\nonumber\\
&\equiv&
\langle e^2v_i{\hat R}v_j\rangle .
\ea
The kernel of the integral operator $Q_{ij}$ depends essentially
on the condition of the specimen surface, i.e. on the
probability of the specular reflection of charge carriers.

The electric field $E_x(x)$ should be determined from the Poisson
equation
%3.14
\be
\dv\bs{E}=4\pi e\langle\psi\rangle,
\ee
which in conductors with a high charge carriers density
reduces to the condition of electrical neutrality of a
conductor
%3.15)
\be
\langle\psi\rangle=0.
\ee
The condition of charge conservation, following from the
continuity equation
%3.16)
\be
-i\omega e\langle\psi\rangle+\dv\bs{j}=0;
\ee
and the macroscopic boundary condition (the absence of the
current through the surface $x_{\rm s}=0$) enable us to relate the
electric field $E_x(x)$ with the other components of the field.
The equality to zero of the current $j_x$ for any $x$,
%3.17
\be
j_x(k)=\sigma_{xx}(k)E_x(k)+\sigma_{x\alpha}(k)E_\alpha(k)=0;
\quad \alpha=(y,z),
\ee
together with the Equation (3.11) allows us to find the
Fourier-transform $\bs{E}(k)$, and then, by using the inverse
Fourier transformation, to obtain the distribution of the
electric field in the conductor
%3.18)
\be
E_\alpha(x)=\frac{1}{2\pi}\int\limits_{0}^{\infty}\mbox{d}k\,
E_\alpha(k)\exp\{-ikx\}.
\ee

%3.1
\subsection{Normal Skin Effect}
The penetration of an electromagnetic field into the
conductor under the following conditions: when the current
density $\bs{j}(\bs{r})$ is determined by the value of the electric field
$\bs{E}(\bs{r})$ in the same point $\bs{r}$, is known as the normal skin
effect.

In the absence of open electron orbits, the high-frequency
current may be produced mainly by conduction electrons that
are removed from the surface at greater distance than the
electron orbit diameter and do not collide with the sample
surface. This takes place in sufficiently strong magnetic
fields parallel to the sample surface ($\phi=0$) when the
curvature radius $r$ of the charge carriers trajectory is
much less than the skin-layer depth. In this case the
relation between the current density and the electric field
can be treated as local, as in the case of the normal
skin-effect, although we may take any proportion between
$l$ and $\delta$.

The depth of the skin-layer is determined by the roots of
the dispersion equation
%3.19
\be
\mbox{det}\,\left\{\left(k^2-\frac{\omega^2}{c^2}\right)
-\frac{4\pi i\omega}{c^2}{\tilde\sigma}_{\alpha\beta}(k)\right\}
=0,
\ee
where
%3.20
\be
{\tilde\sigma}_{\alpha\beta}(k)=\sigma_{\alpha\beta}(k)-
\frac{\sigma_{\alpha x}(k)\sigma_{x\beta}(k)}{\sigma_{xx}(k)};
\quad
\alpha,\beta=(y,z).
\ee

At $kr\ll 1$ the asymptotic expression for $\sigma_{ij}(k)$ is of the same
form as that in a uniform electric field, hence $\sigma_{zz}$ can be
described by Equations (2.31)--(2.33), where the frequency of collisions
$1/\tau$ should be replaced by $\nu$. In a strong magnetic field the
asymptotic value of $\sigma_{yy}$ is proportional to
$\sigma_0(\eta\tan\theta)^2$, because ${\bar v}_y={\bar v}_z\tan\theta$.
However, in this approximation the Hall field is large and
${\tilde\sigma}_{zz}$ is of the order of $\sigma_0$ ($\sigma_0$ is the
conductivity in the layers-plane at $H= 0$ in an uniform electric field). As
a result, at $kr\ll1$ the electric field $E_y$ attenuates at distance
%3.21
\be
\delta_\bot\cong\delta_0=c(2\pi\omega\sigma_0)^{-1/2}
\ee
for any proportion between the free path length and the
skin-layer depth.

At $\eta\ll1$ each of the components $\sigma_{zx}$ and $\sigma_{xz}$ is
proportional to $\eta^2$ or to higher powers of $\eta$, so that
${\tilde\sigma}_{zz}\cong\sigma_{zz}$.  For small $\theta$ the asymptotic
value of $\sigma_{zz}(k)$ is about $\sigma_0\eta^2$ and the depth
$\delta_\|$ is $1/\eta$ times greater than $\delta_\bot$ (as for normal skin
effect conditions), if the Fermi surface corrugation is not too small and
$\eta\ge\delta_0\omega/c$. At $\omega\gg\sigma_0\eta^2$ and $\eta
r\le\delta_0$ the skin-layer depth
%3.22)
\be
\delta_\|=\frac{\delta_0^2\omega}{c}
\left(1+\frac{r^2\omega^2}{c^2}\right)^{-1/2}
\ee
increases with increasing magnetic field and attains the
value $\omega\delta_0^2/c$.

When $\theta$ is not small, there is a sequence of values
$\theta=\theta_{\rm c}$, at which the asymptotic behaviour of $\sigma_{zz}$,
(and hence of ${\tilde\sigma}_{zz}$) changes essentially. For
$\tan\theta\gg1$ this sequence repeats periodically but the period as well
as the values $\theta_{\rm c}$ differ slightly from those in the case of
static fields.  This results from the fact that the stationary phase points
$\bs{kv}=\omega$ do not coincide with the turning points on an electron
orbit where $v_x$ disappears.  However the phase velocity of the wave
$v_\phi=\omega/k=(\omega\tau)^{-1/2}\omega c/\omega_0\eta$ is much less than
the Fermi velocity $v_{\rm F}$, hence the period of changing of the
asymptotic value of ${\tilde\sigma}_{zz}(k,\theta)$ is determined, to a good
accuracy, by the Fermi surface diameter and has the form (2.35). At
$\theta=\theta_{\rm c}$ the asymptotic value for $\sigma_{zz}$ decreases
significantly for small $\eta$, $\gamma=(\Omega_0\tau\cos\theta)^{-1}$,
$\omega/\Omega_0$ and $kr$, viz.
%3.23)
\be
{\tilde\sigma}_{zz}(k,\eta,\theta_{\rm c})=
\sigma_0\eta^2\{\eta^2f_1(\theta_{\rm c})+\gamma^2f_2(\theta_{\rm c})
+(kr)^2f_3(\theta_{\rm c})\},
\ee
where $f_i$ are the functions of $\theta$ of the order of unity.

The penetration depth for the electric field $E_z$ grows
substantially for $\theta=\theta_{\rm c}$. In the angular dependence of the
impedance there is a series of narrow spikes at  $\theta=\theta_{\rm c}$,
which in pure conductors ($l\eta^2>\delta_0$) diminish with increasing
magnetic field, whereas at $l\eta^2<\delta_0$ they grow proportionally to
$l\delta_0/r\eta$ if $l\eta<r<\delta_0/\eta$.  At certain frequencies (not
too high sufficiently), when the displacement current is small compared to
the conduction current the solution of the dispersion equation (3.19) at
$\theta=\theta_{\rm c}$ can be represented by the following interpolation
formula
%3.24
\be
\delta_\|=l\left(\frac{r^2+\delta_0^2\eta^{-2}}{r^2+l^2\eta^2}\right)^{1/2}.
\ee

In the case of low electrical conductivity across the layers, when the
condition $\omega/c>\sigma_0\eta^2(\eta^2+r^2/l^2)$ is met, the skin depth
$\delta_\|$ has the form
%3.25)
\be
\delta_\|=\frac{\delta_0}{\eta^2}\left\{
1+\left(\frac{r}{l\eta}\right)^2
+\left(\frac{r\omega}{c\eta}\right)^2
\right\}^{-1/2}
\left\{
1+\left(\frac{r\eta}{\delta_0}\right)^2
\right\},
\ee
and in a strong magnetic field, when
$r<(l^2\eta^2+\delta_0^2/\eta^2)^{1/2}$, the electric field decay length
across the layers is again $\delta_0/\eta^2$. In the region of sufficiently
strong magnetic fields, when at $\theta=\theta_{\rm c}$ the relation
$\delta_0/\eta\ll r\ll\delta_\|$ is valid, the impedance as a function of a
magnetic field has a minimum value because for $r\gg l\eta$ the skin depth
%3.26)
\be
\delta_\|=l\frac{r\eta}{\delta_0}
\ee
is inversely proportional to the magnetic field magnitude
[27, 28].

At $\delta_\bot\ll r\ll\delta_\|$ the decay length pertaining to the
electric field $E_z(x)$ is, as before, weakly dependent on the character of
charge carriers interaction with the conductor surface, whereas the
penetration depth for the electric field $E_y(x)$ is very sensitive to the
state of the surface if $\delta_\bot$ is less or comparable with the charge
carriers mean free path.  In this range of magnetic fields the normal skin
effect is realizable only for $\delta_\bot\gg l$, when the local connection
between the current density and the electric field takes place at an
arbitrary orientation of the magnetic field. Asymptotic expression for
${\tilde\sigma}_{yy}(k)$ at $kl\ll 1$ coincides with $\sigma_0$ up to
numerical factor of the order unity and, hence, $\delta_\bot$ is of the same
order of magnitude as $\delta_0$.  However, the penetration depth for the
electric field $E_z(x)$ depends essentially on the orientation of the
magnetic field. The solution of the dispersion equation (3.19) has the form
%3.27)
\ba
k&=&\frac{(2\pi\omega)^{1/2}(1+i)}{2c}
\Biggl\{
\sigma_0^{-1}+\sigma_{zz}^{-1}
\nonumber\\
&\pm&
\left[
(\sigma_{zz}^{-1}-\sigma_0^{-1})^2-
\left(
\frac{2H\sin\theta\sin\phi}{Nec}
\right)^2
\right]^{1/2}\Biggr\}^{-1/2}.
\ea
Inessential numerical factors of the order of unity
depending on the concrete form of the electron energy
spectrum are omitted in Equation (3.27). When $\theta$ differs
essentially from $\pi.2$, in an extremely strong magnetic field
($\gamma\ll\eta^2$) the propagation of helical waves is possible. For
$\phi\cong1$ one of the roots of the dispersion equation describes the
attenuation of the electric field along the layers at
distances of the order of
%3.28
\be
\delta_\bot=\delta_0
\left\{
1+\frac{\sigma_{zz}}{\sigma_0\gamma^2}
\right\}^{1/2}.
\ee

It is easily seen that the depth of penetration for the
field $E_y$ grow proportionally to $H$, when $\gamma\ll\eta$. At
$\gamma\gg\eta^2$ the electric field along the normal attenuates at
distances
%3.29
\be
\delta_\|=\delta_0
\left(
\frac{\sigma_0}{\sigma_{zz}}
\right)^{1/2},
\ee
i.e. at distances of the order of $\delta_0/\eta$, as in the absence of
a magnetic field.

The specific dependence of the attenuation length for the
field $E_z(x)$ takes place at $\theta=\pi/2$, when, apart from the charge
carriers drift along $\bs{H}$-direction, in the $xy$-plane there
are many possible directions of drift for charge carriers
belonging to open Fermi surface cross-sections. In this
case the dependence of $\sigma_{zz}$ on the magnitude of a strong
magnetic field ($\gamma_0\ll 1$), given by the following interpolation
expression
%3.30
\be
\sigma_{zz}=\sigma_0\gamma_0^2\eta^2
(\gamma_0^2+\eta)^{-1/2},
\ee
is the same for any orientation of a magnetic field in the
$xy$-plane.

Using Equations (3.29) and (3.30), it is easy to
demonstrate that at $\eta^{1/2}\ll\gamma_0\ll1$ the attenuation length
$\delta_\|$ increases with increasing magnetic field as $H^{1/2}$, and at
$\eta^2\ll\gamma_0\le\eta^{1/2}$ the length of electric field attenuation
along the normal to the layers $\delta_\|\cong\delta_0/\gamma_0\eta^{3/4}$
grows linearly with a magnetic field.

%3.2
\subsection{Anomalous Skin Effect}
When the frequency of the electromagnetic wave increases, a
local connection between the current density and the
electric field $E_y(x)$ may be broken and the Maxwell equations
appear to be integral even in the Fourier representation.
In the presence of a magnetic field a strict solution of
the problem has been suggested by Hartmann and Luttinger
for some special cases [29]. If the numerical factors of
the order of unity are not used, the dependencies of the
surface impedance and other characteristics of waves on
physical parameters can be obtained by means of the correct
estimation of the contribution given by the integral term
in Equation (3.12) to the Fourier transform of the
high-frequency current. In a magnetic field applied parallel to
the sample surface, the contribution to the current from
charge carriers colliding with the surface is essential at
$\delta_\bot\le r$. When the reflection of charge carriers at the specimen
boundary is close to specular (the width of the scattering
indicatrix $w$ is much less than $r^{3/2}/l\delta_\bot^{1/2}$), the
contribution to the high-frequency current from electrons ``slipping'' along
the sample surface is great and the asymptotic expression for
${\tilde\sigma}_{yy}(k)$ at large
$k$ has the form
%3.31
\be
{\tilde\sigma}_{yy}(k)=\frac{\omega_0^2}{\Omega(kr)^{1/2}(w+r/l)}.
\ee
By means of the dispersion equation (3.19) the length of
electric fields decay can be easily determined
%3.32)
\be
\delta_\bot=\delta_0^{6/5}r^{-1/5}
\left(
w+\frac{r}{l}
\right)^{2/5};
\quad
\delta_\|=\frac{\delta_0}{\eta}.
\ee

In this region of weak magnetic fields ($\delta_\bot\ll r\ll l$) the
impedance has a minimum at $r = wl$, the width of the scattering indicatrix
being determined by the position of the minimum.

In a magnetic field applied parallel to the sample surface
under the conditions of anomalous skin-effect, when the
skin depth is the smallest length parameter, i.e. both $\delta_\bot$,
and $\delta_\|$ are much less than $r$ and $l$, the universal
relation between $\delta_\bot$ and $\delta_\|$ takes place at
$w\ll r^{3/2}/l\delta_\|^{1/2}$, viz.
%3.33)
\be
\delta_\bot=\delta_\|\eta^{4/5}.
\ee

If $w\gg r^{3/2}/l\delta_\bot^{1/2}$ and $\delta_\bot\ll r\ll l$, the
high-frequency current is produced mainly by charge carriers that do not
collide with the sample surface and the relation between $\delta_\bot$ and
$\delta_\|$ has the form (3.2).

In the intermediate case, when
$r^{3/2}/l\delta_\|^{1/2}\ll w\ll r^{3/2}/l\delta_\bot^{1/2}$, only
$\delta_\bot$ is essentially dependent on $w$ for $w\ge r/l$:
%3.34
\ba
\delta_\|&=&r^{1/3}
\left(
\frac{\delta_0}{\eta}
\right)^{2/3};
\nonumber\\
\delta_\bot&=&(wl)^{2/5}\delta_0^{4/5}r^{-1/5}.
\ea

In the absence of open electron orbits the information on
the skin-layer field can be electronically transported
deeper into the conductor in the form of narrow field
spikes, predicted by Azbel' [30]. The electron transport of
the electromagnetic field and the screening of the incident
wave at the surface $x_{\rm s}=0$ is due mainly to the charge carriers
that are in phase with the wave and move nearly parallel to
the sample surface. At $\eta\le \delta/r$ field spikes are produced with
the participation of almost all charge carriers [31].
Without reference to collisions, the intensities of the
spikes at distances, that are multiples of the electron
orbit diameter along the $x$ axis, are in the same range.
Allowance for the scattering of conduction electrons in the
bulk of a conductor, results in the attenuation of the
field in the spikes at distance of the order of the mean
free path. Thus under the anomalous skin effect conditions
there are two scale lengths of wave decay. Apart from the
damping effect over the skin depth the electromagnetic
field penetrates into the conductor to distances of the
order of the mean free path $l$.

In the case when $\eta\gg\delta/r$, spike formation is provided by a small
fraction of order $(\delta/r\eta)^{1/2}$ of the charge carriers whose orbit
diameter scatter near the extremal section is comparable to
the skin layer depth. As a result, the spike intensities
decrease as the distance from the face $x_{\rm s}=0$ increases, and
apart from the factor $\exp\{-x/l\}$ every subsequent spike acquires
the small factor $(\delta/r\eta)^{1/2}$.

While the angle $\theta$ approaches $\pi/2$, closed electron orbits
become strongly elongated in the $x$-direction. When the
orbit diameter along the $x$-axis exceeds the free path
length $l$, the spike mechanism of the electromagnetic
field penetration into the conductor is replaced by the
electron transport of the field in the form of Reuter--Sondheimer
quasi-waves.

%3.3
\subsection{Weakly Damping Reuter--Sondheimer Waves}
Consider the electron transport of the electromagnetic
field at $\theta=\pi/2$. In order to find the field in a depth of the
conductor by means of the inverse Fourier transformation
(3.18) we continue $\sigma_{ij}(k)$ analytically into the whole complex
$k$-plane and supplement the integration contour with an arc
of an infinitely large radius. The skin layer depth is
determined by poles of the integrand in Equation (3.18),
whereas weakly attenuated waves are connected to the
integration along the cut-line drawn from the branching
point of the function $E_j(k)$. It can be easily seen that at
however small $\eta$ the components of the tensor $\sigma_{ij}(k)$ has a
power $1/2$ singularity:
%3.35, 3.36)
\ba
\sigma_{zz}(k)&=&
\frac{\omega_0^2\eta^2}{\nu}\{(\alpha_+^2-1)^{-1/2}+(\alpha_-^2-1)^{-1/2}\};
\\
\Delta\sigma_{yy}(k)&=&\nu
\left(
\frac{\omega_0}{kv}
\right)^2
\left\{
\left(
\frac{kv}{\nu}
\right)^2
+1
\right\}^{1/2},
\ea
where $\omega_0$ is the plasma frequency, $\alpha_\pm=i(kv\pm\Omega_0)/\nu$
unessential numerical factors of the order of unity are omitted.

The kernel of the integral operator $Q_{ij}(k,k')$ as a function of $k$
has such a singularity as well.

At distances from the sample surface that are greater than
either the curvature radius of the electron trajectory $r=v/\Omega_0$
or the electron displacement $2\pi v/\omega$ per period of the wave the
electromagnetic field decreases proportionally to $x^{-3/2}\exp(-x/l)$. For
$\Omega_0\gg\omega$ the slowly decreasing electric field $E_z(x)$ oscillates
with $H$ at large $x$:
%3.37)
\ba
E_z(x)&=&E_z(0)\eta^{-4/3}
\left(
\frac{\omega_0}{c}
\right)^{4/3}
\left(
\frac{v}{\omega}
\right)^{2/3}
r^{-1/2}x^{-3/2}
\nonumber\\
&\times&
\exp
\left\{
\frac{ix}{r}-\frac{x}{l}
\right\}; \quad r\ll x\ll\frac{r}{\eta}.
\ea

At $\eta\ll1$ the attenuation of the electric field $E_y(x)$ over the
charge carriers free path length has the form
%3.38
\be
E_y(x)=E_y(0)
\left(
\frac{\omega_0}{c}
\right)^{4/3}
\left(
\frac{v}{\omega}
\right)^{2/3}
x^{-3/2}
\exp
\left\{
-\frac{x}{l}
\right\}; \quad \frac{v}{\omega}\ll x\ll\frac{v}{\omega\eta},
\ee
and does not contain the magnetic field magnitude.

The oscillatory dependence of $E_y(x)$ upon the magnetic field
magnitude occurs only in the case, when the charge with
velocity $v_y$ is not an integral of motion, and manifests
itself in small corrections proportional to $\eta^2$. Numerical
factors of order unity, being dependent on the concrete
form of the electron dispersion law, are omitted in
Equations (3.37) and (3.38).

In a sufficiently strong magnetic field ($r\ll l\eta$) the asymptotic
behaviour of the weakly damping field $E_z(x)$ changes
essentially with displacement from the sample surface. In
the absence of a magnetic field $\sigma_{yy}(k)$ and $\sigma_{zz}(k)$ have
logarithmic singularities at $k_1=i\nu/v_1$ and $k_2=i\nu/v_2$, except were
$\eta$ is not small.  $v_1$ is the electron velocity in the reference point (in
the $x$-direction) of the Fermi surface and $v_2$ is the velocity projection
$v_x$ in the Fermi surface saddle point, where the connectivity of the line
$v_x=\mbox{const}$ changes [32]. At small values of $\eta$ these branching
points for the components of the high-frequency conductivity tensor approach
each other, and at $\eta=0$ the logarithmic singularity is replaced by the
power $1/2$ singularity of the form (3.35), (3.36). Let the integration
contour in the $k$-plane at small $\eta$ be drawn along the cut-line from
the branching points $k_1$ and $k_2$ parallel to the imaginary axis, to go
round both branching points.  Then the electric field $E_z(x)$ at large
distances from the skin layer takes the following form.
%3.39
\ba
E_z(x)&=&
\int\limits_{k_1}^{k_1+i\infty}
\frac{\mbox{d}k\,\exp(ikx)}%
{k^2-\omega^2/c^2-4\pi i\omega c^{-2}\sigma_{zz1}(k)}\nonumber\\
&+&
\int\limits_{k_2}^{k_2+i\infty}
\frac{\mbox{d}k\,\exp(ikx)}%
{k^2-\omega^2/c^2-4\pi i\omega c^{-2}\sigma_{zz2}(k)}
\ea

The integral along the line connecting the branching points $k_1$ and $k_2$
may be neglected; $\sigma_{zz1}(k)$ is the value of the function
$\sigma_{zz}$ on the left side of the cut-line drawn from the point $k_1$,
$\sigma_{zz1}$ is the value on the right side of the cut-line from the point
$k_2$.  $v_1$ is assumed to be larger than $v_2$.  Making use of the
dispersion law (2.21), we obtain the following expressions for the diagonal
components of the high-frequency conductivity tensor at $k_1\le k\le k_2$
%3.40, 3.41)
\ba
\sigma_{yy}(k)&=&
\frac{\omega_0^2\eta}{\pi^3}
\int\limits_{0}^{\pi}\mbox{d}\alpha\,
\int\limits_{0}^{\pi/2}
\frac{\mbox{d}\phi\,\sin^2\phi}{\nu+ikv\cos\phi(1+\eta\cos\alpha)^{1/2}};
\\
\sigma_{yy}(k)&=&
\frac{\omega_0^2\eta}{\pi^3}
\int\limits_{0}^{\pi}\mbox{d}\alpha\,\sin^2\alpha\,
\int\limits_{0}^{\pi/2}
\frac{\mbox{d}\phi}{\nu+ikv\cos\phi(1+\eta\cos\alpha)^{1/2}};\nonumber\\
\ea
It is easily seen that at $\eta=0$ the component of the
high-frequency conductivity $\sigma_{zz}(k)$ is proportional to
$(\nu+ikv)^{-1/2}$ and $\sigma_{yy}(k)$ is proportional to
$(\nu+ikv)^{1/2}$, i.e. at $k=i\nu/v$ both of them have the power $1/2$
singularity,. When the Fermi surface corrugation is not small ($\eta\cong1$)
instead of the power $1/2$ singularity the logarithmic singularity appears
for $k=i\nu/v(1+\eta)^{1/2}$ and $k=i\nu/v(1-\eta)^{1/2}$. The integral with
respect to $\phi$ has the power $1/2$ singularity at
$k=i\nu/v(1+\eta\cos\alpha)^{1/2}$.  Interchanging the order of integration
with respect to $\alpha$ and $k$ in Equation (3.39), we obtain the following
expression for the weakly damping component of the electric field at large
distances from the skin layer
%3.42
\ba
E_y(x)&=&E_y(0)
\left(
\frac{c}{\omega_0}
\right)^{4/3}
\left(
\frac{v}{\omega}
\right)^{2/3}
x^{-3/2}
\left(
\frac{\nu}{v}
\right)^{1/2}
\nonumber\\
&\times&
\int\limits_{0}^{\pi}\mbox{d}\alpha\,
\exp
\left\{
-\frac{\nu x}{v(1+\eta\cos\alpha)^{1/2}}
\right\}
\ea

At great distances from the sample surface the electric
field along the normal to the layers can be described by
the same formula, if the additional factor $\eta^{-4/3}\sin^2\alpha$ is
written in the integral over $\alpha$. At $x\gg v/\omega\eta$ the integrand
in Equation (3.42) is a rapidly oscillating function and the major
contribution to the integral is given by the small vicinities near the
stationary phase points $\alpha=(0,\pi)$. After simple calculations we have
%3.43)
\ba
E_y(x)&=&E_y(0)
\left(
\frac{c}{\omega_0}
\right)^{4/3}
\left(
\frac{v}{\omega}
\right)^{2/3}
x^{-2}\eta^{-1/2}
\nonumber\\
&\times&\left[
\exp\left\{-
\frac{\nu x}{v(1+\eta)^{1/2}}+
\right\}
\exp
\left\{
-\frac{\nu x}{v(1-\eta)^{1/2}}
\right\}
\right];\nonumber\\
&&
x\gg \frac{v}{\omega\eta}.
\ea

In the formulas given above unessential numerical factors
of order unity are omitted.  The factor at the exponent in
Equation (3.43) is inversely proportional to $x^2$, as in
ordinary metals. Such asymptotic behaviour of the electric
field in the layered conductors takes place only in the
range of large frequencies when $\omega\tau\gg 1/\eta$. The difference in
asymptotic behaviour of the electric fields at such
frequencies can be understood by watching the wave phase,
which is carried away from the skin layer by conduction
electrons with different velocity projections $v_x$. At a
moment $t$ and a distance $x$ electrons carry information
on the electromagnetic wave whose phase is late for the
quantity $\omega\Delta t=\omega x/v_x$. After averaging over different
values of $v_x$ we have
%3.44
\be
\bs{E}(x)=\int\mbox{d}v_x\,
\exp
\left\{
-i\omega t+\frac{i\omega x}{v_x}
\right\}.
\ee

It can be easily seen that the slowly damping wave,
propagating with the velocity of electrons from the Fermi
surface reference point $v_1$, is formed by charge carriers
whose velocity $v_x$ differs from $v_1$, by the quantity
$\Delta v_x\le v_1^2/\omega x$.  If $v_1-v_2\cong v\eta$ is less than
$\Delta v_x$, i.e.  $x\le v \omega \eta$, then Equation (3.38) takes place,
and in the opposite case, when $\Delta v_x\ll v\eta$, electrons from the
small vicinity of the Fermi surface saddle and reference points produce
weakly damping waves of the form (3.43).

In a magnetic field, the charge carriers that belong to one
of the sides of the central open Fermi surface cross
section, (at which the velocity $v_x$ alters periodically with
time in the interval between $v_2$ and $v_1$) move more rapidly
into the bulk of the sample. The weakly damping waves
propagate with velocity, equal to the extremal value ${\bar v}_x$,
and can be described by Equations (3.37) and (3.38).

Weakly damping waves are of the analogous form, in a
magnetic field $\bs{H}=(H\cos\phi, H\sin\phi\cos\theta,
H\sin\phi\cos\theta)$ deflected from the layers-plane.  If ??? and $\phi$
differ from zero essentially, the weakly damping wave at $\theta=\pi/2$
propagates with velocity ${\bar v}_x$, which is equal to the velocity of
drift for charge carriers on the open Fermi surface cross-section containing
the reference point with respect to the axis $p_x$.  The asymptotic
behaviour of the electric field $E_y(x)$ can be described by Equation
(3.38), and the oscillatory dependence on the magnetic field, applied
orthogonal to the axis of the corrugated cylinder, manifests itself, as
before, only in small corrections proportional to $\eta^2$.  However at
$\phi\ll1$ and $\theta=\pi/2$ the asymptotic behaviour of $E_z(x)$ and
$E_y(x)$ at great distances from the sample surface can change essentially.

In a magnetic field applied orthogonal to the surface ($\phi=0$)
the electrons with closed orbits and also a considerable
part of charge carriers on open orbits near the self-intersecting
orbit, participate in the formation of the
weakly damping waves at distances $x\ll v/\omega\eta$. If the charge
carriers dispersion law used is, as before, Equation (2.21), $\sigma_{ij}(k)$
takes the form
%3.45
\be
\sigma_{ij}(k)=\frac{2e^2}{(2\pi h)^3}
\sum\limits_{n}\int\mbox{d}p_x\,
2\pi m^*\frac{v_i^{-n}v_j^n}{\nu+ikv_x+in\Omega}.
\ee

At a distance $x\gg v/\omega\eta$ the faster wave is formed by charge
carriers with the closed orbit near the Fermi surface
reference point and the factor at the exponent in the
asymptotic expression for the electric field decreases
proportionally to $x^2$ with increasing $x$. If $\eta^{1/2}\ll\gamma_0$,
i.e. an electron has no time to make a complete revolution along the orbit
during the free path time, then not only the field $E_y$ but also the field
$E_z$ are independent of $H$ at such great distances.

At low temperatures, when the smearing $k_{\rm B}T$ of the Fermi
distribution function for charge carriers is less than the
spacing between quantum energy layers $\hbar\Omega$, the magnetic
susceptibility as well as other kinetic characteristics
oscillates with the inverse magnetic field in the
quasiclassical region. In this case the amplitude of the
quantum oscillations of the magnetic susceptibility may
considerably exceed the monotonically changing part and
even equal the magnitude $1/4\pi$. If the sample surface
coincides with a plane of symmetry for the crystal, the
main axes for the magnetic susceptibility tensor, coincide
with the axes $y$ and $z$ accurately. The Maxwell equations
in this case take the form
%3.46, 3.47
\ba
\left\{
\sigma_{yy}(k)+\frac{\omega}{4\pi i}-\frac{k^2c^2}{4\pi i \omega}
(1-pi\chi_{zz})
\right\}
E_y(k)\nonumber\\
+\sigma^*_{yz}(k)E_z(k)&=&-E'_y(0);\\
\sigma^*_{zy}(k)E_y(k)+
\nonumber\\
\left\{
\sigma^*_{zz}(k)
+\frac{\omega}{4\pi i}-
\frac{k^2c^2}{4\pi i\omega}(1-4\pi\chi_{yy})
\right\}E_z(k)
&=&
-E'_z(0),
\ea
where
$$
\sigma_{\alpha\beta}^*(k)=
\sigma_{\alpha\beta}(k)
-\frac{4\pi\sigma_{\alpha
x}(k)\sigma_{x\beta}(k)}{4\pi\sigma_{xx}(k)-i\omega},
$$
$(\alpha,\beta)=(y,z)$.

It is easy to make sure that the oscillating part of the
magnetization
%(3.48)
\ba
\bs{M}_{\rm osc}&=&
\frac{\partial}{\partial\bs{H}}
\biggl\{
2\mbox{Re}
\sum\limits_{q=1}^{\infty}
\frac{(-1)^q}{q^2}
\frac{eH}{2\pi^2 c(2\pi h)2}
\int\mbox{d}\varepsilon\,
\frac{\partial f_0}{\partial\varepsilon}
\int\mbox{d}p_{\rm H}\,
h\Omega\nonumber\\
&\times&
\exp
\left(
\frac{iqcS(\varepsilon,p_{\rm H})}{eHh}
\right)
\biggr\}
\ea
decreases with increasing $\theta$, so that the amplitude of the
magnetic susceptibility is maximum for the magnetic field
directed along the normal to the layers ($\theta=0$), and $M_{\rm osc}$ is
proportional to $\eta$ for $\theta=\pi/2$. As a result, when the electric
field of the incident wave is linearly polarized along the
normal to the layers, the surface impedance undergoes small
quantum oscillations with an inverse magnetic field. The
electric field in the layers plane attenuates at a distance
$$
\delta=
\left\{
\frac{vc^2(1-4\pi\chi)}{\omega\omega_0^2}
\right\}^{1/3},
$$
if $\chi=\chi_{zz}<1/4\pi$, and the impedance undergoes giant oscillations at
$(1-4\pi\chi^{\rm max})\ll1$. In the opposite case
$((1-4\pi\chi^{\rm max}))<0$ homogeneous state is unstable for some values
of a magnetic field which leads to the formation of magnetic domains.

The considered cases of electromagnetic waves propagation
in the layered conductors with the quasi-two-dimensional
electron energy spectrum prove that they possess a great
variety of specific high-frequency properties, that will,
undoubtedly, be used in modern electronics.

Such a strong dependence of the intensity of the wave on
its polarization allows us to utilize even thin plates of
layered conductors (whose thickness is much more than skin
depth but less or of the order of the free path length) as
filters allowing the wave to pass with a certain
polarization.

%4.
\section{ACOUSTIC TRANSPARENCY OF LAYERED CONDUCTORS}
When acoustic waves propagate in a layered conductor placed
in a magnetic field, the quasi-two dimensional nature of
the charge carriers energy spectrum is expected to be
pronounced. Being very sensitive to the form of electron
energy spectrum, the magnetoacoustic effects [33--36] have
been used successfully for restoration of the Fermi
surface, and in the low-dimensional conductors they are
worthy of the special examination.

In conducting crystals apart from the sound waves
attenuation related to the interaction between thermic
phonons and coherent phonons with the frequency $\omega$, there
are many mechanisms of electron absorption of acoustic
waves. The most essential of them is the so-called
deformation mechanism connected with charge carriers energy
renormalization under strain
%4.1
\be
\delta\varepsilon=\lambda_{ij}u_{ij}.
\ee
where $\lambda_{ij}$ is the deformation potential tensor.

In a magnetic field the induction mechanism connected with
electromagnetic fields generated by sound waves is also
essential. These fields should be derived from the Maxwell
equations (3.3), (3.4) and connection of the current
density with the strain tensor $u_{ij}$ and electric field
%4.2)
\be
{\tilde{\bs{E}}}
=\bs{E}+
\frac{(\dot{\bs{u}}\times\bs{H})}{c}
+\frac{m\ddot{\bs{u}}}{e}
\ee
can be found with the aid of the solution of the Boltzman
kinetic equation. The field $\tilde{\bs{E}}$ is determined in the
concomitant system of axes which moves with the velocity
$\dot{\bs{u}}$.

The last term in the Equation (4.2) is connected with the
Stewart--Tolmen effect.

In a weakly deformed crystal the complete set of equations
for this problem is suggested by Silin [37] for isotropic
metals and generalized by Kontorovich [38] to the case of
an arbitrary dispersion law of charge carriers. A complete
set of nonlinear equations for this problem, valid for any
wave intensity, was derived by Andreev and Pushkarov [39].

Together with the Maxwell equations and Boltzman kinetic
equation it is necessary to consider the dynamics equation
of the elasticity theory for the ionic displacement $\bs{u}$.
%4.3
\be
-\omega^2\rho u_i=\lambda_{ijlm}\frac{\partial u_{lm}}{\partial x_j}+F_i.
\ee
Here $\rho$ and $\lambda_{ijlm}$ are the density and elastic tensor of the
crystal. The equation contains the force $\bs{F}$ applied to the
lattice from the electron system excited by the acoustic
wave which is taken to be monochromatic with the frequency
$\omega$. If the lattice strain is small, the force is
%4.4
\be
F_i=c[\bs{jH}]_i+\frac{m}{e}i\omega j_i
+\frac{\partial}{\partial x_k}\langle \Lambda_{ik}\psi\rangle,
\ee
where
$\Lambda_{ik}(\bs{p})=\lambda_{ik}(\bs{p})-
\langle\lambda_{ik}(\bs{p})\rangle/\langle 1 \rangle$.

In the linear approximation of the deformation tensor the
change of charge carrier dispersion law (1.1) can be
described with the aid of the deformation potential $\lambda_{ij}$
whose components depend on the quasimomentum $\bs{p}$ only and
coincide in order of magnitude with the characteristic
energy of the electron system, viz., the Fermi energy $\varepsilon_{\rm F}$.

Confining ourselves only to the linear approximation in a
weak perturbation of conduction electrons under deformation
of the crystal in kinetic equation, we obtain for $\psi$ the
following expression:
%4.5)
\be
\psi=\hat{R}\{\Lambda_{ij}(\bs{p})\dot{u}_{ij}+e\tilde{\bs{E}}\bs{v}\},
\ee
where $\hat{R}$ is the resolvent of Equation (2.2) when
$1/\tau\to\nu=1/\tau-i\omega$.

Let us consider an acoustic wave propagating in
$x$-direction orthogonal to a magnetic field
$\bs{H}=(0,H\sin\theta,H\cos\theta)$.  Using the Fourier method, derive from
the Maxwell equations and Equation (4.3) a set of equations for the Fourier
components of the electric field $E_i(k)$ and ionic displacements $u_i(k)$:
%4.6
\ba
\frac{4\pi i\omega}{c^2}j_\alpha(k)&=&
k^2E_\alpha(k)
-\left(\frac{\omega}{c}\right)^2E_\alpha(k),
\quad \alpha=y,z
\nonumber\\
j_x(k)&=&0,
\nonumber\\
-\omega^2\rho u_i(k)&=&
-\lambda_{iklx}k^2u_l+\frac{im\omega}{e}j_i(k)
+\frac{1}{c}[\bs{j}(k)\bs{H}]_i+ik\langle\Lambda_{ix}\psi\rangle.
\nonumber\\
\ea
Using the kinetic equation solution, we can conveniently
express the parameters $j_i(k)=\langle ev_i\psi(k)\rangle$ and
$\langle\Lambda_{ix}\psi(k)\rangle$, which characterize the system response
to the acoustic wave, in the form
%4.7
\ba
j_i(k)&=&\sigma_{ij}(k)\tilde{E}_j(k)+a_{ij}(k)k\omega u_j(k),
\nonumber\\
\langle\Lambda_{ix}\psi(k)\rangle&=&
b_{ij}(k)\tilde{E}_j(k)+c_{ij}(k)k\omega u_j(k),
\ea
where the Fourier components of the conductivity tensor and
of acousto-electronic coupling tensors are
%4.8
\ba
\sigma_{ij}(k)=\langle e^2v_i\hat{R} v_j\rangle;&&
a_{ij}(k)=\langle ev_i\hat{R}\Lambda_{jx}\rangle;
\nonumber\\
b_{ij}(k)=\langle e\Lambda_{ix}\hat{R}v_j\rangle;&&
c_{ij}(k)=\langle \Lambda_{ix}\hat{R}\Lambda_{jx}\rangle.
\ea
By substituting Equations (4.8) into the equation set
(4.6), we obtain a system of linear algebraic equations in
$u_i(k)$ and $\tilde{E}_i(k)$. After the inverse Fourier transform of the
solutions of the obtained equation system, the problem of
the distribution of the electric and strain fields in a
conductor will be solved completely.

%4.1
\subsection{The Rate of Sound Attenuation}
The condition for the existence of nontrivial solution of
the set of equations for $u_i(k)$ and $\tilde{E}_i(k)$ (which is equal to
zero of the system determinant) represents the dispersion relation between
the wavevector $\bs{k}$ and the frequency $\omega$. The imaginary part of
the root of the dispersion equation determines the decrements of the
acoustic and electromagnetic waves and the real part describes the
renormalization of their velocities related to the interaction between the
waves and conduction electrons.

However the sound attenuation rate can be also determined
by means of the dissipation function $Q$ which is
proportional to the variation with time of the entropy of a
conductor [40]. Taking into account only the electron
absorption of acoustic waves we have for the dissipation
function
%4.9
\be
Q=\langle\psi\hat{W}_{\rm col}\{\psi\}\rangle;
\ee
and for the sound damping decrement
%4.10
\be
\Gamma=\left\langle\frac{|\psi|^2}{\rho u^2s\tau}\right\rangle,
\ee
where $s$ is the sound velocity, the collision integral is
taken in the $\tau$-approximation.

In ordinary metals the electromagnetic fields generated by
sound are essential in the range of strong enough magnetic
fields when the radius of curvature of the electron
trajectory is much less than the mean free path and also
than the sound wavelength, i.e.  $kr\ll1$. If the charge carriers
trajectories are bent so that
%4.11
\be
1\ll kr\ll kl,
\ee
the absorption of sound wave energy in a metal is
determined mainly by the deformation mechanism. In
low-dimensional conductors the role of electromagnetic
fields generated by a sound wave turns out to be essential
in a wider range of magnetic fields, including a magnetic
field which satisfies the condition (4.11) [41]. In this
range of magnetic fields, the energy absorption coefficient
$\Gamma$ of the acoustic wave in an ordinary (quasi-isotropic)
metal oscillates with variation of the reciprocal magnetic
field. The amplitude of these oscillations is small
compared with the slowly varying component of $\Gamma$, and the
period is determined by the extreme diameter of the Fermi
surface. This effect, which was predicted by Pippard [33],
is associated with a periodic repetition of the conditions
of absorption of the acoustic wave energy by electrons on a
selected orbit, when a number of acoustic wavelengths
fitting in this orbit changes by unity. Under conditions of
strong anisotropy in the energy-momentum relation for
charge carriers, Pippard's oscillations are formed not by
small fractions of electrons, but by almost all charge
carriers on the Fermi surface. As a result, the amplitude
of periodic variations of $\Gamma$ increases sharply compared
with the case of quasi-isotropic metal, and these
variations acquire the form of resonance peaks [42, 43].

If the acoustic wave polarization is aligned with its
wavevector ($\bs{u}=(u,0,0)$) the equations system after the exclusion of
$\tilde{E}_x$ takes the form
%4.12)
\ba
\left(
\tilde{a}_{yx}k\xi+\frac{iH_z}{c}
\right)
\omega u
+(\xi\sigma_{yy}-1)\tilde{E}_y
+\xi\tilde{\sigma}_{yz}\tilde{E}_z
&=&0;
\nonumber\\
\left(
\tilde{a}_{zx}k\xi+\frac{iH_z}{c}
\right)
\omega u
+(\xi\sigma_{zz}-1)\tilde{E}_z
+\xi\tilde{\sigma}_{zy}\tilde{E}_y
&=&0;
\nonumber\\
(\omega^2-s^2k^2)\rho u +
\left[
ik\tilde{c}_{xx}+
\frac{1}{c}(\tilde{a}_{yx}H_z-\tilde{a}_{zx}H_y)k\omega u
\right]
\nonumber\\
+
\left[
ik\tilde{b}_{xy}+
\frac{1}{c}(\tilde{\sigma}_{yy}H_z-\tilde{\sigma}_{zy}H_y)\tilde{E}_y
\right]
\nonumber\\
+
\left[
ik\tilde{b}_{xz}+
\frac{1}{c}(\tilde{\sigma}_{yz}H_z-\tilde{\sigma}_{zz}H_y)\tilde{E}_y
\right]
&=&0,\nonumber\\
\ea
where
\ba
\tilde{\sigma}_{\alpha\beta}=\sigma_{\alpha\beta}
-\frac{\sigma_{\alpha x}\sigma_{x\beta}}{\sigma_{xx}},&&
\tilde{a}_{\alpha j}=a_{\alpha j}
-\frac{a_{xj}\sigma_{\alpha x}}{\sigma_{xx}},
\nonumber\\
\tilde{b}_{i\beta}=
b_{i\beta}-\frac{b_{ix}\sigma_{x\beta}}{\sigma_{xx}},
&&
\tilde{c}_{ij}=c_{ij}-\frac{b_{ix}a_{xj}}{\sigma_{xx}},\quad
\alpha,\beta=y,z.
\nonumber\\
s=
\left(
\frac{\lambda_{xxxx}}{\rho}
\right)^{1/2},
&&
\xi=\frac{4\pi i\omega}{k^2c^2-\omega^2}.
\nonumber
\ea
For $\omega\tau\ll1$ one root of the dispersion equation is close to
$\omega/s$, so we seek its solution in the form
%4.13
\be
k=\frac{\omega}{s}+k_1.
\ee

For $k_1$ we have
%4.14)
\ba
k_1&=&\frac{ik^2}{2\rho s}\frac{1}{(1-\xi\tilde{\sigma}_{yy})}
\biggl\{
\xi(\tilde{a}_{yx}\tilde{b}_{xy}-\tilde{c}_{xx}\tilde{\sigma}_{yy})
\nonumber\\
&+&
\tilde{c}_{xx}-i(\tilde{a}_{yx}-\tilde{b}_{xy})\frac{H_z}{kc}
+\tilde{\sigma}_{yy}\frac{H^2_z}{k^2c^2}
\biggr\}_{k=\omega/s}.
\ea

The acousto-electronic tensors components oscillate with
the magnetic field. When $1\ll kr\ll1/eta$ the spread of electron orbit
diameters, $\Delta D\cong2r\eta$, is much smaller than the acoustic
wavelength, and the amplitude of the oscillations may be
comparable to the slowly varying parts of these functions.
This leads to weak damping of acoustic waves under these
conditions except for the values of a magnetic field at
which the magnetoacoustic resonance occurs.

For example, for $\sigma_{yy}$ and $a_{yj}$ we have
%4.15)
\ba
\sigma_{yy}(k)&=&\frac{G}{kD}(1-\sin kD);\nonumber\\
a_{yx}(k)&=&-i\frac{G\Lambda_{xx}}{evkD}\cos kD,
\ea
were
$$
G=\frac{4vD_{\rm p}e^2\tau}{ac(2\pi\hbar)^2},\quad
D=\frac{cD_{\rm p}}{eH\cos\theta}
$$
and $D_{\rm p}$ is the diameter of the Fermi
surface along the $p_y$-axis, $v$ and $\Lambda_{jx}$ are the electron
velocity and value of $\lambda_{jx}(\bs{p})$ at the reference point along
the $p_y$-axis.

It is easily seen that the parameter $\tilde{\sigma}_{yy}$ is largely
controlled by the component $\sigma_{yy}$ and the denominator in the
expression (4.14) decreases considerably at $kD=2\pi(n+1/4)$, and terms
of higher order with respect to the small parameters $1/kD$ and
$\gamma=r/l$ should be retained in the asymptotic expression for
$\sigma_{yy}$-component of the tensor of high-frequency conductivity.
This leads to the sharp increase of $\Gamma$, and the height of
resonant peaks
%4.16
\be
\Gamma_{\rm res}=\frac{\omega\tau}{r}
\ee
is proportional to $H$ if $l\ll kr^2$.

Out of the resonance, in a wide range of magnetic fields
until $\sin kD$ differs essentially from unity, there is no need
to take into account small corrections in the formula for
$\sigma_{yy}$ and the sound damping decrement has the form
%4.17
\be
\Gamma=\frac{\omega\tau}{r}
\left[
\left(\frac{r}{l}\right)^2+(rt\eta)^2
\right].
\ee

In this case the rate of sound attenuation decreases with
increasing of the magnetic field magnitude. The attenuation
length for a longitudinal wave is the largest when $kD$ is
close to $2\pi(n-1/4)$. As a result, between resonant values of the
sound decrement, which repeat with the period
%4.18
\be
\Delta\left(\frac{1}{H}\right)=
\frac{2\pi e\cos\theta}{kcD_{\rm p}}
\ee
the anomalous acoustic transparency should be observed with
the same period (Fig. 4).

\begin{figure}
\centering
\includegraphics[width=10cm]{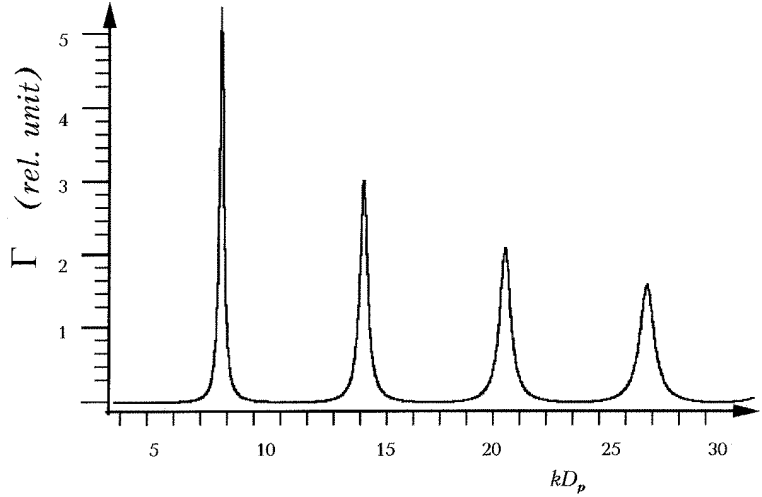}
\caption{The magnetic field dependence for the attenuation rate of
the longitudinal sound wave at $kr\gg1$.}
\end{figure}

One can easily derive explicit expressions for $\Gamma$ at any $kr\eta$
making use of an example of a layered conductor whose
electron spectrum has the form
%4.19)
\be
\varepsilon(\bs{p})=
\frac{p_x^2+p_y^2}{m}
+\eta\frac{\hbar}{a}v_0\cos\left(\frac{ap_z}{\hbar}\right),\quad
v_0=\frac{2\varepsilon_{\rm F}}{m},
\ee
and assume that the magnetic field is perpendicular to the
layers [43]. In this case, at the lowest order in the small
parameters $\gamma$ and $(kr)^{-1}$ the conductivity component is of the
form
%4.20
\be
\sigma_{yy}=\frac{2Ne^2\tau}{\pi mkr_0}
[1-J_0(kR\eta)\sin(2kr_0)],
\ee
where $N$ is the charge carriers density, $r_0=v_0/\Omega$,
$\Omega=eH/mc$, $R=2hc/eHa$ and $J_0$
is the Bessel function.

For $kR\eta\gg 1$, the Fermi surface corrugation is essential, and the
acoustic absorption is similar to that in an ordinary
(nearly isotropic) metal:
%4.21)
\be
\Gamma=\frac{Nm\omega n_0}{4\pi\rho s^2}\Omega\tau
\left[
1+\left(\frac{2}{\pi kR\eta}\right)^{1/2}
\cos\left(kR\eta-\frac{\pi}{4}\right)
\sin(2kr_0)
\right]_{k=\omega/s}
\ee
for $kR\eta\ll1$, peculiar properties of a quasi-two-dimensional
conductor are manifested clearly, and $\Gamma$ is described by
the expression
%4.22
\ba
\Gamma&=&\frac{Nm\omega v_0}{4\pi\rho s^2}\Omega\tau
\nonumber\\
&\times&
\mbox{Re}\,
\Biggl[
(\pi\gamma)^2+\frac{(kR\eta)^2}{2}+i\mu[1+\sin(2kr_0)]
\nonumber\\
&\times&\biggl(
1-\sin(2kr_0)+\frac{(\pi\gamma)^2}{2}+\frac{(kR\eta)^2}{2}
\nonumber\\
&+&
\frac{1}{2}\left(\frac{3}{4kr_0}\right)^2+i\mu\biggr)^{-1}
\Biggr]_{k=\omega/s},
\ea
where
$$
\mu=\frac{\pi v_0c^2\omega}{2s^3\omega^2_0\Omega\tau}
$$
and $\omega_0$ is the plasma frequency. If it is
comparable to that of ordinary metals ($10^{15}$--$10^{16}$~s$^{-1}$), the
parameter $\mu$ in the range of ultrasonic frequencies is fairly small, and
the function $\Gamma(1/H)$ has giant resonant oscillations. This shape of
$\Gamma(1/H)$ is usual for any electron spectrum described by Equation (1.1).

It should be noted that the resonance attenuation of
acoustic waves in ordinary metals in a magnetic field is
observed only if the charge carriers drift along the
direction of the wave vector [35].

In the case of transverse acoustic wave polarization, $\bs{u}=(0,u_y,u_z)$,
the external magnetic field $\bs{H}=(0,H_y, H_z)$ is contained only in
expressions for acousto-electric coefficients, hence
$$
E_\alpha=\frac{m\omega^2}{e}u_\alpha+\xi j_\alpha.
$$

Having excluded $\tilde{E}_\alpha$ using Equation (4.8), we obtain
%4.23
\ba
j_y(1-\xi\tilde{\sigma}_{yy})-j_z\xi\tilde{\sigma}_{yz}
&=&
\left(
k\omega\tilde{a}_{yy}+\frac{m\omega^2}{e}\tilde{\sigma}_{yy}
\right)u_y
\nonumber\\
&+&
\left(
k\omega\tilde{a}_{yz}+\frac{m\omega^2}{e}\tilde{\sigma}_{yz}
\right)u_z,
\nonumber\\
-j_y\xi\tilde{\sigma}_{zy}+j_z(1-\xi\tilde{\sigma}_{zz})
&=&
\left(
k\omega\tilde{a}_{zy}+\frac{m\omega^2}{e}\tilde{\sigma}_{zy}
\right)u_y
\nonumber\\
&+&
\left(
k\omega\tilde{a}_{zz}+\frac{m\omega^2}{e}\tilde{\sigma}_{zz}
\right)u_z.
\ea

Taking a combination of these equations with the elasticity
equations (4.3), we obtain the equation system, whose
self-consistency condition
%4.24
\be
\left|{\scriptsize
\begin{array}{cccc}
1-\xi\tilde{\sigma}_{yy}\!\!\!\!&\!\!\!\!
-\xi\tilde{\sigma}_{yz}\!\!\!\!&\!\!\!\!
\chi_{yy}\!\!\!\!&\!\!\!\!
\chi_{yz}
\\
-\xi\tilde{\sigma}_{zy}\!\!\!\!&\!\!\!\!
1-\xi\tilde{\sigma}_{zy}\!\!\!\!&\!\!\!\!
\chi_{zy}\!\!\!\!&\!\!\!\!
\chi_{zz}
\\
(i\omega m/e)+ik\xi\tilde{b}_{yy}\!\!\!\!&\!\!\!\!
ik\xi\tilde{b}_{yz}\!\!\!\!&\!\!\!\!
(\omega^2-s^2_yk^2)\rho+\phi_{yy}\!\!\!\!&\!\!\!\!
\phi_{yz}
\\
ik\xi\tilde{b}_{zy}\!\!\!\!&\!\!\!\!
(i\omega m/e)+ik\xi\tilde{b}_{zz}\!\!\!\!&\!\!\!\!
\phi_{zy}\!\!\!\!&\!\!\!\!
(\omega^2-s^2_zk^2)\rho+\phi_{zz}
\end{array}}
\right|=0
\ee
yields damping parameters of the acoustic wave and the
co-moving electromagnetic wave. Here $s_y=(\lambda_{yzyx}/\rho)^{1/2}$ and
$s_z=(\lambda_{zxzx}/\rho)^{1/2}$ are the velocities of $y$- and
$z$-polarized sound, respectively;
%4.25)
\ba
\chi_{\alpha\beta}&=&
-k\omega\tilde{a}_{\alpha\beta}
-\frac{m\omega^2}{e}\tilde{\sigma}_{\alpha\beta},
\nonumber\\
\phi_{\alpha\beta}&=&
ik
\left[
k\omega\tilde{c}_{\alpha\beta}+\frac{m\omega^2}{e}\tilde{b}_{\alpha\beta}
\right],
\ea
and the components of the elastic tensor, $\lambda_{yxzx}$ and
$\lambda_{zxyx}$, equal to zero if, for example, the $xy$ plane is a crystal
symmetry plane [44]. Otherwise these components should be taken into
account, but they do not essentially affect the result. This crystal
symmetry is implied in Equation (1.1).

The pronounced anisotropy of the electron spectrum in
layered conductors leads to different attenuation lengths
of sound with polarizations perpendicular and parallel to
the layers, in the limit of small $\eta$, the displacement of
ions along the normal to the layers decays over a length $\eta^{-2}$
times larger than a wave with $y$-polarization. One can
easily prove that expansions in powers of $\eta$ of the
components of acousto-electronic tensors with one or two
$z$ indices start with terms of second or higher order.
Omitting in Equation (4.24) terms of the order higher than
two with respect to $\eta$, we obtain
%4.26
\ba
[(\omega^2-s_y^2k^2)\rho+\phi_{yy}](1-\xi\tilde{\sigma}_{zz})
\chi_{yy}
\left[
\frac{i\omega m}{e}+ik\xi\tilde{b}_{yy}
\right]&?&
\nonumber\\
{} [(\omega^2-s_z^2k^2)\rho+\phi_{zz}]
\left[
1-\xi\tilde{\sigma}_{zz}-\chi_{zz}\frac{i\omega m}{e}
\right]
&=&0\nonumber\\
\ea
Since Equation (4.26) is factored, acoustic waves with $y$-
and $z$-pola\-ri\-za\-tion do not interfer in this approximation.
By equating to zero the first miltiplicator in (4.26), we
obtain the equation for $k=\omega/s_y+k_2$, from which follows
%4.27
\ba
k_2&=&\frac{i}{2\rho s_y^2(1-\xi\tilde{\sigma}_{yy})}
\biggl[
\xi k\omega(\tilde{a}_{yy}\tilde{b}_{yy}-\tilde{c}_{yy}\tilde{\sigma}_{yy})
\nonumber\\
&+&
\frac{m\omega^2}{e}(\tilde{a}_{yy}+\tilde{b}_{yy})k\omega\tilde{c}_{yy}
+\frac{m^2\omega^3}{ke^2}\tilde{\sigma}_{yy}
\biggr]_{k=\omega/s_y}.
\ea
The denominator in this equation is similar to that in the
equation for $k_1$, so the absorption of the $y$-polarized
wave has the same resonances as the longitudinal wave.

The deviation of the other root of Equation (4.26) from $\omega/s$
is proportional to $\eta^2$ when $\eta\to0$ and described by the
expression
%4.28)
\ba
k_3&=&\frac{i}{2\rho s_z^2}
\Biggl[
\frac{m\omega^2}{e}
\left(
\frac{\tilde{a}_{zz}}{1-\xi\tilde{\sigma}_{zz}}+\tilde{b}_{zz}
\right)
+
\left(
\frac{m\omega^2}{e}
\right)^2
\frac{s_z\tilde{\sigma}_{zz}}{1-\xi\tilde{\sigma}_{zz}}
\nonumber\\
&+&\frac{\omega^2}{s_z}\tilde{c}_{zz}
\Biggr]_{k=\omega/s_z}.
\ea

The transparency of the layered conductor for the acoustic
wave with the polarization parallel to the normal to the
layers occurs only at selected values of a magnetic field
when $\sin kD=-1$. If $\sin kD$ differs essentially from $-1$ the sound
attenuation rate has the form [45]:
%4.29)
\be
\Gamma=\frac{\omega\tau}{r}\eta^2\{
1+\sin kD+kr\eta^2(1-\sin kD)\}.
\ee
At a higher magnetic field, when $kr\ll1$, acousto-electronic
coefficients appear to be very susceptible to the magnetic
field orientation with respect to the layers [46]. If in
the expression for $\Lambda_{zz}$ and $v_z$, i.e.
%4.30, 4.31
\ba
\Lambda_{xx}(\bs{p})&=&\sum\limits_{n=1}^{\infty}
\Lambda_n(p_x,p_y)\cos\frac{anp_z}{\hbar};
\\
v_z(\bs{p})&=&-\sum\limits_{n=1}^{\infty}n\varepsilon_n(p_x,p_y)
\frac{a}{h}\sin\frac{anp_z}{\hbar}
\ea
the functions $\Lambda_n(p_x,p_y)$ and $\varepsilon_n(p_x,p_y)$ decrease
rapidly with $n$, the asymptotic form of the acousto-electronic
coefficients are essentially different at some angles $\theta$ between the
magnetic field and the normal to the layers. These are the values
$\theta=\theta_{\rm c}$ at which $\eta^2$ in the expansion in powers of
$\eta$ equal zero.  For $\tan\theta\gg1$ these terms turn to zero repeatedly
with a period $\Delta(\tan\theta)=2\pi\hbar/D_{\rm p}$, where $D_{\rm p}$ is
the Fermi surface diameter along the $p_y$ axis.  These oscillations are due
to the Larmour precession of electrons in strongly elongated cross-sections
of the Fermi surface which intersect a large number of cells in the
reciprocal lattice, while the oscillation period is associated with a change
of this number by unity.

In the case when the dispersion law is described by
Equation (4.19), $\Gamma(\eta)$ is given by
%4.32
\be
\Gamma=\eta^2\frac{Nm\omega v_0}{4\pi\rho s^2}
\frac{sl}{r^2\omega}
\frac{J_0^2(\zeta)\sin\theta}{1+\alpha^2J_0^2(\zeta)},
\ee
where
$$
\alpha=\eta^2\frac{s\omega_0^2\tau}{c^2\omega},\quad
\zeta=\frac{av_0m}{h}\tan\theta.
$$

\begin{figure}
\centering
\includegraphics[width=10cm]{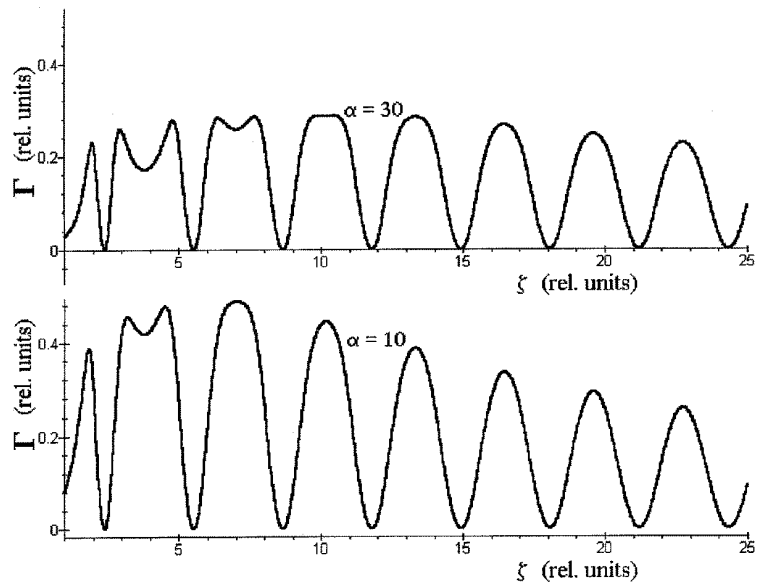}
\caption{The angular dependence of the sound attenuation rate at
$kr\ll1$.}
\end{figure}

If the plasma frequency comparable with the value
$\omega_0\sim10^{15}$--$10^{16}$~s$^{-1}$ for a ``good metal'', the
parameter $\alpha$ is generally not small in spite of a small value of
$\eta$.  This complicates the form of the angular oscillations of $\Gamma$
(Fig.~5) [46].

One can easily find that the last term in expression (4.28)
is a factor of $(v_{\rm F}/s_z)^2$ larger than other term in brackets. If
$\theta$ is different from $\theta_{\rm c}$, the following expression can be
derived for $\Gamma=\mbox{Im}\,k_3$ for $kr\ll 1$ and $\omega\tau\ll1$:
%4.33)
\be
\Gamma=\frac{N\varepsilon_{\rm F}}{\rho s_z^3}\omega^2\tau\eta^2.
\ee
But at $\theta=\theta_{\rm c}$ the acoustic attenuation length
$l_{\rm at}=1/\Gamma$ is considerably larger because $\Gamma$ takes the form:
%4.34)
\be
\Gamma=\frac{N\varepsilon_{\rm F}}{\rho s_z^3}\omega^2\tau\eta^2
\left[
\eta^2+(kr)^2+
\left(
\frac{s_z}{v_{\rm F}}
\right)^2
f(\eta)
\right].
\ee
The latter term in Equation (4.34) is due to the mismatch
between the zeros of the functions $\tilde{a}_{zz}(\theta)/\eta^2$ and
$\tilde{b}_{zz}(\theta)/\eta^2$, on one side, and
$\tilde{c}_{zz}(\theta)/\eta^2$, on the other side, at $\eta\to 0$.

For an electron spectrum of the form (1.1), (4.30), (4.31)
the acousto-electronic coefficients $a_{zz}$ and $b_{zz}$ tend to zero
at $\eta\to 0$ faster than $\eta^2$, i.e. $f(\eta)$ also tends to zero at a
small $\eta$. Strictly speaking, this is the main feature of the electron
spectrum (1.1). For this reason we retain the parameters $\tilde{a}_{zz}$
and $\tilde{b}_{zz}$ in the final formulas for $k_3$, although this does not
correspond to the actual accuracy of the formulas, given the electron
spectrum described by Equation (1.1).

If $\eta$ is not infinitesimal, but satisfies the condition
%4.35)
\be
\frac{\omega c}{\omega_0s_z}(\omega\tau)^{-1/2}\ll\eta\ll 1,
\ee
the term $\xi\tilde{\sigma}_{zz}$ in the denominator of Equation (4.28)
cannot be omitted. For $kr\gg1$ the damping rate of the sound with
$z$-polarization may have resonances if
%4.36)
\be
\frac{\omega c}{\omega_0s_z}(\omega\tau)^{-1/2}\ll\eta
\ll (kr)^{-1}\ll 1.
\ee

The condition of the magnetoacoustic resonance in rather
strict for tetrathiafulvalene salts, which have been
extensively investigated recently. In these compounds the
mean free path is $10^{-3}$--$10^{-2}$~cm, and resonances can be observed at
acoustic frequencies of the order of $10^9$~s$^{-1}$. But
the effect of field orientation on the sound absorption can
be observed in such layered materials at acoustic
frequencies of the order of $10^8$~s$^{-1}$ because for $kr\ll1$
the ratio of the electron mean free path to the acoustic
wavelength is not essential, and only the condition $r\ll l$,
which is fulfilled in a field of 10--20 T, is obligatory.

The specific behaviour of damping of acoustic waves with
different polarization can be used in filters transmitting
waves of a definite polarization, and sound absorption may
be a very accurate tool for studying electron spectra in
layered conductors.

If an electron drifts along the sound wavevector (for
instance, the sound wave propagates along the $y$-axis) the
sound decrement reduces in $(kl\eta)^2$ times for $r/l\ll kr\eta\ll1$ [47].
The solution of the kinetic equation in this case takes the form
%4.37)
\be
\psi=
\{
\exp(\nu T+i\bar{\bs{k}}\bar{\bs{v}}T)-1
\}^{-1}
\int\limits_{t}^{t+T}\mbox{d}t'\,g(t')
\exp\{i\bs{k}[\bs{r}(t')-\bs{r}(t)]\},
\ee
where $g(t)=\Lambda_{ji}(t)k_iu_j+e\bs{v}(t)\tilde{E}$.

At $1\ll kl\eta\ll l/r$ in the expansion in the powers of $\nu T$ and
$$
\bar{\bs{k}}\bar{\bs{v}}T=\int\limits_{0}^{T}\mbox{d}t\,
\bs{kv}(t)
$$
of
the factor in front of the integral, the terms proportional to
$\bar{\bs{k}}\bar{\bs{v}}T$ are the
most essential. Finally, in the case when the charge carriers drift along
$\bs{k}$ with the velocity
$\bar{v}_y=\bar{v}_z\tan\theta\cong\eta v\tan\theta$, in the expression for
the rate of sound attenuation $\nu T$ should be replaced by
$kr\eta\tan\theta$.  If $kr\eta\tan\theta\gg1$, i.e.  during the free path
time an electron is capable of drifting along the sound wavevector at
distance which exceeds significantly the sound wavelength; the
magnetoacoustic resonance predicted and studied theoretically by Kaner,
Privorotsky and one of the authors of this paper [35] takes place. The
resonance occurs at $\bar{\bs{k}}\bar{\bs{v}}T=2\pi n$ and in contrast to
the case of an ordinary metal the amplitude of the resonant oscillations is
determined by the parameter $kr\eta$ rather than by $kr$.

The formulas given above are valid when $\cos\theta\gg cD_{\rm p}/eHl$. If
$\theta$ is close to $\pi/2$, i.e. $\cos\theta$ is so small that an electron
has no time to make a total rotation along the orbit in a magnetic field,
then the components of the tensors $\tilde{\sigma}_{yy}$ and
$\tilde{\sigma}_{zz}$ are close to their values in the absence of a magnetic
field. This results from the fact that in the quasi-two-dimensional
conductor only the projection $H_z$ affects the charge carriers dynamics,
and at $\eta\ll1$ the component $H_y$ manifests itself only in small
corrections in the parameter $\eta$.

At $\theta=\pi/2$ the dependence of the sound damping decrement on the
magnetic field magnitude is present only in the terms that
vanish when $\eta\to 0$, and the magnetoacoustic effects are
pronounced in the case of the shear wave with the ionic
displacement along a normal to the layers only. In the
range of a sufficiently strong magnetic field ($kr\ll1$) the
attenuation rate for the wave with such polarization
depends essentially on the magnitude of a magnetic field
and its orientation with respect to the layers, and in the
$\theta$-dependence of $\Gamma$ sharp peaks and dips appear. At
$\tan\theta\gg1$ they repeat periodically with the period
$\Delta(\tan\theta)=2\pi\hbar/D_{\rm p}$.  The concrete form of the
$\theta$-dependence of $\Gamma$ is analogous to the angular dependence of
the electromagnetic impedance for $kr\ll 1$.

When the acoustic waves propagate along the normal to the
layers, the Maxwell equations have the form
%4.38
\ba
\{1-\xi\tilde{\sigma}_{xx}(k)\}\tilde{E}_x
-\xi\tilde{\sigma}_{xy}(k)\tilde{E}_y
&=&
\xi\tilde{a}_{xj}(k)u_j-(u_yH_z-u_zH_y)\frac{i\omega}{c}
\nonumber\\
&-&\frac{m\omega^2u_x}{e};
\nonumber\\
-\xi\tilde{\sigma}_{yx}(k)\tilde{E}_x
\{1-\xi\tilde{\sigma}_{yy}(k)\}\tilde{E}_y
&=&
\xi\tilde{a}_{yj}(k)u_j+u_xH_z\frac{i\omega}{c}
\nonumber\\
&-&\frac{m\omega^2u_y}{e}.
\ea

It is easy to see that the electric field and the
components of the matrix $\tilde{\sigma}_{\alpha\beta}$ do not disappear
when $\eta\to0$ and, consequently, the induction mechanism of the sound
waves attenuation is more significant. The drift of conduction electrons
along the $z$-axis does not take place only for the magnetic field
orientation in the layers-plane.  If $\theta$ is not equal to $\pi/2$, at
$kr\eta\ll 1$ the displacement of charge carriers along the wavevector for
the period of motion in a magnetic field is much less than the sound
wavelength.  The acousto-electronic coefficients are of the same order of
magnitude as the analogous values for the case of weak spatial dispersion
being reduced in $kl\eta$ times, if $kl\eta\gg1$.

The dependence of $\Gamma$ on $H$ occurs only in the range of
magnetic fields when $kr\eta\gg1$. At $\tan\theta\ll leH/cD_{\rm p}$ an
ordinary magnetoacoustic resonance takes place and this is connected with
the charge carriers drift along the sound wavevector. At $\theta=\pi/2$
electrons drift in the $xy$-plane only, that is the direction orthogonal to
the wavevector.  In this case the sound attenuation rate oscillates with
$1/H$ which is analogous to the Pippard oscillations in metals
%4.39)
\be
\Gamma(H)=\frac{\eta\omega\tau}{r}
\left\{
1-(kr\eta)^{-1/2}\beta\sin
\left(
\frac{kc\Delta D_{{\rm p}x}}{2eH}+\frac{\pi}{4}
\right)
\right\}.
\ee
Measurements of the period of these oscillations
%4.40)
\be
\Delta\left(
\frac{1}{H}
\right)
=\frac{4\pi e}{kcD_{{\rm p}x}},
\ee
enable the corrugation of the Fermi surface to be
evaluated. Here $D_{{\rm p}x}$ is the difference between the maximum
and minimum diameters along the axis $p_x$ at $p_y=0$. The
condition $kr\eta\gg1$ is very strict and can be satisfied in the
range of a magnetic field where $r\ll l$ only for $\eta\ge 1/10$. Therefore,
there are no grounds to expect that the clear dependence of
$\Gamma$ on the magnitude and orientation of a magnetic field can
be observed in the layered conductors.

%4.2
\subsection{Fermi-Liquid Effects}
Charged elementary excitations in conductors form a Fermi
liquid, and their energy spectrum is determined by the
distribution function for quasiparticles. As a result, the
response of the electron system in solids to an external
perturbation depends to a considerable extent on the
correlation function describing the electron--electron
interaction [48, 49]. Usually, the inclusion of the
Fermi-liquid interaction of charge carriers leads to a
renormalization of kinetic coefficients calculated on the
assumption that conduction electrons form a Fermi gas. In
some cases, however, the Fermi-liquid interaction approach
leads to specific effects such as spin waves in nonmagnetic
metals [50] and ``softening'' of metals in a strong magnetic
field [51]. As a rule, in stationary fields Fermi-liquid
effects results only in the renormalization of the charge
carriers energy within the gas-approximation. For this
reason, the analysis of the galvanomagnetic phenomena, when
the charge carriers are assumed to form a Fermi-gas with
an arbitrary electron energy spectrum, is equivalent to the
consideration of the problem in the Fermi-liquid theory.

We consider the propagation of acoustic waves and co-moving
electromagnetic waves with reference to the
electron--electron interaction [52--54]. Charge carriers
are supposed to form, not a Fermi-gas but a Fermi-liquid in
which the correlation effects are essential.

Now the energy of charge carriers is a function of the
density of elementary excitations. If the temperature is
not too low, smearing of the equilibrium Fermi distribution
function for charge carriers significantly exceeds the
spacing between energy levels quantized by a magnetic
field. In this case the energy of elementary excitations
carrying a charge in the quasiclassical approximation has
the form
%4.41
\be
\varepsilon=\varepsilon_0(\bs{p})+\lambda_{ij}(\bs{p})u_{ij}
+\Psi(\bs{p},\bs{r},t),
\ee
where $\varepsilon_0(\bs{p})$ is the charge carriers energy in undeformed
crystal in the gas approximation; the second term takes
into consideration the renormalization caused by the
deformation; and the function
%4.42
\be
\Psi(\bs{p},\bs{r},t)=\int
\Phi(\bs{p},\bs{p}')\delta f(\bs{p}',\bs{r},t)
\frac{2\mbox{d}^3p'}{(2\pi\hbar)^3}
\ee
accounts for the correlation effects associated with the
electron--electron interaction.

Here $\delta f=f(\bs{p},\bs{r},t)-f_0\{\varepsilon_0(\bs{p})\}$ is the
nonequilibrium correction to the Fermi distribution function
$f_0\{\varepsilon_0(\bs{p})\}$ for charge carriers in the undeformed
conductor.

The Landau correlation function  $\Phi(\bs{p},\bs{p}')$ can be expanded in
the complete set of orthonormal functions $\phi_n(\bs{p})$:
%4.43 4.44
\ba
\Phi(\bs{p},\bs{p}')&=&\sum\limits_{n=0}^{\infty}
\Phi_n\phi_n(\bs{p})\phi_n(\bs{p}');
\\
-\int\phi_n(\bs{p})\phi_m(\bs{p})
\frac{\partial f_0(\varepsilon_0)}{\partial\varepsilon_0}
\frac{2\mbox{d}^3p}{(2\pi\hbar)^3}&=&
\langle\phi_n(\bs{p})\phi_m(\bs{p})\rangle
\nonumber\\
&=&\delta_{nm},
\ea
and the nonequilibrium correction $\delta f(\bs{p},\bs{r},t)$ should be
found by means of the solution for the kinetic equation
%4.45
\be
\frac{\partial f}{\partial t}+
\frac{\partial f}{\partial \bs{r}}\frac{\mbox{d}\bs{r}}{\mbox{d}t}
+
\frac{\partial f}{\partial \bs{p}}\frac{\mbox{d}\bs{p}}{\mbox{d}t}=
W_{\rm col}\{f\}.
\ee

The collision integral $W_{\rm col}\{f\}$ disappears being applied to the
Fermi function $f_0(\varepsilon)$ depending on the charge carriers energy
with regard to the correlation effects. Below we shall
consider the collision integral in the $\tau$-approximation,
i.e.
%4.46)
\be
W_{\rm col}\{f\}=
\frac{f_0(\varepsilon)-f}{\tau}=\frac{\Psi}{\tau}
\frac{\partial f_0(\varepsilon)}{\partial\varepsilon} .
\ee

The equation of charge carriers motion in this case
is of the form
%4.47)
\be
\frac{\mbox{d}\bs{p}}{\mbox{d}t}=
e\bs{E}+\frac{e}{c}\frac{\partial\varepsilon}{\partial(\bs{p}\times\bs{H})}
-\frac{\partial\varepsilon}{\partial\bs{r}},
\ee
where the last tern accounts not only for the force of the
deformation but also for the Fermi-liquid interaction of
charge carriers.

In the linear approximation in a weak perturbation of
charge carriers the kinetic equation (4.45) takes the form
%4.48
\ba
\left\{
\frac{\partial \Psi}{\partial t}+e\bs{Ev}
+\Lambda_{ij}\frac{\partial u_{ij}}{\partial t}
-\frac{\partial\psi}{\partial t}-\bs{v}\frac{\partial\psi}{\partial\bs{r}}
-\frac{\partial\psi}{\partial t_{\rm H}}
\right\}
&\times&
\nonumber\\
\frac{\partial f_0(\varepsilon_0)}{\partial\varepsilon_0}
&=&
\frac{\psi}{\tau}\frac{\partial f_0(\varepsilon_0)}{\partial\varepsilon_0}
\nonumber\\
\ea
or
%4.49
\ba
\frac{\partial\psi}{\partial t}+\bs{v}\frac{\partial\psi}{\partial\bs{r}}
+\frac{\partial\psi}{\partial t_{\rm H}}+\frac{\psi}{\tau}
=
\frac{\partial\Psi}{\partial t}+e\tilde{\bs{E}}Bv
+\Lambda_{ij}\frac{\partial u_{ij}}{\partial t},
\ea
where
%4.50
\ba
\bs{v}&=&\frac{\partial\varepsilon_0}{\partial\bs{p}},\nonumber\\
\frac{\partial\psi}{\partial t_{\rm H}}
&=&
\frac{e}{c}\frac{\partial\varepsilon_0}{\partial(\bs{p}\times\bs{H})}
\ea
the function $\Psi(\bs{p},x)$ is found
with the help of the solution for the following integral equation:
%4.51)
\ba
\Psi(\bs{p},x,t)&=&
\int\Psi(\bs{p},\bs{p}')
\{\Psi(\bs{p}',x,t)-\psi(\bs{p}',x,t)\}\nonumber\\
&\times&
\frac{\partial f_0(\varepsilon_0')}{\partial \varepsilon_0'}
\frac{2\mbox{d}^3p'}{(2\pi\hbar)3},
\ea
where $\varepsilon_0'=\varepsilon_0(\bs{p}')$.

Using the Fourier representation
%4.52
\be
\Psi(\bs{p},\bs{r})=\sum\limits_{n=0}^{\infty}\int\mbox{d}k\,
\Psi_n(k)\phi_n(\bs{p})\exp\{i\bs{kr}\},
\ee
we obtain the following system of the algebraic equations
for the Fourier transforms $\Psi_n(k)$:
%4.53)
\ba
&\displaystyle\Psi_n(k)\{1+\Phi_n^{-1}\}
+
i\omega\langle\phi_n(\bs{p})\hat{R}
\sum\limits_{m}\phi_m(\bs{p})\rangle\Psi_m(k)=
\nonumber\\
&\displaystyle
-ik_ju_i(k)\langle\phi_n(\bs{p})\Lambda_{ij}(\bs{p})\rangle
+
\langle\phi_n(\bs{p})\hat{R}
[e\bs{v}\tilde{\bs{E}}(k)
-k_j\omega\Lambda_{ij}(\bs{p})u_i(k)]
\rangle,
\nonumber\\
\ea
where
%4.54)
\be
\hat{R}g=\int\limits_{-\infty}^{t}\mbox{d}t'\,g(t)
\exp\{i\bs{k}[\bs{r}(t')-\bs{r}(t)]+\nu(t'-t)\}.
\ee

The value of $\omega\tau$ is smaller than unity even in pure
conductors at low temperatures in a wide acoustic frequency
range, and the integral term in Equation (4.53) can he
taken into account in the perturbation theory. In the
asymptotic approximation in the small parameter $\omega\tau$, the
Fourier transform of the kinetic equation solution $\psi(k)$ is of
the form
%4.55)
\ba
\psi(k)&=&
\hat{R}\{e\tilde{E}_j(k)v_j+k_i\omega\Lambda_{ji}u_j(k)\}
-i\omega\hat{R}\sum\limits_{n}\Phi_n(1+\Phi_n)^{-1}
\nonumber\\
&\times&
[
\langle\phi_n\hat{R}e\tilde{E}_jv_j\rangle
-ik_iu_j\langle\phi_n\Lambda_{ji}\rangle
+\omega\langle\phi_n\Lambda_{ji}\rangle k_iu_j
]\phi_n
\ea
and the acousto-electronic coefficients can be found
easily. In the case, when $\bs{k}=(k,0,0)$ and displacement of ions is in
the $xy$-plane, they are given by
%4.56, 4.57, 4.58, 4.59
\ba
\sigma_{ij}(k)
&=&
\langle e^2v_i\hat{R}v_j\rangle
\nonumber\\
&-&
i\omega e^2\sum\limits_{n=1}^{\infty}\Phi_n(1+\Phi_n)^{-1}
\langle v_i\hat{R}\phi_n\rangle\langle\phi_n\hat{R}v_j\rangle;
\\
a_{ij}(k)
&=&
\langle ev_i\hat{R}\Lambda_{jm}\rangle
-
\sum\limits_{n=1}^{\infty}\Phi_n(1+\Phi_n)^{-1}
\langle ev_i\hat{R}\phi_n\rangle
\nonumber\\
&\times&
\left\{
\langle\phi_n\lambda_{jm}\rangle\frac{k_m}{k}
+i\omega\langle\phi_n\hat{R}\Lambda_{jm}\rangle
\right\};
\\
b_{ij}(k)
&=&
\langle e\Lambda_{i}\hat{R}v_j\rangle
\nonumber\\
&-&
i\omega e\sum\limits_{n=1}^{\infty}\Phi_n(1+\Phi_n)^{-1}
\langle\Lambda_{ix}\hat{R}\phi_n\rangle
\langle\phi_n\hat{R}v_j\rangle;
\\
c_{ij}(k)
&=&
\langle e\Lambda_{ix}\hat{R}\Lambda_{jx}\rangle
-
\sum\limits_{n=1}^{\infty}\Phi_n(1+\Phi_n)^{-1}
\nonumber\\
&\times&
\langle e\Lambda_{ix}\hat{R}\phi_n\rangle
\{\langle\phi_n\Lambda_{jx}\rangle
+i\omega\langle\phi_n\hat{R}\Lambda_{jx}\rangle\}.
\ea

Using Equations (4.56)--(4.59) we can easily determine the
rate of absorption for the acoustic wave. For brevity of
computations only, we assume that $\phi_1(-\bs{p})=\phi_1(\bs{p})$ and
$\phi_2(-\bs{p})=-\phi_2(\bs{p})$, while $\phi_n$ with $n>2$ are equal to
zero.  Taking into account that $\varepsilon(-\bs{p})=\varepsilon(\bs{p})$,
at $\eta\ll kr\eta\ll1$ we obtain for $\sigma_{yy}$ the following expression:
%4.60)
\be
\sigma_{yy}^{\rm liquid}=\sigma_{yy}^{\rm gas}(1-L),
\ee
where
%4.61)
\be
L=
\frac{i\omega\Omega\langle1\rangle}{\pi\nu kv}
\left\{
\frac{\Phi_1}{1+\Phi_1}\phi_1^2(1+\sin kD)
+\frac{\Phi_2}{1+\Phi_2}\phi_2^2(1-\sin kD)
\right\},
\ee
and $\phi_1$, $\phi_2$ and $v$ are the values of the functions $\phi_i(t)$
and the velocity modulus in the point where $\bs{kv}=\omega$.

At $kr\ll1$ the angular dependence of neither the electromagnetic
impedance nor the sound attenuation rate undergo
substantial changes due to the allowance for the
Fermi-liquid interaction between charge carriers.

Naturally, the acoustic transparency and the sound
attenuation rate of layered conductors with a
quasi-two-dimensional electron energy spectrum depend on
the intensity of the Fermi-liquid interaction between
charge carriers. The inclusion of the Fermi-liquid
interaction significantly affects the shape of the
resonance curve but the period of oscillations of $\Gamma$ with
$1/H$ and the positions of sharp peaks in the angular
dependence $\Gamma(\theta)$ remain unchanged when Fermi-liquid effects
are taken into account.

The magnitude of the Fermi-liquid interaction between
charge carriers can be determined from measured
electromagnetic and acoustic impedances either for
different wave frequencies or at sufficiently low
temperatures, when effects of the charge carriers energy
quantization are manifested clearly.

%5
\section{POINT-CONTACT SPECTROSCOPY OF LAYERED\\ CONDUCTORS}
%5.1
\subsection{Point-Contact Investigation of Electron Energy Spectrum}
In 1965 Sharvin [55] studied the dynamics of conduction
electrons by using a magnetic field for longitudinal
focusing of carriers injected into a metal from a point
contact.  Figure 6 shows the schematic diagram of the
circuit for longitudinal electron focusing, in which an
uniform magnetic field $\bs{H}$ is directed along the line
connecting two point contacts, viz., the emitter {\it E}
and the collector {\it C} situated at the opposite surfaces
of a thin plate. The longitudinal electron focusing was
first observed by Sharvin and Fisher [56].

\begin{figure}
\centering
\includegraphics[width=10cm]{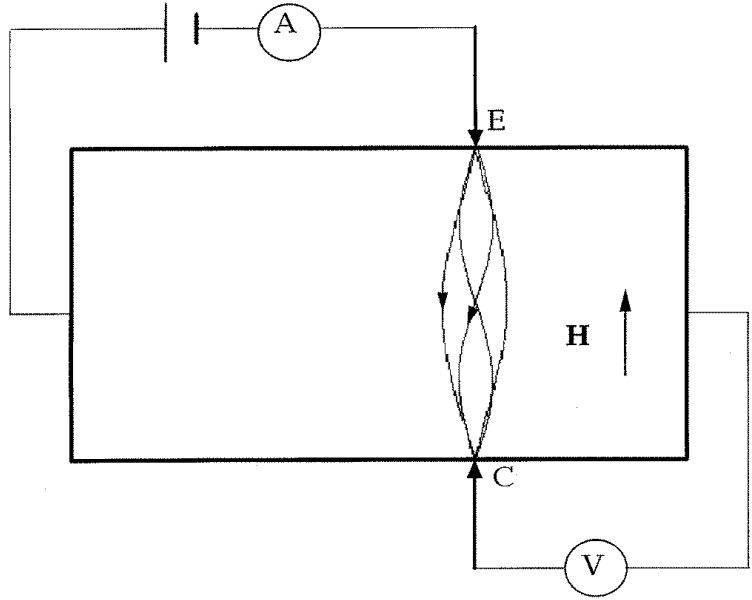}
\caption{The schematic diagram of the circuit for observing
longitudinal electron focusing.}
\end{figure}

Another possibility for observing focused electron beams in
metals is associated with the geometry of the experiments
in which the magnetic field $\bs{H}$ is directed at a right
angle to the line connecting the contacts at the same
surface of the sample (transverse electron focusing). This
idea was proposed by Pippard [57] and first realized
experimentally by Tsoi [58]. The diagram of a circuit for
transverse electron focusing is shown in Fig.~7.

\begin{figure}
\centering
\includegraphics[width=10cm]{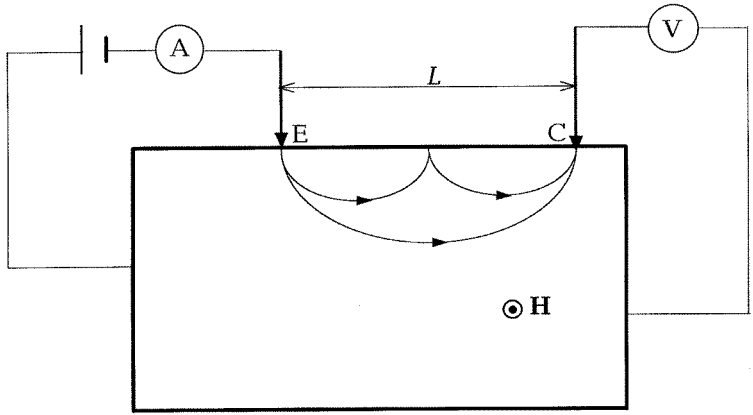}
\caption{The schematic diagram of the circuit for observing
transverse electron focusing.}
\end{figure}

It was noted even in first experimental [55, 58] and
theoretical [59, 60] publications that electron focusing, a
ballistic effect in its origin, is extremely sensitive to
the energy-momentum relation for charge carriers, and the
electron focusing signal might have extrema due to
electrons belonging to open Fermi surface cross-sections
[60]. A peculiar feature of the further analysis is the
quasi-two-dimensional nature of the electron energy
spectrum which is responsible for a significant difference
in both the amplitude and the shape of electron focusing
for the layered conductors and for metals with weakly
anisotropic conducting properties [61]. This difference is
associated with a small displacement of electrons in the
direction perpendicular to the layers over the time of
their motion from one point contact to another, and as a
result, with the dependence of the electron focusing signal
on the relation between the above displacement and point
contact diameters.

Let us consider a standard experimental geometry for
electron focusing observations, when the current point
contact (emitter {\it E}) and the measuring point contact
(collector {\it C}) are mounted either on the same surface
or on the opposite surfaces of a plate. The quantity under
investigation is the difference (measured by a voltmeter)
in electrochemical potentials of the collector and a
periphery point of the sample as a function of the
magnitude and direction of the magnetic field $\bs{H}$.

In order to determine quantity $U$, measured with the aid
of potential point contacts, let us consider a stationary
nonequilibrium electron state described by the distribution
function
%5.1)
\be
f(\bs{p},\bs{r})=f_0(\varepsilon)
-e\frac{\partial f_0}{\partial \varepsilon}
(x_{\bs{p}}-\phi(\bs{r}))
\ee
which satisfies the Boltzman equation (2.2) supplemented by
the boundary condition on the sample surface $\bs{R}\in\Sigma$
%5.2)
\ba
f(\tilde{\bs{p}},\bs{R})
\!\!\!&=&\!\!\!
f(\bs{p},\bs{R})
\nonumber\\
\!\!\!&+&\!\!\!
\int\mbox{d}^3\bs{p}'\,\Theta(-v_n')W(\bs{p},\bs{p}')
[f_{\bs{p}'}(\bs{p}',\bs{R})
-f(\bs{p},\bs{R})]\Theta(\bs{R}\notin S_i) \nonumber\\
\!\!\!&+&\!\!\!
\sum\limits_{i}f^{(i)}(\bs{p},\bs{R})\Theta(\bs{R}\in S_i),
\ea
where $v_n=\bs{vn}$; $\bs{n}$ is the interior normal to the
specimen boundary; $\Theta(x)$ is, as before, the Heaviside
function, $\Theta(\bs{R}\in S_i)$ is a unit function that differs
from zero for values of $\bs{R}$ belonging to the plane $S_i$ of
the $i$-th contact opening; $\Theta(\bs{R}\notin
S_i)=1-\Theta(\bs{R}\in S_i)$; $f^{(i)}(\bs{p},\bs{R})$ is the
electron-distribution function on the plane $S_i$. The electron
monenta $\tilde{\bs{p}}$ and $\bs{p}$ are connected through the
specular reflection conditions:

\begin{figure}
\centering
\includegraphics[width=10cm]{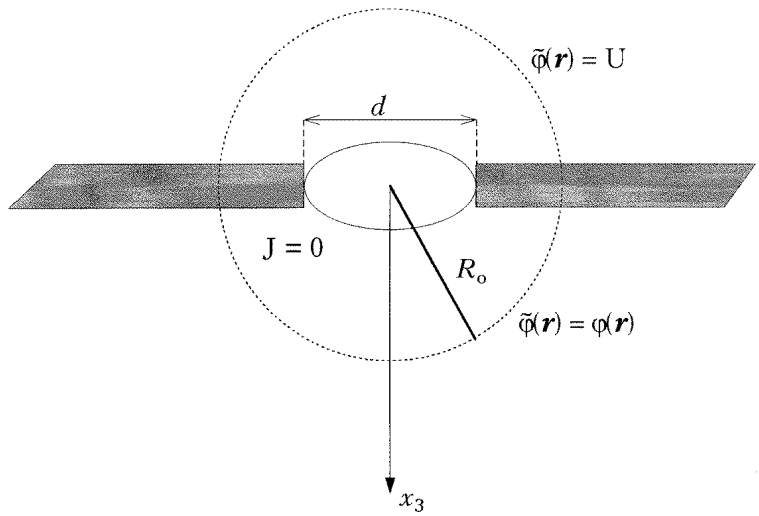}
\caption{A model of the measuring contact (collector {\it C})
shaped as a circular orifice of a diameter $d$.}
\end{figure}

%5.3
\be \varepsilon(\bs{p})=\varepsilon(\tilde{\bs{p}});\quad
[\bs{pn}]=[\tilde{\bs{p}}\bs{n}]. \ee The Equation (5.2) takes
into account the surface scattering of electrons and the injection
of charge carriers into the sample through the current and
potential contacts.  The kernel of the integral operator
$W(\bs{p},\bs{p}')$ (the surface scattering indicatrix) depends on
the nature of the surface scattering (see [62--64]). The function
$f^{(i)}(\bs{p},\bs{R})$ depends on the contact shape and the
nature of its contamination, and must be found by solving a
kinetic equation in the contact region. The condition (5.2)
automatically ensures the current does not flow across the surface
and potential contacts, and is conserved in the plane of the
current orifice. The electric field
$\bs{E}(\bs{r})=-\nabla\phi(\bs{r})$ in the sample is determined
from the electroneutrality condition (3.15).

In the case of a bulk layered conductor and the current
contacts of a large area which are placed at the opposite
surfaces of the sample we can ignore the surface effects
and not consider the boundary condition (5.2). It is well
known [65], that at distances from the sample boundaries
larger than the maximum mean free path of electrons $l$, an
uniform electric field $\bs{E}(\bs{r})=\mbox{const}$ automatically
satisfies the equation of electrical neutrality (3.15), and the
relation between a current density $\bs{j}$ and an electric field
$\bs{E}$ has an ordinary form (2.1).

Let a point contact {\it C} with an ideal voltmeter (of an
infinite resistance) connected to it, be situated at the
metal surface. The diameter $d$ of the contact is presumed
to be much less than $l$. A transient current which appears
at the initial moment will nullify as the potential far
from the contact (at distances $|\bs{R}_0|\gg d$) in the bulk
reaches a certain value $U$ (Fig.~8). We shall assume that the
second potential contact is placed at the specimen periphery
($\bs{r}\to\infty$, $x>0$) where $f(\bs{p})=f_0(\varepsilon)$ and
$\phi=0$.  Then the potential difference, as measured by voltmeter,
will be equal to $U$.  The condition of the absence of the current
$J$ through the measuring contact orifice can be written by
representing the distribution function in the form
%5.4, 5.5
\ba
f(\bs{p},\bs{r})&=&f_0(\varepsilon)
-e\frac{\partial f_0}{\partial\varepsilon_p}
\left\{
\begin{array}{ll}
\tilde{\chi}_{\bs{p}}(\bs{r})-\tilde{\phi}(\bs{r}),&v_3<0\\
U-\tilde{\phi}(\bs{r}),&v_3>0
\end{array}
\right.
;
\\
J&=&
\frac{2e^2}{(2\pi h)^3}
\int\limits_{S_{\rm c}}\mbox{d}^2\bs{r}\,
\langle
v_3(U-\tilde{\chi}_{\bs{p}}(\bs{r}))\Theta(v_3)
\rangle=0.
\ea
In this chapter the angle brackets
indicate the integration over Fermi surface
$$
\langle\ldots\rangle=
-\int\mbox{d}^3\bs{p}\frac{\partial f_0}{\partial\varepsilon}
\ldots\, ,
$$
$v_3$ is the velocity component perpendicular to the contact
plane (Fig. 8); $\tilde{\chi}_{\bs{p}}$ and $\tilde{\phi}(\bs{r})$
represent the function $\chi_{\bs{p}}$ and the potential $\phi$
being perturbed by the measuring contact.  The integration in
Equation (5.5) is carried out over the area of the contact orifice,
$S_{\rm c}=\pi d^2/4$. At distances $|\bs{R}_0|\gg d$ from the
contact we have
%5.6
\be
\tilde{\chi}_{\bs{p}}\cong
\left\{
\begin{array}{ll}
\chi_p,&x_3>0\\
0,x_3<0
\end{array}
\right. ;
\quad
\tilde{\phi}(\bs{r})\cong
\left\{
\begin{array}{ll}
\phi(\bs{r}),&x_3>0\\
U,x_3<0
\end{array}
\right. .
\ee
These relations are valid under the condition
$d\ll |\bs{R}_0|\ll L_0$, where $L_0$ is the characteristic spatial
scale of variation of the function $\chi_{\bs{p}}$ far from the
contact {\it C}.  When $d$ tends to zero, from Equations (5.5) and
(5.6) we obtain the following expression for the quantity $U$ [66]:
%5.7
\be
U=
\frac{\langle v_3\chi_{\bs{p}}(\bs{L})\Theta(-v_3)\rangle}%
{\langle v_3\Theta(-v_3)\rangle},
\ee
where the vector $\bs{L}$ defines the position of the measuring
contact at the surface.

The function $\chi_{\bs{p}}(\bs{r})$ in the solution (5.1) of the
kinetic equation (2.2) in the approximation of the relaxation time
$\tau$ can be presented in the form
%5.8)
\ba
\chi_{\bs{p}}(\bs{r})
&=&
F(\bs{r}-\bs{r}(t))\exp\left(\frac{\lambda-t}{\tau}\right)
\nonumber\\
&+&
\int\limits_{\lambda}^{t}\mbox{d}t'\,
\phi(\bs{r}+\bs{r}(t')-\bs{r}(t))
\exp\left(\frac{t'-t}{\tau}\right),
\ea
where $\lambda(\bs{r},\bs{p})\le t$ is the instant of time when an
electron is reflected from the specimen boundary at the point
$\bs{R}$ given by equation
%5.9
\be
\int\limits_{\lambda}^{t}\mbox{d}t'\,\bs{v}(t')
=\bs{r}(t)-\bs{r}(\lambda)=\bs{r}-\bs{R},
\ee
$F(\bs{r}-\bs{r}(t))$ is an arbitrary function of characteristics
whose value is preserved along the trajectory of motion of charge
carriers between two collisions with the surface. The
condition (5.2) enables us to obtain an explicit expression
for the function $F$.  Representing the distribution
function of electrons leaving the emitter in the form
$$
f^{(E)}(\bs{p},\bs{r})=f_0(\varepsilon)
-\frac{\partial f_0}{\partial\varepsilon}(\chi_{\bs{p}}^*-\phi(\bs{r})),
$$
we can  write the following expression for the value of the
function $\chi_{\bs{p}}$ in the plane $S_{\rm c}$,of the collector
orifice [67]:
%5.10
\be
\chi_{\bs{p}}(\bs{L})=\chi_{\bs{p}}^*
\sum\limits_{n}q^{n-1}\Theta(\bs{L}-\Delta \bs{R}_n\in S_{\rm E})
\left(
\exp-\frac{\Delta T_n}{\tau}
\right)
+\Delta\chi_{\bs{p}}(\bs{R}).
\ee
where
$$
q(\bs{p})=1-\int\mbox{d}^3\bs{p}'\,
\Theta(-v_n')W(\bs{p},\bs{p}')
$$
is the specular reflection parameter.

{\looseness=3
Equation (5.10) contains the summations over the number $n$
of collisions with the surface of electrons injected from
the emitter. The displacement of these electrons in the
plane parallel to the boundary during the time $\Delta T_n$ of
motion from contact to contact is $\Delta R_n$. The first term on
the right-hand side of Equation (5,10) is the nonequilibrium
component of the distribution function for electrons moving
strictly along ballistic trajectories. This term determines
the part of the signal on the measuring contact, which
depends nonmonotonically on the magnetic field and contains
singularities (extremes and kinks) known as electron
focusing lines. The second term in Equation (5.10) results from
electron scattering in the bulk as well as from scattering
on the surface of the conductor ($\Delta\chi_{\bs{p}}\to 0$, if
$\tau\to\infty$, $q=1$). This term is responsible for the emergence
of background in the $U(H)$ signal which varies smoothly with
$\bs{H}$.  In this version, we will consider a more informative
(from the experimental point of view) situation when $\Delta
T_n\ll\tau$ and $1-q\ll 1$, therefore we shall not adduce the
explicit form of the function $\chi_{\bs{p}}(R)$ whose contribution
to the electron focusing signal $U(H)$ can be discounted when the
above inequalities are satisfied.  In the case of pure contacts,
whose diameter $d$ is much less than $l$, and relatively weak
magnetic field, considered below, the affect of the field on
electron trajectories can be discounted over a distance of the
order $d$. In this case the function $\chi_{\bs{p}}^*$ is
independent of the coordinate and momentum and is equal to the
voltage $V$ applied to the contact [68].

}

In the case when the separation $L$ between the point
contacts is much greater than the contact diameter $d$, the
asymptotic behaviour of the integral in Equation (5.7) is
conditioned by the fact that only a small faction of charge
carriers starting from the emitter can reach the collector.
These are electrons that belong to a small region of the
Fermi surface near the points $p=p_0(\bs{H},n)$ specified by
equations:
%5.11)
\be
\Delta \bs{R}_n(\bs{p}_0)=\bs{L}.
\ee
The approximation of the step function
$\Theta(\bs{R}\in S_{\rm E})$ by the function
$\exp[-|\bs{R}|^2/(d/2)^2]$ for $d/L\to0$ allows us to find the
asymptotic value of the quantity $U$ by using the Laplace method.
Retaining the main term in the asymptotic series, it is convenient
to write the amplitude of the electron focusing signal in the form

%5.12)
\be
U(H)=V\sum\limits_{n=n_0}^{N}
\exp\left(-
\frac{\Delta T_n(p_0)}{\tau}
\right)
A(\bs{p}_0),
\ee
where $n_0(\bs{H})$, $N(\bs{H})$ are the minimum and maximum
numbers of collisions with the boundary by which an electron can
reach the collector for a given value of the magnetic field.
$A(\bs{p}_0)$ is the partial contribution to the electron focusing
signal from the electrons having a momentum $\bs{p}_0(\bs{H},n)$,
and getting from the current contact to the measuring one without
scattering.  The form of function $A$ is determined by the mutual
orientation of the crystal surface, line $\bs{L}$ connecting the
contacts, and the magnetic field vector $\bs{H}$.

We shall consider below the two main experimental
geometries when the axes of the contacts are either
parallel or orthogonal to the layers with a high electrical
conductivity, in addition to the expression for the
function $A(\bs{p}_0)$ which is valid for an arbitrary form of the
functions $\varepsilon_0$ and $\varepsilon_1$ characterizing the
Fermi surface (1.1), we shall give the results of calculations
based on a commonly used simple model (2.21).  This model enables
us to obtain the magnetic field dependence of $U$ in an explicit
form.

{\it 1. The contacts are on the crystal boundary
perpendicular to the layers with a high electrical
conductivity.} In this geometry, it is possible to observe
the transverse electron focusing by a magnetic field
applied parallel to the boundary (Fig. 9{\it a}). If the
vector $\bs{H}=(0,0,H)$ is oriented at an angle $\theta$ to the
plane of the layers, charge carriers perform a periodic motion over
the surface. Their displacement along the boundary $\Delta\bs{R}$
during the time $\Delta T$ between two collisions with the boundary
is the same for all segments of the trajectory:
%5.13
\ba
\Delta\bs{R}_n&=&
n\Delta\bs{R}=(\Delta y_n,\Delta z_n)=
\frac{nc}{eH}
\left(
D_x,
\frac{\partial S_{\rm seg}}{\partial p_z}
\right)=\bs{L};
\nonumber\\
\Delta T_n&=&
\frac{nc}{eH}
\frac{\partial S_{\rm seg}}{\partial\varepsilon_{p}},
\ea
where $n$ is the number of a collision with the boundary,
$D_x(p_y,p_z)$ is the chord of the section of the Fermi surface
$\varepsilon(\bs{p})=\varepsilon_{\rm F}$ cut by the plane
$p_z=\mbox{const}$, which is parallel to the normal
$\bs{n}=(1,0,0)$ to the boundary; $S_{\rm seg}$ is the area of the
segment cut by the chord $D_x$ on this section; and
$\bs{L}=(0,L_y,L_z)$.

\begin{figure}
\centering
\includegraphics[width=10cm]{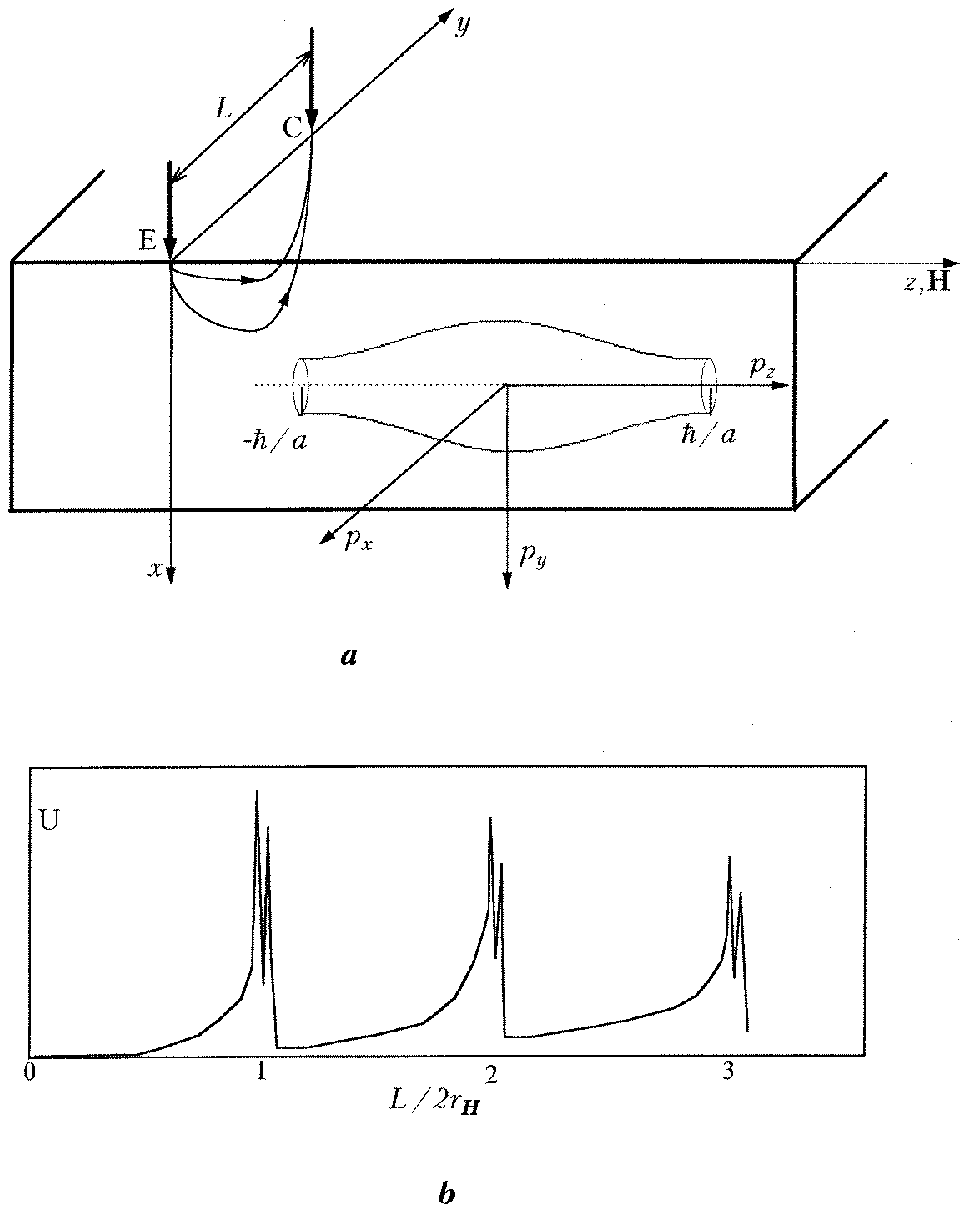}
\caption{Experimental geometry ({\it a}) and the shape of the
transverse ($\bs{L}\bot\bs{H}$) electron focusing peaks ({\it b})
in the case when the line $\bs{H}$ connecting the contacts on the
same face of the crystal lies in the plane of the layers, and the
magnetic field vector $\bs{H}$ is orthogonal to the layers.}
\end{figure}

{\looseness=2
The minimum number of the collisions $n_0$ with which an
electron can reach the collector is connected with the
maximum displacement $R_m=|\Delta R^{\rm max}|$ through the
inequalities $L/R_m<n_0<L/R_m-1$, and the maximum number of the
collisions $N$ tends to infinity and corresponds to charge carriers
sliding over the crystal surface at very small angles.

}

{\looseness=2
The magnitude of the electron focusing amplitude is
extremely sensitive to the relation between two small
parameters $\mu=d/L$ and $\eta=\varepsilon_1/\varepsilon_{\rm F}$,
that characterize the relative number of electrons participating in
the formation of the signal and anisotropy of the energy spectrum,
respectively. For $\mu\ll\eta$, only the charge carriers having
momenta from the narrow strips on the Fermi surface
($\Delta p_za/\hbar\simeq\mu$) can reach the collector ($a$ is the
distance between the layers).  In this case, the computation of the
quantity $A$ in Equation (5.12) leads to the following results:

}
%5.14)
\be
A(\bs{p}_0)=
\frac{D_x^2}{\langle v_x\Theta(-v_x)\rangle}
\left\{
\begin{array}{ll}
\displaystyle \frac{\pi}{4}\mu^2D^{-1},&D\ne0\\[8pt]
\displaystyle\frac{\sqrt{\pi}}{2}
\Gamma\left(\frac{1}{4}\right)\mu^{3/2}Q^{-1}; &D=0;\;Q\neq0
\end{array}
\right.
\ee
where
\ba
D&=&\frac{\partial D_x}{\partial p_y}
\frac{\partial^2S_{\rm seg}}{\partial p_z^2}+
\left(
\frac{\partial D_x}{\partial p_z}
\right)^2;
\nonumber\\
Q&=&
\frac{\partial^2S_{\rm seg}}{\partial p_z^2}
\left|
D_x\frac{\partial^2D_x}{\partial p_y^2}
\right|^{1/2}.
\nonumber
\ea
The peaks of the electron focusing signal correspond to the
values of the field $H_n$ satisfying the condition (5.11) for
which $D=0$.

If the large-diameter contacts satisfy the condition
$1\gg\mu\gg\eta$ and the angle $\theta$ between the vector $\bs{H}$
and the Fermi surface cylinder axis is less than $\eta$, the drift
$\Delta z_n$ along the $\bs{H}$-direction is smaller than the
contact diameter $d$ for all electrons leaving the emitter.  In
other words, electrons with any value of the component $p_z$ can
get from contact to contact. An increase in the number of electrons
hitting the collector results in a stronger signal as compared to
the situation when $\mu\ll\eta$. Assuming that the Fermi surface is
cylindrical ($\varepsilon(\bs{p})=\varepsilon_0(p_x,p_y)$) in the
main approximation in the parameter $\eta/\mu\ll1$, we obtain the
following expression for the function $A$:

%5.15)
\be
A=\frac{D_x^2}{\langle v_x\Theta(-v_x)\rangle}
\left\{
\begin{array}{ll}
\displaystyle\sqrt{\pi}\mu
\left|
\frac{\partial D_x}{\partial p_{y0}}
\right|^{-1},
&\displaystyle
\frac{\partial D_x}{\partial p_{y0}}\ne0\\[8pt]
\displaystyle\frac{\Gamma(1/4)}{2^{1/4}}\mu^{1/2}
\left|
D_x\frac{\partial^2 D_x}{\partial p^2_{y0}}
\right|^{-1/2},
&\displaystyle
\frac{\partial D_x}{\partial p_{y0}}=0
\end{array}
\right.
\ee

For $\bs{H}\bot\bs{L}$ (Fig. 9{\it a}), the electron focusing
signal has peaks for those values of a magnetic field $H=H_n$ at
which electrons, reaching the collector, have the maximum
displacement along the surface and correspond to the
extremal Fermi surface diameter ($D_x^{\rm ext}=D_x(p_{0e})$)
[58--60].  However, separate observation of transverse electron
focusing peaks associated with charge carriers belonging to
different extremal cross sections of Fermi surface is possible only
for contacts with a small diameter ($\mu\ll\eta$).

For a quadratic and isotropic energy-momentum relation for
electrons in the layers (2.21) and for $\bs{H}\bot\bs{L}\bot p_z$
Equation (5.14) and (5.15) take the simple form
%5.16, 5.17
\ba
\mu\ll\eta\nonumber\\
A&=&
\frac{p_{\rm F}h}{m\varepsilon_1a}
\left\{
\begin{array}{ll}
\displaystyle
\mu^2
\frac{\xi_n^3}{\arcsin(\xi_n)\sqrt{1-\xi_n^2}},
&\xi_n<1\\[8pt]
\displaystyle
\Gamma\left(
\frac{1}{4}
\right)
\pi^{3/2}\mu^{3/2},
&\xi_n=1
\end{array}
\right.
\\
\mu\gg\eta\nonumber\\
A&=&
\left\{
\begin{array}{ll}
\displaystyle
\frac{\sqrt{\pi}}{4}\mu
\frac{\xi_n^3}{\sqrt{1-\xi_n^2}},
&\xi_n<1\\[8pt]
\displaystyle
\Gamma\left(
\frac{1}{4}
\right)
2^{9/2}\mu^{1/2},
&\xi_n=1
\end{array}
\right.
\ea
where $n\ge n_0$; $\xi_{n_0}<1$, and $\xi_{n_0-1}>1$. The quantity
$\xi_n$ in Equation (5.16) has the two values
$\xi_n=L/2n r_{\rm H}(1\pm\eta)$ ($r_{\rm H}=cp_{\rm F}/eH$;
$p_{\rm F}=\sqrt{2m\varepsilon_{\rm F}}$) for the same value of
$n$, which correspond to the contributions from electrons belonging
to the maximum and minimum Fermi surface cross-sections.  We must
put $\xi_n=L/2nr_{\rm H}$ in Equation (5.17) which is valid to
within the terms proportional to $\eta/\mu\ll1$.

The period of motion along the trajectory is a multiple
valued function of $\xi_n$
($\Delta T^{(1)}=(T_{\rm H}/\pi)\arcsin\xi_n$;
$\Delta T^{(2)}=T_{\rm H}-\Delta T^{(1)}$; $T_{\rm H}=2\pi mc/eH$),
since the same displacement over the surface corresponds to the two
possible electron orbits with the height of segment $\Delta
x^{(1,2)}=r_{\rm H}(1\pm\sqrt{1-\xi_n^2})$.

Figure 9{\it b} is plotted as a result of numerical
calculations based on the Equations (5.12) and (5.16) and
shows the $U(H)$ dependence for $\mu\ll\eta$. The split structure of
the electron focusing peaks is due to a small difference
between the extremal diameters of the cross-sections cut by
$p_z=0$ and $p_z=\pm\hbar/a$.

The angle $\theta$ between the vector $\bs{H}$ and Fermi surface
cylinder axis significantly affects the form of electron
trajectories. As the value $\theta$ increases, the orbits become
elongated along the normal to the sample boundary and
acquire additional indentations associated with Fermi
surface corrugation (see Fig.~2{\it b}). For certain
magnetic field orientations, electron trajectories acquire
saddle points (halt points), and the period of motion tends
logarithmically to infinity (Equation (2.18)). With
increasing $\theta$, the electron drift along the magnetic field
is enhanced due to an increase in the period of motion
along the trajectory as well as an increase in possible
values of the electron velocity components along the vector
$\bs{H}$ [69].

In accordance with the evolution of electron orbits
described above, the rotation of the line connecting the
contacts in the plane of the surface leads to a
nonmonotonic dependence of the electron focusing signal
amplitude on the angle $\theta$ [60] and to the emergence of
additional electron focusing peaks associated with the
emergence of new extremal diameters on the Fermi surface
cross section satisfying Equation (5.11).

\begin{figure}
\centering
\includegraphics[width=10cm]{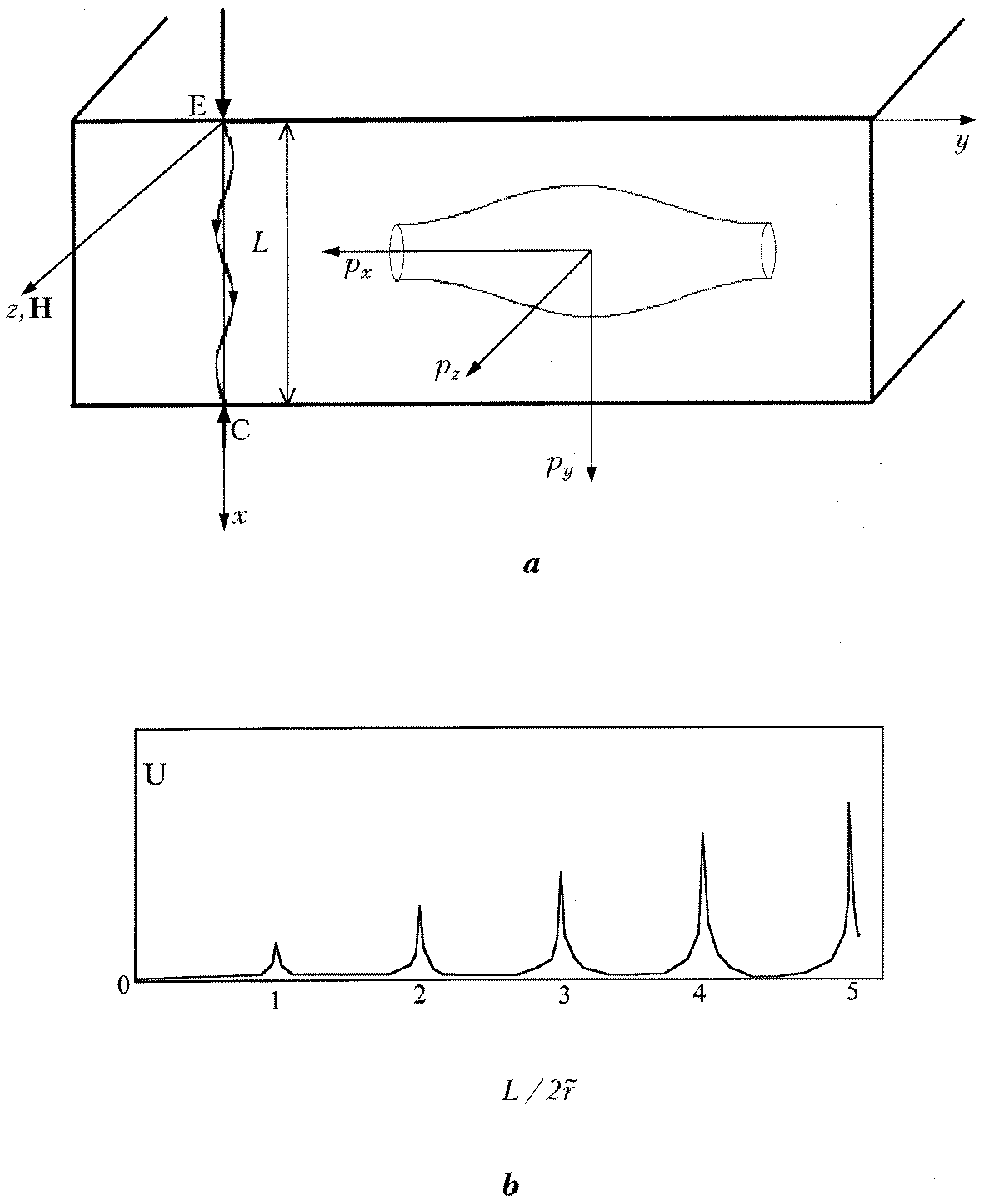}
\caption{Experimental geometry ({\it a}) and the shape of the
transverse ($\bs{L}\bot\bs{H}$) electron focusing peaks ({\it b})
in the case when the line $\bs{L}$ connects the contacts at
opposite faces of a thin plate, and the magnetic field vector
$\bs{H}$ lies in the plane of the layers.}
\end{figure}

In the limiting case when the vector $\bs{H}$ lies in the plane
of the layers ($\theta=\pi/2$), charge carriers belonging to open
Fermi surface cross sections move along periodic trajectories into
the bulk of the sample, which allows the observation of the effect
analogous to the longitudinal electron focusing with the help of
the point contact located on the second surface of thin plate with
the thickness $L<l$ (Fig.~10{\it a}). If both of the point
contacts are on the $x$-axis, the electron from the emitter can
reach the collector only if its displacement in the direction
orthogonal to this axis over the time $\Delta T$ of motion from one
surface of the plate to another is less then the contact
diameter $d$. For $d\to 0$ we have the conditions:
%5.18)
\ba
\Delta\bs{R}&=&
\left(
\frac{c}{eH}(p_{x0}(\lambda+\Delta T),v_{z0}\Delta T)
\right)=0,
\nonumber\\
\Delta x&=&\frac{c}{eH}
(p_{y0}(\lambda+\Delta T)-p_{y0}(\lambda))=L,
\ea
where $\lambda$ characterizes the position of an electron on the
Fermi surface at the moment of its ``start'' from the
emitter. When the condition $2\pi\hbar ck/eHa=L$ ($k=1,2,\ldots$)
is satisfied, the time $\Delta T=L/\bar{v}_y$ is multiple of the
period of motion $T_{\rm op}$.  In this case all the electrons have
zero displacement along the $y$-axis, and the amplitude of the
electron focusing signal attains its maximum value. For
$\bs{L}\bot p_z\bot\bs{H}$, only the electrons that do not interact
with the sample surface can get from contact to contact along the
ballistic trajectory, and we must put $n_0=N=1$ in Equation (5.12).
The partial contribution to the amplitude $A(\bs{p})$ is associated
with electrons for which the focusing conditions (5.18) are
satisfied:
%5.19
\ba
A(p)&=&
\frac{(m_{zz})\bar{v}^2_y}{\langle v_x\Theta(v_x)\rangle}
\frac{eH}{c}
\nonumber\\
&\times&
\left\{
\begin{array}{ll}
\displaystyle
\frac{\pi}{2}\mu^2\frac{L}{|v_x-v_x'|},
&\Delta T\ne kT_{\rm op}
\\[8pt]
\displaystyle
\frac{\sqrt{\pi}}{4}\mu T
\left[
1+
\left(
m_{zz}v_x\frac{1}{T}\frac{\mbox{d}T}{\mbox{d}p_z}
\right)^2
\right]^{-1/2},
&\Delta T=kT_{\rm op}
\end{array}
\right.
\nonumber\\
\ea
where
$$
\bar{v}_y=\frac{L}{\Delta T};\quad
m^{-1}_{zz}=\frac{\partial\varepsilon}{\partial p_z^2};\quad
\bs{v}'=\bs{v}(\lambda+\Delta T);
$$
$T_{\rm op}$ is the period of electron motion along an open
trajectory.

For the energy-momentum relation (2.21), Equation (5.19)
takes the form
%5.20
\be
A=
\left\{
\begin{array}{ll}
\displaystyle \frac{\mu^2}{\pi}
\frac{L}{r_{\rm H}}
\frac{1}{\eta'\sin\pi\gamma'},
&\gamma\ne k\\
\displaystyle\frac{\mu}{\sqrt{\pi}},
&\gamma=k\;(k=0,1,\ldots)
\end{array}
\right.
\ee
where $\gamma=L/(v_{\rm F}T_{\rm op})$;
$T_{\rm op}2\pi hmc/eHp_{\rm F}a$.

Note that the necessary condition for the nonmonotonic
dependence of the electron focusing signal on $H$ is that
the size, characteristic for the electron trajectory in the
direction of the ``openness'' of the Fermi surface
($\Delta y_m\simeq\tilde{r}\eta$; $\tilde{r}=\pi\hbar c/eHa$), must
be much larger than the contact diameter. In strong fields, when
$\tilde{r}$ becomes smaller than $d/\eta$, the signal $U$ depends
smoothly on $H$.  Figure 10{\it b} shows the results of numerical
calculations based on Equations (5.12) and (5.20). It can be seen
that electron focusing peaks are equidistant in the magnetic field
and have a symmetric shape.

\begin{figure}
\centering
\includegraphics[width=10cm]{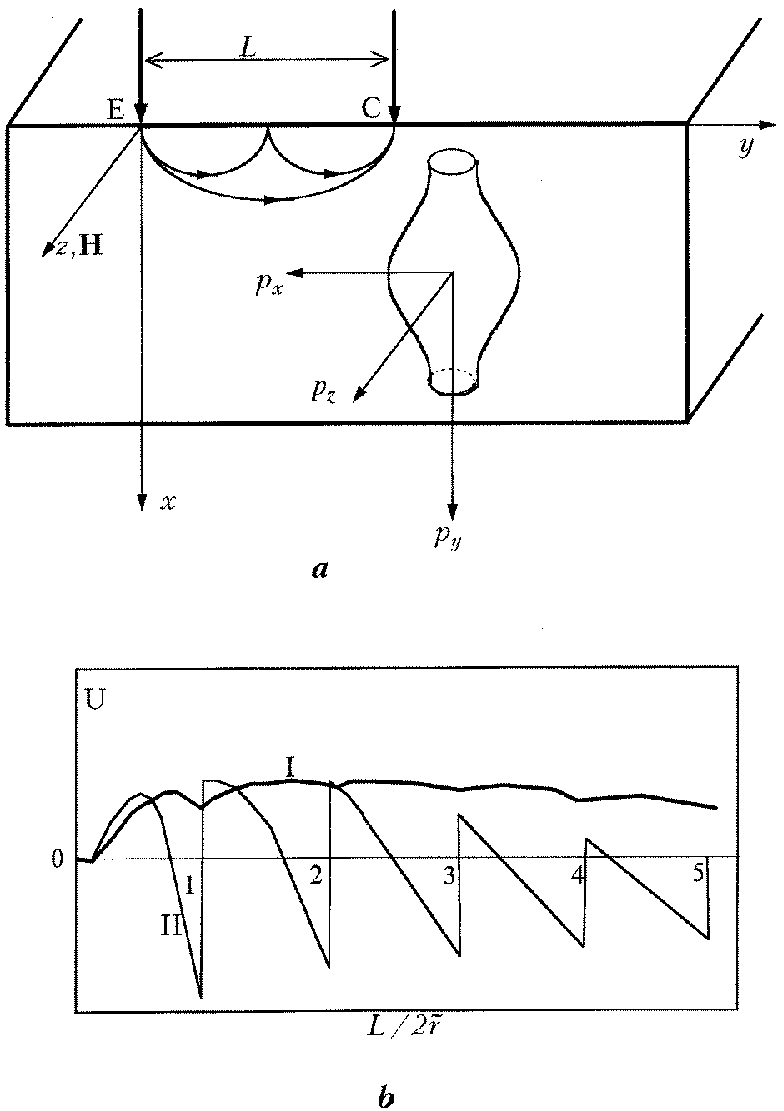}
\caption{Experimental geometry ({\it a}) and the shape of the
transverse ($\bs{L}\bot\bs{H}$) electron focusing signal (curve
{\it I}) and its derivative with respect to $\bs{H}$ (curve {\it
II}) in the case when the line $\bs{L}$ connecting the contacts on
the same face of the sample and the magnetic field vector $\bs{H}$
lies in the plane of the layers.}
\end{figure}

{\it 2. The crystal surface bearing contacts coincides with
a high-con\-du\-cti\-vity plane.} In this case, a thin layer
(whose thickness is of the order of $\tilde{r}\eta$) of ``jumping''
electron trajectories exists in a magnetic field $\bs{H}\bot p_y$
parallel to the boundary, in the transverse electron
focusing geometry (Fig. 11{\it a}), such electrons ensure a
ballistic transport of charge between point contacts
arranged at the same face of the crystal, and the function
$A$ is defined by relations (5.15). In the case under
investigation, the values of a magnetic field for which the
separation between contacts is multiple to the maximum jump
over the surface are preferred [60, 61], i.e.,
%5.21)
\be
L=\frac{2\pi hnc}{eH_na},\quad(n=1,2,\ldots).
\ee

For $H=H_n$, the $U(H)$ dependence has kinks rather than extrema
(curve {\it 1}in Fig.  11{\it b}) since electrons having
the maximum displacement along the surface over a period
approach the surface at small angles, and their
contribution to the emf measured by collector is small in
accordance with Equation (5.7). In the Fermi surface model
(2.21) in which the cross section $p_z=\mbox{const}$ does not
contain extremal diameters (and hence ordinary electron focusing
peaks are absent), the partial amplitude $A$ for $\bs{H}\bot\bs{L}$
and $H\neq H_n$ is given by
%5.22
\be
A=\mu^2
\frac{\zeta^2_n\sin\pi\zeta_n}%
{2F(\sqrt{\eta},\pi\zeta_n/2)\sqrt{1-\eta\sin^2(\pi\zeta_n/2)}},
\quad\zeta_n<1,
\ee
where
$$
\zeta_n=\frac{L}{2n\tilde{r}};\quad
\tilde{r}=\frac{\pi hc}{eHa}; \quad
n\ge n_0; \quad
\zeta_{n_0}<1;\quad
\zeta_{n_0-1}>1.
$$

The contribution of electrons, sliding along the surface
with a small velocity component normal to the boundary,
cannot be described by Equation (5.15) and must be taken
into account with the help of the next approximation in the
small parameter $\mu$:
%5.23
\ba
A^{\rm max}&=&
\frac{\sqrt{\pi}(2\pi\hbar/a)}{\langle v_x\Theta(-v_x)\rangle}
\mu^3
\frac{\dot{v}_xL}{nv_y^2(\partial^2S/\partial p_z^2)}
\nonumber\\
&=&
\pi^{-3/2}\mu^3
\frac{1}{K(\eta)\sqrt{1-\eta}},\quad v_x=0, \zeta_n=1,
\ea
where $K(k)$, $F(k,\varphi)$ are complete and incomplete elliptic
integrals of the first kind. The dot in Equation (5.23)
denotes the differentiation of electron velocity of motion
along the trajectory with respect to time. All the Fermi
surface characteristics in Equation (5.23) are taken at the
point $\bs{p}_0$ at which
%5.24
\be
v_x(\bs{p}_0)=v_z(\bs{p}_0)=0.
\ee
Apparently, it would be more convenient in this case to
investigate experimentally the derivative $\partial U/\partial H$
of the electron focusing signal with respect to a magnetic field,
which has jumps at the values $H_n$ determined by Equation
(5.21). Figure 11{\it b} shows the magnetic field
dependence of the electron focusing signal (curve {\it 1})
and its derivative (curve {\it 2}) for the simple model of
the Fermi surface (2.21).

In a magnetic field orthogonal to the crystal surface,
electrons injected by the emitter move along helical
trajectories to the bulk of the crystal. Setting the
collector and emitter at opposite faces of a thin plate and
directing the vector $\bs{H}$ along the line connecting the
contacts, we can observe the effect of the longitudinal
electron focusing (Fig. 12{\it a}). It can be seen that
with such a geometry, only the electrons for which the time
of motion $\Delta T=L/\bar{v}_z$ is a multiple of their period of
motion in a magnetic field ($\Delta T=kT_{\rm H}$; $k=1,2,\ldots$)
can get from contact to contact.

The peaks of the longitudinal electron focusing signal
correspond to the value of the field at which electrons
have the extremal value of the time-averaged velocity
component $\bar{v}_z$:
%5.25
\be
H_n=\frac{kc}{eL}
\left(
\frac{\partial S}{\partial p_{z0}}
\right)_{\rm extr},
\ee
where
$$
p_{z0}=-\frac{\pi\hbar}{2a}
\left(
\frac{\partial S}{\partial p_{z0}}
\right)_{\rm extr}
$$
is the extremal value of the derivative of the Fermi
surface cross-sections perpendicular to the axis.

In the case when the Fermi surface (1.1) is a body of
revolution, we obtain the following expression for the
partial amplitude:
%5.26
\ba
A&=&
\frac{m^*m_{zz}v_z^2}{\langle v_z\Theta(v_z)\rangle}
\nonumber\\
&\times&
\left\{
\begin{array}{ll}
\displaystyle
\sqrt{\pi}\mu\frac{v_z}{v_\bot},
&\displaystyle
\frac{\partial v_z}{\partial p_{z0}}=m^{-1}_{zz}\neq0;
\\[8pt]
\displaystyle
\frac{1}{2}
\Gamma
\left(
\frac{1}{4}
\right)
\mu^{1/2}
\left(
\frac{a}{\hbar}m_{zz}\sqrt{v_{\bot v_z}}
\right)^{-1},
&\displaystyle
\frac{\partial v_z}{\partial p_{z0}}=m^{-1}_{zz}=0,
\\[8pt]
&\displaystyle
p_{z0}=-\frac{\pi\hbar}{2a}
\end{array}
\right.
\nonumber\\
\ea
where
$$
v_\bot=\sqrt{v_x^2+v_y^2};\quad
m^*=\frac{1}{2\pi}\frac{\partial S}{\partial\varepsilon_p}.
$$

\begin{figure}
\centering
\includegraphics[width=10cm]{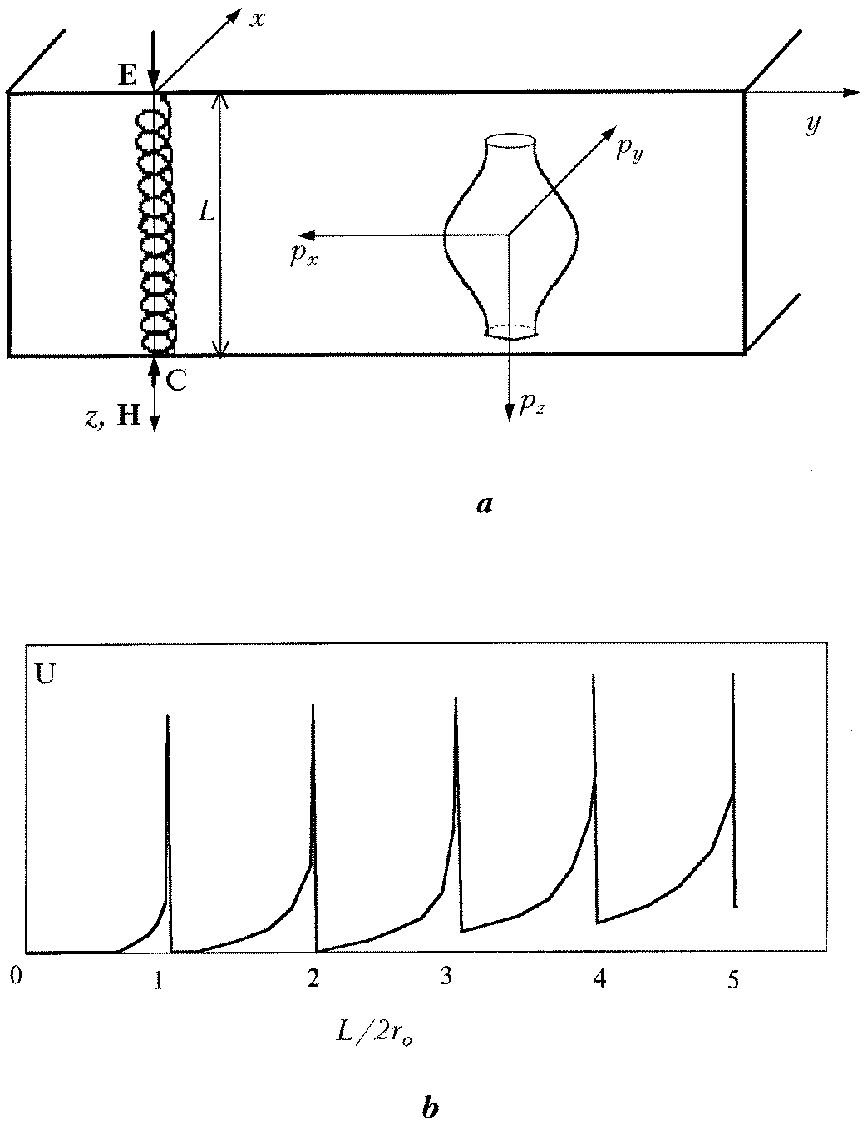}
\caption{Experimental geometry ({\it a}) and the shape of the
longitudinal ($\bs{L}\|\bs{H}$) electron focusing peaks ({\it b})
in the case when the line $\bs{L}$ connecting the contacts on the
same face of the crystal and the magnetic field vector $\bs{H}$ is
orthogonal to the plane of the layers.}
\end{figure}

If the function $\varepsilon_0$ is quadratic and isotropic. Equation
(5.26) can be transformed to
%5.27
\be
A=\sum\limits_{k=k_0}^{\infty}
\left\{
\begin{array}{ll}
\displaystyle
\frac{\sqrt{\pi}}{2}\mu\frac{\varepsilon_1a}{v_{\rm F}\hbar}
\frac{\lambda_k^3}{[(1-\eta\sqrt{1-\lambda_k^2})(1-\lambda_k^2)]^{1/2}},
&\lambda_k<1;
\\[8pt]
\displaystyle
\frac{\Gamma(1/4)}{4}\mu^{1/2}
\left(
\frac{a\varepsilon_1}{\hbar v_{\rm F}}
\right)^{1/2},
&\lambda_k=1
\end{array}
\right.
\ee
where
$$
\lambda_k=\frac{L}{2\pi kr_0};\quad
r_0=\frac{\varepsilon_1 aT_{\rm H}}{2\pi\hbar};\quad
\lambda_{k_0}<1;\quad
\lambda_{k_0-1}>1.
$$

The time of motion along the trajectory for an electron
reaching the collector over $k$ complete periods is
$\Delta T=Lh/\varepsilon_1 a\lambda_k$.  The explicit form of the
magnetic field dependence for the longitudinal electron focusing
signal can be easily obtained, if we substitute Equation (5.27)
into Equation (5.12) and put $n_0=N=1$ in it. The results of
numerical calculations are presented in Fig.~12{\it b}.

Thus, the magnetic field dependence $U(H)$ of the electron
focusing signal in the layered conductors is determined by
the size of the point contacts and their orientation,
relative to crystallographic axes. If the contacts are
arranged on the crystal surface perpendicular to the layers
with a high electrical conductivity, electron focusing can
be observed only in a magnetic field directed along this
surface. Setting both of the point contacts at the same
sample boundary, we can determine the diameters of closed
Fermi surface cross-sections from the positions of
transverse electron focusing peaks on the magnetic field
scale. The shape of the lines ($U(H)$ dependence) depends
significantly on the relation between two small parameters:
the ratio $\eta$ of the conductivity across the layers to the
conductivity in the layers-plane and the ratio $\mu$ of the
diameter of contacts to the separation between them.

In the case when vector $\bs{H}$ lies in the layers-plane, the
electrons injected by emitter are displaced along an open
trajectory to the bulk of the sample. In the thin plate
(the thickness of which is much smaller than the mean free
path), such electrons can be focused by a magnetic field at
the collector arranged on the sample face opposite to the
emitter. However, electron focusing in such geometry can be
observed only for very small contacts for which $\mu\ll\eta$.

If the plane of the surface bearing the contacts is
parallel to high conductivity layers, and the magnetic
field is orthogonal to the Fermi surface axis electrons
belonging to open Fermi surface cross sections move along
periodic trajectories along the crystal boundary. For
values of the field $H$ corresponding to the separation
between the electrodes that are multiple of the electron
displacement in an open trajectory over period $T_{\rm op}$, the
derivative $\mbox{d}U/\mbox{d}H$ of the electron focusing signal
undergoes jumps. The presence of singularities in the derivative
$\mbox{d}U/\mbox{d}H$ and not in the signal $U(H)$ can be explained
by the fact that electrons, for which the time between two
consecutive collisions with the boundary is close to $T_{\rm op}$,
reach the collector at small angles and their contribution to
$U(\bs{H})$ is small.

The effect of longitudinal electron focusing can be
observed in a magnetic field orthogonal to the surface. The
values of $H$ for which the signal $U(\bs{H})$ has peaks can be
used for determining the extremal value of the electron
velocity along the Fermi surface cylinder axis.

Thus, the electron focusing in the layered organic metals
being a source of information on their Fermi surface,
enable us to obtain, along with the extremal diameters and
velocities, the period of constant energy surfaces in the
direction of the ``openness''.  Another important advantage
of the electron focusing is the applicability of this
method to the analysis of surface scattering of electron.

%5.2
\subsection{Resistance of Point-Contact between Layered Conductors}
Two bulk metallic electrodes, contacting one another
(contiguously) over a small area, form an electrical
contact of small dimensions -- a point contact. When an
electric current is passed through such a system, it is
concentrated in a narrow region (system with concentration
of current), reaching densities of $10^{13}$--$10^{14}$~A/m$^2$.
The metal in this narrow region is not overheated
due to the effective heat flow to the banks (the
electrodes) of the contact, provided that mean free path
for energetic relaxation of electrons $l_\varepsilon$ is greater
than the characteristic (the largest) dimension of the narrow
region. Two possible asymptotic regimes can be
distinguished for the electric current through a point
contact: the ballistic regime (clean conductor) when the
mean free path for elastic relaxation $l_i$ is much greater
than $d$ and the diffusion regime (impure conductor) when
$l_i\ll d$ (where $d$ is the characteristic dimension of the
contact).

Let us consider a junction in the form of a single -- sheet
hyperboloid of revolution (Fig. 13{\it a})
%5.28
\be
F(x_1,x_2,x_3)=
\frac{\rho^2}{\nu^2}-\frac{x_3^2}{\beta^2}-1=0
\ee
subjected to a voltage $V$ and placed in a magnetic field
$\bs{H}$ directed at an angle $\theta$ with respect to the
$x_3$-axis ($\rho^2=x_1^2+x_2^2$) This model allows us to examine
the limiting cases:  the model of aperture with an infinitely thin
flat partition, dividing the two metallic half-spaces ($\beta\to0$)
(see Fig.~13{\it b}) and the model of a long channel (the length
of the channel is much larger than its diameter $d=2\nu$), filled
with metal and connecting the bulk metallic banks ($\nu\to0$).

In the semiclassical approximation the current
%5.29
\be
\bs{I}=\frac{2e^2}{(2\pi\hbar)^3}
\int\limits_{S_{\rm c}}\int\mbox{d}^3\bs{p}\,\bs{v}f(\bs{p})
\ee
flowing through a contact area $S_{\rm c}=\pi\nu^2$ should be
determined by means of the nonequilibrium electron-distribution
function $f(\bs{p},\bs{r})$ satisfying the Boltzman equation (2.2).

\begin{figure}
\centering
\includegraphics[width=10cm]{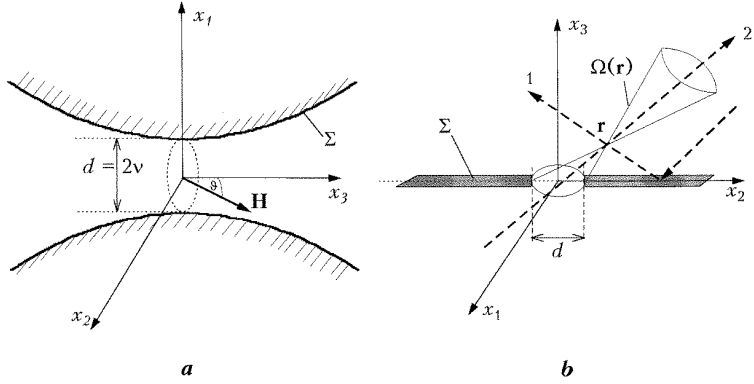}
\caption{A model of the point-contact in the form of a
single-sheet hyperboloid of revolution ({\it a}); its limiting
case ($\beta=0$) is a circular orifice ({\it b}); $\Omega(\bs{r})$
is the solid angle within which velocities of the electrons
passing the contact along the ballistic trajectories fit.}
\end{figure}

In the zeroth approximation in electron--phonon collision
integral, the distribution function
$f(\bs{p},\bs{r})=f_{\bs{p}}^{(0)}(\bs{r})$ can be presented in the
form [70]
%5.30)
\ba
f_{\bs{p}}^{(0)}(\bs{r})&=&
\alpha_{\bs{p}}(\bs{r})
f_0
\left(
\varepsilon(\bs{p})+e\Phi(\bs{r})-\frac{eV}{2}
\right)
\nonumber\\
&+&
(1-\alpha_{\bs{p}}(\bs{r}))
f_0
\left(
\varepsilon(\bs{p})+e\Phi(\bs{r})+\frac{eV}{2}
\right),
\ea
where $f_0$ is the Fermi distribution function, $\Phi(\bs{r})$ is
the electrical potential, and $\alpha_{\bs{p}}(\bs{r})$ signifies
the probability for an electron with the momentum $\bs{p}$ to
arrive at the point $\bs{r}$ from $+\infty$.  For
$eV\ll\varepsilon_{\rm F}$, the function  $\alpha_{\bs{p}}(\bs{r})$
satisfies a field-independent kinetic equation which, for the sake
of simplicity, can be written in the approximation of the mean
relaxation time $\tau$
%5.31
\be
\frac{\partial\alpha_{\bs{p}}(\bs{r})}{\partial t}+
\frac{\alpha_{\bs{p}}-\alpha}{\tau}=0.
\ee
The boundary condition to Equation (5.31) ensures that the
current does not flow through the surface $\Sigma$
%5.32
\be
\alpha_{\bs{p}}(\bs{r})=\alpha_{\tilde{\bs{p}}}(\bs{r}),\quad
\bs{r}\in\Sigma
\ee
and that equilibrium is restored in the electron subsystem
at the contact banks $|\bs{r}|\to\infty$
%5.33)
\be
\alpha_{\bs{p}}(\bs{r}\to\infty)=\Theta(x_3).
\ee
The momenta $\bs{p}$ and $\tilde{\bs{p}}$ satisfy Equations (5.3).
In order to avoid cumbersome calculations, we shall consider the
specular reflection of electrons at the boundary (Equation (5.32)).

Here
%5.34)
\be
\alpha(\bs{r},\bs{H})=\frac{\langle\alpha_{\bs{p}}\rangle}{\langle1\rangle};
\quad
\langle\ldots\rangle=
\frac{eH}{c}
\int\limits_{\varepsilon(\bs{p})
=\varepsilon_{\rm F}}\mbox{d}t\,\mbox{d}p_{\rm H}\,\ldots\, ,
\ee
$t$ is, as below, the time of  the electron motion along its
trajectory in a magnetic field, and $p_{\rm H}$ is the momentum
projection on the direction of the vector $\bs{H}$.  The
integration in Equation (5.34) is carried out over the open
constant-energy surface within the limits of the Brillouin zone.

The distribution of the electric potential $\Phi(\bs{r})$ in a
sample satisfies the Poisson equation which can be reduced to the
electroneutrality condition (3.15) if a Debye screening
radius is small in comparison with the contact diameter. It
was shown in [70, 71] that having the probability
$\alpha_{\bs{p}}(\bs{r})$ determined, we can calculate the contact
resistance for low voltages ($V\to0$)
%5.35
\be
R^{-1}=
\frac{I}{V}=
\frac{2e^2}{(2\pi\hbar)^3}
\int\limits_{S_{\rm c}}\mbox{d}S\,
\langle\bs{v}\alpha_{\bs{p}}\rangle,
\ee
and the electrical potential distribution
%5.36
\be
\Phi(\bs{r})=\frac{V}{2}(2\alpha(\bs{r},\bs{H})-1).
\ee

The solution of Equation (5.31) can be formally presented
as
%5.37)
\be
\alpha_{\bs{p}}(\bs{r},\bs{H})=\frac{1}{\tau}
\int\limits_{-\infty}^{t}\mbox{d}t'\,
\exp
\left(
\frac{t-t'}{\tau}
\right)
\alpha(\bs{r}-\bs{r}(t)+\bs{r}(t')).
\ee

If the electric potential, and hence (in view of Equation
(5.36)) the mean probability $\alpha$, varies smoothly at
distances comparable with the characteristic size
$\bs{r}(t)-\bs{r}(t-\tau)$ of the electron trajectory in the
corresponding direction, the integral equation (5.37) for
$\alpha_{\bs{p}}(\bs{r})$ can be replaced by a differential
equation for its mean value $\alpha$.  We expand
$\alpha(\bs{r}-\bs{r}(t)+\bs{r}(t'))$ in the integral in Equation
(5.37) into a series near the point $\bs{r}(t)=\bs{r}(t')$:
%5.38, 5.39
\ba
\alpha_{\bs{p}}(\bs{r})&=&
\alpha-\frac{\partial\alpha}{\partial\bs{r}}
\nonumber\\
&+&\frac{\partial^2\alpha}{\partial x_i\partial x_k}
\int\limits_{-\infty}^{t}\mbox{d}t'\,
\exp
\left(
\frac{t-t'}{\tau}
\right)
v_i(t')(x_k(t)-x_k(t'))\ldots\, ,
\nonumber\\
\\
g(\bs{p},\bs{H})&=&
\int\limits_{-\infty}^{t}\mbox{d}t'
\exp
\left(
\frac{t-t'}{\tau}
\right)
\bs{v}(t').
\ea

Averaging Equation (5.38) over the directions of the
momentum $\bs{p}$ and going over to dimensionless variables
%5.40)
\be
x_i'=\frac{2x_i}{d}
\left(
\frac{\sigma_{33}}{\sigma_{ii}}
\right)^{1/2}
\ee
we obtain the following equation for $\alpha(\bs{r},\bs{H})$
%5.41
\be
\Delta\alpha(\bs{r}')
+\sum\limits_{i\ne k}C_{ik}^{(\rm s)}
\frac{\partial^2\alpha}{\partial x_i'\partial x_k}=0,
\ee
where
%5.42, 5.43
\ba
C_{ik}&=&\frac{\sigma_{ik}}{\sqrt{\sigma_{ii}\sigma_{kk}}},\\
\sigma_{ik}&=&
\frac{2e^2}{(2\pi\hbar)^3}\langle v_ig_k\rangle
\ea
are the electrical conductivity tensor components in a bulk
conductor and $C_{ik}^{(\rm s)}$ is the symmetric part of $C_{ik}$
(5.42).  Equation (5.41), which coincides with the continuity
equation $\div\bs{j}=0$, must be supplemented by boundary
conditions.  Using Equations (5.32) and (5.33), we obtain
%5.44, 5.45
\ba
\alpha(\bs{r}\to\infty)&=&\Theta(x_3);\\
\bs{n}'\nabla\alpha(\bs{r}')+
\sum\limits_{i\ne k}n_i'C_{ik}
\left.
\frac{\partial\alpha}{\partial x_k'}
\right|_{\bs{r'}\in\Sigma'} &=&0,
\ea
($\bs{n}'=\nabla F(\bs{r}')/|\nabla F(\bs{r}')|$ is the vector normal to
the metal boundary $\Sigma'$ (5.28)) in coordinates (5.40). The
solution of boundary-value problem (5.41), (5.44), (5.45) would
lead to a rigorous criterion for the applicability of the
expression (5.38) for the function $\alpha_{\bs{p}}(\bs{r})$. In
spite of the considerable simplification due to making use of the
representation (5.38) for the function  $\alpha_{\bs{p}}(\bs{r})$,
the problem of evaluating the electric field in a metal remains
quite complicated and can be solved analytically only if the
coefficients $C_{ik}$ are small. In the main approximation
($C_{ik}=0$) the boundary value problem (5.41)--(5.45) is reduced
to the Neumann's problem with zeroth boundary condition for the
normal derivative $\partial\alpha/\partial n'$ on the hyperboloid
surface (5.28) which is no longer a figure of revolution in
coordinates (5.40).  For this reason, it is convenient to solve the
Equation (5.41) by going to the general ellipsoidal coordinates
$\xi$, $\eta$, $\zeta$
%5.46)
\ba
(x_1')^2&=&
\frac{(a^2+\xi)(a^2+\eta)(a^2+\zeta)}{(a^2-b^2)a^2};
\nonumber\\
(x_2')^2&=&
\frac{(b^2+\xi)(b^2+\eta)(b^2+\zeta)}{(b^2-a^2)b^2};
\nonumber\\
(x_3')^2&=&
\frac{\xi\eta\zeta}{a^2b^2};
\ea
$$
-a^2\le\zeta\le -b^2;\quad
-b^2\le\eta\le0;\quad
\xi\ge0;
$$
$$
a^2=c_1^2+\varkappa^2;\quad
b^2=c_2^2+\varkappa^2;\quad
c_i^2=\frac{\sigma_{33}}{\sigma_{ii}};\quad
\varkappa=\frac{\beta}{\nu}.
$$
For the sake of definiteness, we suppose that
$\sigma_{11}\le\sigma_{22}$ and hence $a\ge b$. In these
coordinates the contact surface corresponds to $\eta=-\varkappa^2$,
and we obtain for the probability $\alpha(\bs{r}')$:
%5.47)
\be
\alpha(\bs{r}')=
\Theta(z')-
\frac{1}{2K(k)}\mbox{sign}\,(x_3')
F\left(
\arctan\frac{a}{\sqrt{\xi}},k
\right),
\ee
where $F(\phi,k)$ and $K(k)$ are incomplete and complete elliptical
integrals of the first kind, $k^2=1-b^2/a^2$.

Substituting the expansion (5.38) and the function $\alpha$
(5.45) into Equation (5.35) for the electric current, we
have the final result for the resistance [72,7 3]:
%5.48
\be
R^{-1}(h)=
\sqrt{\sigma_{11}\sigma_{22}}\mbox{d} a\Psi(\phi,k).
\ee
Here,
$$
\Psi(\phi,k)=
F(\phi,k')
\left[
\frac{\bs{E}(k)}{\bs{K}(k)}-1
\right]
+E(\phi,k')-
\frac{\varkappa}{a}\frac{c_2}{c_1}
$$
%5.49
\be
k'=\sqrt{1-k^2};\quad
\phi=\arcsin
\left(
\frac{ac_1^2}{bc_2^2}
\right);\quad
d=2\nu
\ee
and $\bs{E}(k)$ is a complete elliptic integral of the second kind.

Let us first consider the case when $\bs{H}=0$ and a contact is
formed by conductors with the small elastic mean free path
($l_i\ll d$). In this case the function $\bs{g}(\bs{p},\bs{H}=0)$
can be written in the form:
%5.50)
\be
\bs{g}(\bs{p})=\tau\bs{v}.
\ee
In the absence of a magnetic field the conductivity tensor
(5.43) is diagonal and (5.47) is the exact solution of
Equation (5.41). Using the expression (5.47) for the
probability $\alpha$ we can write the resistance for a point
contact in the form of aperture $r^2=x_1^2+x_2^2\le d^2/4$ of
diameter $d$ in a plane of insulating partition
($\beta=\varkappa=0$) (see Fig.~13{\it b}) as
%5.51
\be
R^{-1}(0)=
\frac{\pi}{2}\sqrt{\sigma_{11}\sigma_{22}}\mbox{d}K^{-1}
\left(
\sqrt{1-
\left(
\frac{\sigma_{22}}{\sigma_{11}}
\right)^2}
\right);\quad \sigma_{11}\ge\sigma_{22}.
\ee
Equation (5.51) indicates that the point contact resistance
%5.52
\be
R^{-1}=
\left\{
\begin{array}{ll}
d\sqrt{\sigma_{\bot}\sigma_\|};
&\sigma_{33}=\sigma_\bot;\;
\sigma_{11}=\sigma_{22}=\sigma_\|,
\\[6pt]
\displaystyle
\pi\mbox{d}\sigma_\|
\ln^{-1}
\left(
\frac{\sigma_\|}{\sigma_\bot}
\right);
&\displaystyle
\sigma_{33}=\sigma_{11}=\sigma_{\|};\;
\sigma_{22}=\sigma_\|;\;
\frac{\sigma_\|}{\sigma_\bot}\gg1,
\end{array}
\right.
\ee
where
$$
\sigma_\bot=
\frac{e^2\tau m\pi\varepsilon_1^2a}{(2\pi)^3\hbar^4};\quad
\sigma_\|=\frac{8\pi^2e^2\tau\varepsilon_{\rm F}}{(2\pi\hbar)^2a}
\left(
1-\frac{\varepsilon_1}{\varepsilon_{\rm F}}
\right),
$$
depends not only on the electrical conductivity along the
contact axis, but also on the conducting properties of the
sample in the direction perpendicular to this axis. This is
due to the three-dimensional nature of current flow in the
point-contact region. It should be noted that for
$\sigma_{33}=\sigma_\|$ and $(\sigma_\|/\sigma_\bot)\to\infty$
(strictly two-dimensional conductivity), the resistance (5.52)
contains the logarithmic divergence. In this case we must take into
account the sample size $D$, and for
$(\sigma_\|/\sigma_\bot)\gg(D/d)\gg1$ the resistance is
$R(0)\sim(\sigma_\| d)^{-1}\ln(D/d)$.

In pure metals the ballistic mode of current flow through
the contact, for which $d\ll l_i$ can be realized. In this case we
can discount the second term in Equation (5.31). Its
solution is $\alpha=\mbox{const}(t)$. Using the boundary condition
(5.33) we found $\alpha_{\bs{p}}(\bs{r})$ for the contact in the
form of aperture
%5.53
\be
\alpha_{\bs{p}}(\bs{r})=\Theta(\bs{v}\in\Omega(\bs{r}));
\quad
x_2<0,
\ee
where $\Omega(\bs{r})$ is the solid angle enclosing the velocities
of electrons passing through the aperture and the point $\bs{r}$
(see Fig. 13{\it b}).

Substituting the solution (5.53) into Equation formula
(5.35), we obtain the following expression for the
resistance:
%5.54
\be
R^{-1}(0)=
\left\{
\begin{array}{ll}
\displaystyle
\frac{1}{4\pi^2}\sigma_\|\frac{S_{\rm c}}{l_i}
\bs{E}
\left(
\sqrt{\frac{\varepsilon_1}{\varepsilon_{\rm F}}}
\right);
&\sigma_{33}=\sigma_{11}=\sigma_\|;\;
\sigma_{22}=\sigma_\bot,
\\[8pt]
\displaystyle
\sqrt{\frac{\pi\sigma_\bot\sigma_\|}{2}}
\frac{S_{\rm c}}{l};
&\sigma_{33}=\sigma_\bot;\;\sigma_{11}=\sigma_{22}=\sigma_\|.
\end{array}
\right.
\ee
It is interesting to note that, according to Equations
(5.52) and (5.54), the ratio $R_\|/R_\bot$ for
$\varepsilon_1\ll\varepsilon_{\rm F}$ is inversely proportional to
the square root of the ratio $\sigma_\|/\sigma_\bot$ and depends
weakly on the relation between the contact diameter $d$ and the
electron mean free path $l_i$ ($R_\|$, $\sigma_\|$ and $R_\bot$,
$\sigma_\bot$ are the resistances and the conductivities of the
contact with the axis oriented parallel and perpendicular to the
layers, respectively).

Thus, the electrical conductivity of the point contact
between layered metals is extremely sensitive to the
orientation of the crystals in the contact. However, in
contrast to the case of a bulk sample, the resistance
depends on the metal conductivity in the directions
parallel and perpendicular to the contact axis in view of
the three-dimensional nature of current flow.

In a strong magnetic field ($\gamma_0=cp_{\rm F}/eHl_i\ll1$,
$p_{\rm F}$ is the Fermi momentum) the function
$\bs{g}(\bs{p},\bs{H})$, appearing in the theory of galvanomagnetic
phenomena [69], can be presented in the form of a power series in
$1/H$ whose first terms are given by
%5.55)
\ba
g_x&=&-\frac{c}{eH}(p_y\cos\theta-p_z\sin\theta);
\nonumber\\
g_y&=&v_{\rm H}\tau\sin\theta-\frac{cp_x}{eH}\cos\theta;
\nonumber\\
g_z&=&\frac{cp_y}{eH}\sin\theta+v_{\rm H}\tau\cos\theta,
\ea
for closed electron orbits ($\gamma=\gamma_0\sec\theta\ll1$,
$\bs{H}=(0,H\sin\theta,H\cos\theta)$) and
%5.56
\be
g_x=\frac{c}{eH}(p_y-\bar{p}_y);\quad
g_y=\bar{v}_y\tau;\quad
g_z=v_z\tau
\ee
for open electron orbits ($\theta=\pi/2$, $\gamma_0\ll1$). In
Equations (5.55), (5.56) $v_{\rm
H}=\partial\varepsilon(\bs{p})/\partial p_{\rm H}$ and the bar
indicates the value of a function averaged over the period of
motion in a magnetic field.  In following analysis we shall assume
that $\eta\ll\gamma$ and the inclination $\theta$ of the magnetic
field does not coincide with the value $\theta_{\rm c}$ for which
the conductivity across the layers decreases considerably
($\sigma_{zz}(\theta_{\rm c})\approx\eta^4$, see Equations (2.33)).
In this case, as it follows from Equations (2.27), and (5.42), the
coefficients $C_{xz}$ and $C_{yz}$ are small and the terms
containing these coefficients in Equation (5.41) can be omitted.
Since the anisotropy of conductivity in the layers-plane is usually
not significant, unjustifiably cumbersome formulas can be avoided
by assuming that the Fermi surface is a figure of revolution. It
was shown for this case (see Equations (2.24), (2.27)) that if the
inequality $\eta\ll\gamma$ is satisfied, then
$\sigma_{xx}\approx\sigma_{yy}=\sigma_\|(H)$ for any angle
$\theta$, while for $\theta=\pi/2$ we have
$\sigma_{xy}=\sigma_{yz}=0$.  The inequality $\eta\ll\gamma$, which
is violated only in very strong fields, indicates that the
conductivity across the layers always remains much lower than the
conductivity along the layers.

We consider separately the two main geometries of the
experiment.

{\it 1. Contact axis is perpendicular to the layers ($x_3=x$;
$x_2=y$; $x_3=z$; $\sigma_{33}=\sigma_{zz}$;
$\sigma_{11}=\sigma_{22}=\sigma_\|$)} (see Fig.~14).

In Equation (5.46) and (5.47) we must replace
$$
c_1^2=c_2^2=L^2=\frac{\sigma_{zz}}{\sigma_\|}<1.
$$

The probability $\alpha$ can be presented in the form [71]:
%5.57
\be
\alpha(\bs{r})=\Theta(z)-\frac{1}{\pi}\mbox{sign}(z)
\arctan\left(\frac{1}{\chi}\right).
\ee
where
%5.58
\be
\chi=
\left\{
\frac{r^2}{2\nu^2}-\frac{1}{2}+
\left[
\left(
\frac{r^2}{2\nu^2}-\frac{1}{2}
\right)^2+
\frac{z^2L^2}{\nu^2}
\right]^{1/2}
\right\}^{1/2}\quad
r=\sqrt{x^2+y^2}.
\ee
\begin{figure}
\centering
\includegraphics[width=10cm]{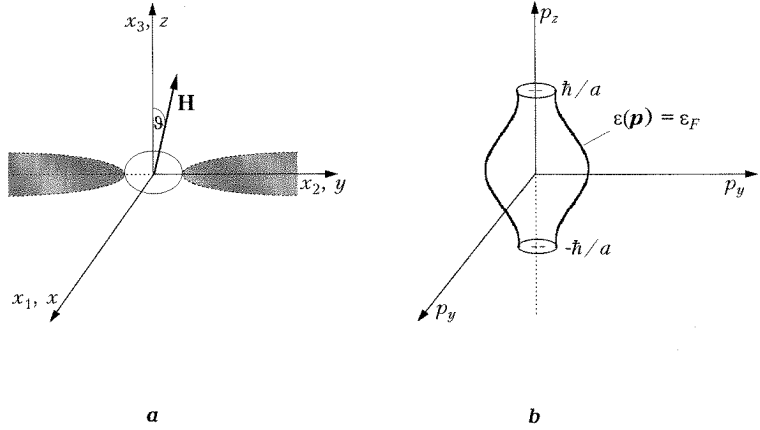}
\caption{The model of the point ({\it a}) in the case when the
contact axis coincides with the axis of the Fermi surface ({\it
b}) of a layered conductor.}
\end{figure}

It follows from Equations (5.36), (5.57) that the
characteristic scale of variation of the electric field
along the $z$-axis is of the order of $d/(L^2+\varkappa^2)^{1/2}$
(where $L^2=\sigma_{zz}/\sigma_\|$, $\varkappa=\beta/\nu$) and may
be either larger or smaller than the contact diameter.
Substituting the function $\alpha_{\bs{p}}(\bs{r})$ in the form
(5.38) into the expression (5.35) for the resistance and taking
into account the equalities (5.55) and (5.57), we obtain
%5.59
\be
R^{-1}(H)=\sqrt{\sigma_{zz}\sigma_\|}d\tan\psi;\quad
\psi=\frac{1}{2}\arctan\left(\frac{L}{\varkappa}\right).
\ee
In view of the oscillatory dependence of $\sigma_{zz}$ on the slope
of the magnetic field (Equation (2.33)), the magnetoresistance of
the point contact is also a nonmonotonic function of $\theta$.  It
should be noted that $R(H)$ is highly sensitive to the contact
geometry.  Thus, for a constriction having a form close to aperture
($\varkappa\ll1$), we have
%5.60
\be
R^{-1}(H)=
\left\{
\begin{array}{ll}
\displaystyle
\sigma_{zz}\frac{d^2}{4\beta};& 1\gg\varkappa\gg L,
\\[6pt]
d\sqrt{\sigma_{zz}\sigma_\|};
&1\gg L\gg\varkappa.
\end{array}
\right.
\ee
The magnetic field variation of the parameter $L^2(H)$, viz., the
ratio of electrical conductivities across and along the
layers, depends significantly on the orientation of the
vector $\bs{H}$. If the vector $\bs{H}$ is parallel to the contact
axis ($\sigma_{zz}(\bs{H})=\sigma_\bot=\mbox{const}(\bs{H})$), an
increase in the field causes a decrease in the conductivity
$\sigma_\|(H)$ and an increase in $L$. For $L(0)\ll \varkappa$,
there exist a range of values of $H$ ($L(H)<\varkappa$) in which
$R(H)=\mbox{const}$.  Formally, such a situation corresponds to a
contact in the form of strongly elongated channel, since, although
the real contact length for $\varkappa\ll1$ is much smaller than
its diameter, the electron velocity along the channel axis is much
lower than the velocity of diffusive motion along a normal to this
axis of the Larmour orbit center. In stronger fields, when $L(H)$
becomes larger than the parameter $\varkappa$ characterizing the shape of
the constriction, the quasi-two-dimensional current flow is
replaced by the three-dimensional flow, and the contact resistance
depends linearly on the magnetic field as in the case of an
isotropic metal [13].

The magnetic field, which is orthogonal to the contact
axis, suppresses the electric conductivity $\sigma_{zz}$ across the
layers, leaving the in-plane conductivity $\sigma_\|$ practically
unchanged. This corresponds to a decrease in parameter $L$
with increasing $H$.  Hence for $L(0)\ll\varkappa$, the contact resistance
in strong magnetic fields is proportional to $H^2$ for
$\gamma_0\le\eta^{1/2}$ (see Equation (2.14)). If, however,
$L(0)\gg\varkappa$ (although $L(0)\simeq\eta\ll1$), the quadratic
dependence $R(H)$ is preceded by the linear magnetoresistance
$L(H)>\varkappa$.

\begin{figure}
\centering
\includegraphics[width=10cm]{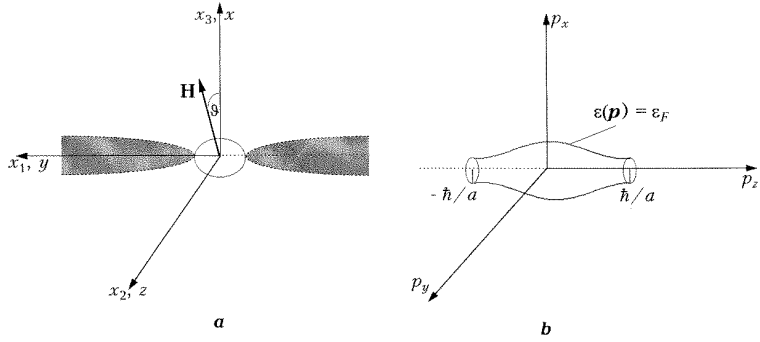}
\caption{The model of the point-contact ({\it a}) in the case when
the contact axis is parallel to the layers ({\it b}).}
\end{figure}

{\it 2. Contact axis is parallel to the layers ($x_3=x$)
(Fig.~15).}

If the magnetic field is perpendicular to the axis of
the cylindrical Fermi surface, the coefficients $C_{ik}$ are small
for any angle of inclination of the magnetic field to the contact
axis.  It follows from (5.48) that if the contact axis lies in the
plane of a high electrical conductivity ($\sigma_{33}=\sigma_\|$)
the resistance depends logarithmically on the magnetic field $H$
(except for the case of a strongly elongated constriction
($\varkappa\gg L^{-2}$), for which $R(H)=\mbox{const}$:
%5.61)
\be
R^{-1}(H)=\sigma_\|d
\left\{
\begin{array}{ll}
\displaystyle
\arctan\left(
\frac{1}{\varkappa}
\right)
\ln^{-1}
\left(
\frac{4}{\varkappa L}
\right);
&\varkappa L\ll1,
\\[8pt]
\displaystyle\frac{1}{2\varkappa};
&\varkappa L\gg1.
\end{array}
\right.
\ee
Thus, the magnetic field dependence $R(H)$ of contacts
oriented at a right angle to the layers with a high
electrical conductivity is extremely sensitive to the shape
of the constriction. If the square root of the ratio
$\sigma_\bot/\sigma_\|=L^2$ is smaller than the ratio
$\varkappa=\beta/\nu$ of the length of the contact to its radius,
the longitudinal resistance ($\theta=0$, $\theta$ is the angle
between the contact axis and vector $\bs{H}$) and is independent of
a magnetic field while the transverse magnetoresistance
($\theta=\pi/2$) is proportional to $H^2$.  If the opposite
inequality $L\gg\varkappa$ is satisfied, the resistance increases linearly
with $H$. A variation of the field changes the magnitude $L^2$
($L^2$ decreases for $\theta=\pi/2$ and increases in all other
cases), thus allowing two types of field dependence $R(H)$ for the
same contact in different ranges of $H$.  As the inclination of the
vector $\bs{H}$ relative to the layers is changed, $R(\bs{H})$
becomes a nonmonotonic function of an angle. In the geometry of the
experiment, when the contact axis and the vector $\bs{H}$ lie in
the plane of the layers with a high conductivity, the resistance
shows a logarithmic dependence on $\bs{H}$.

%5.3.
\subsection{Point-Contact Spectroscopy of Electron--Phonon
Interaction}
The information about the electron--phonon interaction in
layered conductors is of great importance for understanding
the nature of the superconducting state and for analysis of
transport phenomena in these materials. Point-contact
spectroscopy is an effective and reliable method of
studying electron--phonon interaction [74, 75, 68].  Point
contact spectroscopy is based on a strong nonequilibrium in
the electronic system only in a small region of space whose
dimension is less than the inelastic mean free path of
electrons. In a point contact the Fermi surface splits into
two parts with maximum energies differing by the bias
energy $eV$. Effectively, there are two electronic beams
moving in opposite directions with energies differing at
each point of space by exactly the bias energy.
Electron--phonon scattering in contacts results in a back
flow:  some electrons are reflected after entering the
constriction and do not contribute to the contact current.
Each time the bias $eV$ equals the energy of particular
phonon $\hbar\omega_{\bs{p}}$, the current decreases. When averaged
over all phonons, these contributions result in a nonlinear current
in the contact. This technique has been used recently for
reconstructing the point-contact spectroscopy of the
electron--phonon interaction in organic metals
$\beta$-(BEDT-TTF)$_2X$ ($X=I_2$, $I$Au$I$) [76--80]. The
experiments carried out in [76--80] revealed that the form of the
point contact spectrum depends to a considerable extent on the
orientation of the contact axis relative to planes with a high
electrical conductivity.

The inelastic process of the electron--phonon interaction
can be taken into account in the collision integral
%5.62
\ba
W^{\rm (ep)}_{\rm col}\{f(\bs{p})\}\!\!&=&\!\!
\sum\limits_{k}
\int\frac{\mbox{d}^2q}{(2\pi\hbar)^3}W_{\bs{q},k}
\big\{
[f(\bs{p}+\bs{q})(1-f(\bs{p}))(N_{\bs{q},k}+1)
\nonumber\\
\!\!&-&\!\!
f(\bs{p})(1-f(\bs{p}+\bs{q}))N_{\bs{q},k}
]\nonumber\\
\!\!&\times&\!\!
\delta(\varepsilon(\bs{p}+\bs{q})
-\varepsilon(\bs{p})-\omega_{\bs{q},k})
\nonumber\\
\!\!&+&\!\!
[f(\bs{p}-\bs{q})(1-f(\bs{p}))N_{\bs{q},k}
\nonumber\\
\!\!&-&\!\!
f(\bs{p})(1-f(\bs{p}-\bs{q}))(N_{\bs{q},k}+1)
]\nonumber\\
\!\!&\times&\!\!
\delta(\varepsilon(\bs{p}-\bs{q})
-\varepsilon(\bs{p})+\omega_{\bs{q},k})
\big\}.
\ea
Here, the summation is carried out over the numbers $k$ of
the phonon spectrum branches $\omega_{\bs{q},k}$; $W_{\bs{q},k}$ is
the square of the magnitude of the electron--phonon interaction
matrix element.  In general, the system of Equations (2.2), (3.15)
must be supplemented by the kinetic equation for determining the
phonon distribution function $N_{\bs{q},k}$. But usually the state
of the phonon system in the contact does not influence the form of
the point contact spectra so much. This is so in the case of weak
phonon--electron scattering, when the phonon distribution function
$N_{\bs{q},k}$ is independent of bias, $V$ [81].  In the following
consideration we shall assume that phonons are in thermal
equilibrium, i.e.
$$
N_{\bs{q},k}=\left(
\exp\left(
\frac{h\omega_{\bs{q},k}}{T}
\right)-1
\right)^{-1}.
$$

We should write [70, 71] the particular solution
$f_{\bs{p}}^{(1)}(\bs{r})$ of the non-homogeneous Equation (2.2)
using the corresponding Green's function
$g_{\bs{pp}'}(\bs{r},\bs{r}')=g_{-\bs{p}',-\bs{p}}(\bs{r}',\bs{r})$:
%5.63
\be
f_{\bs{p}}^{(1)}(\bs{r})=
\int\mbox{d}\bs{p}'\,\mbox{d}\bs{r}'\,
g_{\bs{pp}'}(\bs{r},\bs{r}')W^{\rm (ep)}_{\rm col}
\{f_{\bs{p}}^{(0)}(\bs{r}')\},
\ee
where $f_{\bs{p}}^{(0)}(\bs{r})$ is the distribution function
(5.30) in zeroth approximation in electron--phonon collision
integral. The function $g_{\bs{pp}'}(\bs{r},\bs{r}')$ must be
determined from the relations
%5.64, 5.65, 5.66
\ba
\bs{v}'\frac{\partial}{\partial\bs{r}'}g_{\bs{pp}'}(\bs{r},\bs{r}')
-\frac{1}{\tau}\{g_{\bs{pp}'}-\langle g_{\bs{pp}'}\rangle\}&=&
-\delta(\bs{p}-\bs{p}')
\nonumber\\
&-&
\delta(\bs{r}-\bs{r}');\\
g_{\bs{pp}'}(\bs{r},\bs{r}'\to\infty)&=&0;\\
g_{\bs{pp}'}(\bs{r},\bs{r}'\in\Sigma)&=&
g_{\bs{p}\tilde{\bs{p}}'}(\bs{r},\bs{r}'\in\Sigma).
\ea
Substituting the value $f_{\bs{p}}^{(1)}(\bs{r})$ in formula
(5.29) we will get the expression for the change in the electric
current $\Delta I$ due to the electron phonon interaction
%5.67
\be
\Delta I=\frac{2e}{(2\pi h)^3}
\int\mbox{d}\bs{p}\,\mbox{d}\bs{r}\,
G_{\bs{p}}(\bs{r})W^{\rm (ep)}_{\rm col}
\{f_{\bs{p}}^{(0)}(\bs{r}')\},
\ee
where
%5.68)
\be
G_{\bs{p}}(\bs{r})=
\int\mbox{d}S\,n_3\int\mbox{d}\bs{p}'\,
\mbox{d}v_3'\,g_{\bs{p}'\bs{p}}(\bs{r},\bs{r}').
\ee
Multiplying Equation (5.64) by $v_3$ and integrating it over
the contact area $S_{\rm c}$ and momentum $\bs{p}$ we get the
following equation and the boundary conditions for the function
$G_{\bs{p}}(\bs{r})$
%5.69, 5.70, 5.71
\ba
\bs{v}\frac{\partial}{\partial\bs{r}}G_{\bs{p}}(\bs{r})
-\frac{1}{\tau}\{
G_{\bs{p}}(\bs{r})-
\langle G_{\bs{p}'}(\bs{r})\rangle\}&=&-\delta(x_3),\\
G_{\bs{p}}(\bs{r}\to\infty)&=&0,\\
G_{\bs{p}}(\bs{r}\in\Sigma)&=&G_{\tilde{\bs{p}}'}(\bs{r}\in\Sigma).
\ea
The probability $\alpha_{\bs{p}}(\bs{r})$ satisfies the Equations
(5.31)--(5.33), which combined with Equations (5.69)--(5.71) yield
the relation [70]
%5.72
\be
G_{\bs{p}}(\bs{r})=\alpha_{-\bs{p}}(\bs{r})-\Theta(x_3).
\ee

The point contact spectrum is the second derivative of the
point contact current $I(V)$ with respect to the voltage
$V$. Substituting Equations (5.30) and (5.47) into the
expression for the point contact current correction (5.67)
and making use of the equality
$$
\frac{\mbox{d}^2I}{\mbox{d}V^2}=
-R^{-2}\frac{\mbox{d}R}{\mbox{d}V},
$$
where $R(V)$ is the dynamic resistance $\mbox{d}I/\mbox{d}V$, we
obtain the relation [68, 70]:
%5.73)
\be
\frac{1}{R}\frac{\mbox{d}R}{\mbox{d}V}=
\frac{32}{3\pi}e\tau_0
\int\limits_{0}^{\infty}\frac{\mbox{d}\omega}{T}\,
\chi
\left(
\frac{\hbar\omega-eV}{T}
\right)G(\omega),
\ee
where
$$
\tau_0=\nu(\varepsilon_{\rm F})\frac{(2\pi\hbar)^3d}{2S_{\rm F}};
\quad
\chi(x)=x(\exp(x)-1),
$$
$\nu(\varepsilon_{\rm F})$ is the density of states at the Fermi
surface; $S_{\rm F}$ is the area of the Fermi surface within the
Brillouin zone. For the isotropic metal the coefficient $\tau_0$ is
equal to $d/v_{\rm F}$ ($v_{\rm F}$ is the Fermi velocity). The
second derivative of the function $\chi(x)$ tends to $\delta(x)$ at
low temperatures. At $T=0$ this point contact spectrum is
%5.74
\be
\frac{1}{R}\frac{\mbox{d}R}{\mbox{d}V}=
\frac{32}{3\pi\hbar}e\tau_0
G\left(\frac{eV}{\hbar}\right).
\ee
The point-contact function of the electron--phonon
interaction $G(\omega)$ in Equation (5.73) is defined as
%5.75, 5.76
\ba
G(\omega)&=&\frac{1}{\langle1\rangle}
\sum\limits_{k}\langle\langle
W_{\bs{p}-\bs{p}',k}K(\bs{p},\bs{p}')
\delta(\omega-\omega_{\bs{p}-\bs{p}',k})
\rangle\rangle;\\
K(\bs{p},\bs{p}')&=&\frac{3\pi}{32}
\int\mbox{d}^3r\,
[\alpha_{\bs{p}}(\bs{r},\bs{H})-\alpha_{\bs{p}'}(\bs{r},\bs{H})]
\nonumber\\
&\times&
[\alpha_{-\bs{p}}(\bs{r},-\bs{H})-\alpha_{-\bs{p}'}(\bs{r},-\bs{H})]
\nonumber\\
&\times&
\left\{
\tau_0\int\mbox{d}S\,
\frac{\langle\bs{v}\alpha_{\bs{p}}(\bs{r})\rangle}{\langle1\rangle}
\right\}.
\ea
The expansion (5.38) and the equality (5.47) enable us to
write an expression for the function $K(\bs{p},\bs{p}')$ [72, 73]
%5.77
\ba
K(\bs{p},\bs{p}')&=&
\frac{3\pi}{32}
\frac{e^2S_{\rm F}Ra^2}{(2\pi\hbar)^3K^2(k)}
\sum\limits_{i=1}^{3}
\left(
\frac{\sigma_{11}\sigma_{22}}{\sigma_{ii}^2}
\right)^{1/2}
\nonumber\\
&\times&\big[
(g_i^{\rm (ev)}(\bs{p})-g_i^{\rm (ev)}(\bs{p}'))^2
-
(g_i^{\rm (od)}(\bs{p})-g_i^{\rm (od)}(\bs{p}'))^2
\big]
\nonumber\\
&\times&
I_i(k,\varkappa),
\ea
where
$g_i^{\rm (ev)}(\bs{p},\bs{H})=g_i^{\rm (ev)}(\bs{p},-\bs{H})$
and
$g_i^{\rm (od)}(\bs{p},\bs{H})=-g_i^{\rm (od)}(\bs{p},-\bs{H})$ are
even and odd parts of the functions $g_i$ (5.55), (5.56);
%5.78
\ba
I_i(k,\varkappa)&=&
\int\limits_{b^2}^{a^2}\mbox{d}\zeta\,
\int\limits_{\varkappa^2}^{b^2}\mbox{d}\eta\,
\int\limits_{0}^{\infty}\mbox{d}\xi\,
\frac{\zeta-\eta}{(\xi+\eta)(\xi+\zeta)}
\nonumber\\
&\times&
\left(
\frac{\xi(a^2+\xi)(b^2+\xi)}%
{(a^2-\eta)(b^2-\eta)\eta(a^2-\zeta)(\zeta-b^2)\zeta}
\right)^{1/2}
\left(
\frac{\partial x_i'}{\partial\xi}
\right)^2.
\nonumber\\
\ea
Here, $x_i'$, $a$, and $b$ are defined by relation (5.46).

Let us first consider the case $\bs{H}=0$ and a point contact in
the form of an aperture ($\varkappa=0$). For an impure conductor
($l_i\ll d$), from Equations (5.77), (5.38), (5.50) we obtain the
form-factor $K(\bs{p},\bs{p}')$
%5.79
\be
K(\bs{p},\bs{p}')=
\frac{3\pi}{32}\frac{\tau}{\tau_0}
\int\mbox{d}^3\bs{r}\,
\left(
(\bs{v}-\bs{v}')\frac{\partial\alpha}{\partial\bs{r}}
\right)^2
\left\{
\int\limits_{S_{\rm c}}\mbox{d}S\,
\langle\bs{v}(\bs{v}\nabla\alpha)\rangle.
\right\}
\ee
The form factor $K(\bs{p},\bs{p}')$ in the case under investigation is given
by
%5.80)
\ba
K(\bs{p},\bs{p}')&=&
\frac{3}{128}\frac{\tau}{\tau_0}\lambda^{-1/2}\bs{K}^{-1}(k)
\frac{1}{\langle v_3^2\rangle}
\nonumber\\
&\times&
\Biggl\{
(v_1-v_1')^2\frac{\sigma_{33}}{\sigma_{11}}
\frac{1}{\lambda-1}
\hat{F}\{f(\lambda)\}
\nonumber\\
&+&
(v_2-v_2')^2\frac{\sigma_{33}}{\sigma_{22}}
\frac{\lambda}{\lambda-1}
\hat{F}\{f^{-1}(\lambda)\}
\nonumber\\
&+&
(v_3-v_3')^2[4\pi\sqrt{\lambda}\bs{K}(k)
\nonumber\\
&-&
\frac{1}{\lambda-1}
(\hat{F}\{f(\lambda)\}+\lambda\hat{F}\{f^{-1}(\lambda)\}]
\Biggr\},
\ea
where
$$
\lambda=\frac{a^2}{b^2};\quad
k^2=1-\frac{1}{\lambda};
$$
\ba
\hat{F}\{f\}&=&
\int\limits_{1}^{\lambda}\mbox{d}\zeta\,
\int\limits_{0}^{1}\mbox{d}\eta\,
\int\limits_{0}^{\infty}\mbox{d}\xi\,
\frac{\zeta-\eta}{(\xi+\eta)(\xi+\zeta)}
\left(
\frac{\xi}{\zeta\eta}
\right)^{1/2}
f(\zeta,\eta,\xi);\nonumber\\
f(\zeta,\eta,\xi)&=&
\left[
\frac{(\lambda-\eta)(\lambda-\zeta)(\xi+1)}{(1-\eta)(\zeta-1)(\xi+\lambda)}
\right]^{1/2}.
\nonumber
\ea

If the contact axis is oriented at right angle to the
planes with a high electrical conductivity (Fig.~15) ($\lambda=1$,
$\sigma_{33}=\sigma_{\bot}$, $\sigma_{11}=\sigma_{22}=\sigma_\|$),
for the Fermi surface (2.21) Equation (5.80) assumes the form
%5.81
\ba
K(\bs{p},\bs{p}')&=&
\frac{3\pi}{128}\frac{\tau}{\tau_0}
\frac{\langle1\rangle}{\langle v_3^2\rangle} \nonumber\\
&\times&
\left\{
\frac{\sigma_\bot}{\sigma_\|}
[(v_1-v_1')^2+(v_2-v_2')^2]+2(v_3-v_3')^2
\right\}.
\nonumber\\
\ea
For the Fermi surface (1.23) formula (5.81) reduces to
%5.82
\be
K(\bs{p},\bs{p}')=
\frac{3l_i}{32d}
\left\{
\frac{1}{p_{\rm F}^2}(\bs{p}_\|-\bs{p}'_\|)^2
+8\pi
\left[
\sin\left(\frac{p_3a}{\hbar}\right)-
\sin\left(\frac{p_3'a}{\hbar}\right)
\right]^2
\right\},
\ee
where $\bs{p}_\|$ is component of $\bs{p}$ parallel to the layers.
In other words, for $\varepsilon_{1}\ll\varepsilon_{\rm F}$ the
intensity of the normalized point contact spectrum (5.79) does not
depend on $\varepsilon_1$, and the contributions to the spectrum
from electrons scattering by phonons are of the same order of
magnitude for both cases:  when the charge carriers velocity
changes in parallel and perpendicular directions to the layers.

If, however, the contact axis is parallel to the layers
(Fig. 15) ($\sigma_{11}=\sigma_{33}=\sigma_\|$;
$\sigma_{22}=\sigma_\bot$), the quasi-two-dimensional nature of
the current flow is responsible for dominating the contribution to
the point contact spectrum from the electron scattering involving a
change in the momentum component tangential to the layers:
%5.83
\be
K(\bs{p},\bs{p}')=
\frac{3\pi\tau}{32\tau_0}
\frac{1}{p_{\rm F}}
(\bs{p}_\|-\bs{p}'_\|)^2;\quad
\frac{\sigma_\bot}{\sigma_\|}\ll1.
\ee
The expression for the $K$-factor for the point contact
function of the electron--phonon interaction describes the
influence of anisotropy in
\linebreak[10000]
energy-momentum relation for
conduction electrons on the point contact spectrum [68].
Substituting the expression for probability (5.53) into
Equation (5.76), with reference to the clean conductor ($l_i\gg d$)
we obtain [68]
%5.84
\be
K(\bs{p},\bs{p}')=
\frac{|v_3v_3'|\Theta(-v_3v_3')}{|v_3\bs{v}'-v_3'\bs{v}|}.
\ee
Thus, an analysis of the intensity of the point contact
spectra of the layered conductor with various orientations
of the contact axis allows us to single out unambiguously,
the lines corresponding to electron relaxation at
two-dimensional phonon modes whose existence in layered
crystals was predicted theoretically [82].

The magnetic field dependence of the point contact spectrum
is contained in $K$-factor (5.76). If the contact axis is
perpendicular to the layers (Fig. 14) ($x_3=z$;
$\sigma_{zz}=\sigma_\bot(H)$,
$\sigma_{yy}=\sigma_{xx}=\sigma_\|(H)$), evaluating the integrals
of products of functions $\alpha_{\bs{p}}$ (5.38) and using the
asymptotic expressions (5.55) and (5.56) for the function $\bs{g}$
in strong magnetic fields, we obtain
%5.85
\be
K(\bs{p},\bs{p}')=
\frac{3\pi}{64}
\frac{\langle1\rangle}{\langle v_zg_z\rangle}
\frac{\tau}{\tau_0}
\left(
2A_{\bs{pp}'}\cos^2\psi+
\frac{\sigma_{zz}}{\sigma_\|}B_{\bs{pp}'}\sin^2\psi
\right),
\ee
where for $\gamma_0\sec\theta\ll1$:
%5.86
\ba
A_{\bs{pp}'}&=&
(v_{\rm H}-v_{\rm H}')^2\cos^2\theta-\gamma_0^2m^{-2}
(p_x-p_x')^2\sin\theta;
\nonumber\\
B_{\bs{pp}'}&=&
(v_{\rm H}-v_{\rm H}')^2\sin^2\theta
\nonumber\\
&-&
\gamma_0^2m^{-2}
[(\bs{p}_\|-\bs{p}_\|')^2\cos^2\theta-(p_z-p_z')^2\sin^2\theta],
\ea
and for $\gamma_0\ll1$, $\theta=\pi/2$
%5.87
\ba
A_{\bs{pp}'}&=&
-\gamma_0^2[(p_y-p_y')-(\bar{p}_y-\bar{p}_y')]^2
\nonumber\\
B_{\bs{pp}'}&=&
[(v_z-v_z')^2+(\bar{v}_y-\bar{v}_y')^2]
\ea
where $\gamma_0=1/\Omega_0\tau$; $\Omega_0=eH/mc$; $m$ is the
minimum cyclotron mass; $\bs{p}_\|$ is the component of $\bs{p}$
parallel to the layers;
$$
\psi=\frac{1}{2}\arctan
\left[
\frac{1}{\varkappa}
\left(
\frac{\sigma_{zz}}{\sigma_\|}
\right)^{1/2}
\right]
$$
It follows from Equation (5.85) that the behaviour of
$K$-factor in a magnetic field depends significantly on the
shape of the contact and the inclination $\theta$ of the field
$\bs{H}$ to the contact axis. In the limiting cases $\theta=0$ and
$\theta=\pi/2$, the $K$-factor does not contain the parameter
$\gamma\ll1$ for both very short ($\varkappa\ll L$) and elongated
($\varkappa\gg L$) contacts.

If the contact axis as well as a magnetic field lie in the
layers-plane for a contact in the form of an orifice ($\varkappa=0$)
(Fig.~10), we obtain simple asymptotic expressions for the
integral $I_i(L,\varkappa)$ (5.78) at $L\ll1$:
%5.88
\be
I_1\approx 2\pi L;\quad
I_2\approx \frac{4\pi L}{3}\ln\left(\frac{1}{L}\right);\quad
I_3\approx \frac{8\pi L}{3}\ln\left(\frac{1}{L}\right);\quad
L^2=\frac{\sigma_{zz}}{\sigma_\|}.
\ee
Substituting the expressions for $I_i$ (5.88) into Equation
(5.77) and noting that the function $\Psi(k)=(\pi/2)\ln^{-1}(1/L)$
for $\varkappa\ll1$, we find
%5.89
\ba
K(\bs{p},\bs{p}')&=&
\frac{3\pi\tau^2}{32\tau_0}
\frac{\langle1\rangle}{\langle v_xg_x\rangle}
\nonumber\\
&\times&
\biggl\{
-\gamma_0^2m^{-2}
[(p_y-p_y')-(\bar{p}_y-\bar{p}_y')]^2
\frac{\sigma_\|}{\sigma_\bot\ln(\sigma_\|/\sigma_\bot)}
\nonumber\\
&+&
\frac{2}{3}[
2(v_z-v_z')^2+(\bar{v}_y-\bar{v}_y')^2
]
\biggr\}.
\ea
It is follows from Equation (5.89) that in the main
approximation in above parameters, the point contact,
spectrum is practically independent of $H$ in a strong
field.

Thus, the point contact spectra differ considerably for
different orientations of the contact axis and vector $\bs{H}$.
If the current passes in the direction with low
conductivity, the electron--phonon collisions, which change
the charge carrier velocity component parallel to the
layers, make a negative contribution to the point contact
spectrum in a longitudinal field. The situation is reversed
as the field $\bs{H}$ is turned through right angles, and the
negative contribution is now made by the scattering
processes which change the electron velocity at right angle
to the layers. These processes play the most important role
and must cause an inversion of the point contact spectrum
in a magnetic field transverse to the axis for $L\ll\varkappa$. The
intensity and sign of the point contact spectrum are
determined by the type of phonons being excited, as well as
by the shape of the contact. When the main role in the
passage of current through the contact is played by
electrons moving parallel to the layers, the point contact
spectrum is independent of $\bs{H}$ in the strong magnetic field.

\section*{CONCLUSION}
Electron phenomena in quasi-two-dimensional conductors with
the metal type of electrical conductivity, analysed above,
shows the rich variety of properties of artificial metals.

In contrast to the case of an ordinary metal, the presence
of an extra small parameter such as the parameter of
quasi-two-dimensionality of the electron energy spectrum,
allows us to study theoretically and in far more detail,
the relaxation phenomena and also to reveal new effects
containing the information about properties of charge
carriers in low-dimensional conductors. Among them there is
the orientation effect (peculiar dependence of kinetic
characteristics of a layered conductor on the orientation
of a strong magnetic field) arising from the sharp
anisotropy of the velocity of charge carriers with the
Fermi energy. The dependencies of magnetoresistance,
surface impedance, and sound attenuation rate on the angle
between the normal to the layers and a magnetic field show
sharp peaks, the spacing between them containing the
information on the form of the Fermi surface. Experimental
investigations of the orientation effect under the
conditions, when the diameter of electron orbit is the
least length parameter of the problem, and enable the
anisotropy and magnitude of the Fermi surface diameters to
be determined.

In ordinary metals only the Fermi surface topology can be
restored by galvanomagnetic measurements, but in order to
determine the sizes of the Fermi surface the response of
electron system to an alternating field must be
investigated.

When electromagnetic and acoustic waves propagate in
layered conductors, when the wavelength is the least length
parameter of the problem, extra possibilities for studying
the properties of charge carriers appear. Even if charge
carriers are incapable of drifting along the direction of
the wave vector, magnetoacoustic resonance takes place, and
between resonant peaks the acoustic transparency is
observed.

The attenuation length for acoustic and electromagnetic
waves in quasi-two-dimensional conductors is very
sensitive to the polarization of the wave. Investigations
of the high-frequency impedance in a magnetic field at
different polarizations of the wave enable the character of
the interaction between charge carriers and the sample
surface to be studied.

The magnetic field dependence $U(H)$ of the electron
focusing signal in layered conductors is determined by the
size of point contacts and their orientation in relation to
crystallographic axes. If contacts are arranged on the
crystal surface perpendicular to the layers with a high
electrical conductivity, one can determine the diameters of
closed Fermi surface cross sections from the positions of
transverse electron focusing peaks on the magnetic field
scale. When the vector $\bs{H}$, lies in the plane of the layers,
singularities in the electron focusing signal occur for the
values of the field $\bs{H}$ corresponding to the separation
between the electrodes multiple to the electron
displacement per period in an open trajectory. The effect
of longitudinal electron focusing can be observed in a
magnetic field orthogonal to the surface. The values of $H$
for which the signal $U(H)$ has peaks can be used for
determining the extremal value of the electron velocity
along the Fermi surface cylinder axis. Thus, the electron
focusing observations in layered organic metals can serve
as a source of information on their Fermi surface.

The magnetic field dependence of the resistance $R(H)$ for
contacts oriented at a right angle to the layers with a
high electrical conductivity is extremely sensitive to the
shape of the constriction and to the ratio of
conductivities across and along the layers.  A variation of
the field changes the ratio of conductivities paving the
way for two types of field dependence $R(H)$ for the same
contact in different ranges of $H$. As the inclination of
the vector $\bs{H}$ in relation to the layers changes, $R(H)$
becomes a nonmonotonic function of the angle. In the
geometry of experiment, when the contact axis and vector $\bs{H}$
lie in the plane of a layer with a high conductivity, the
resistance shows algorithmic dependence on $\bs{H}$.

The point contact spectra differ considerably for different
orientations of both the contact axis and vector $\bs{H}$. If the
current passes in the direction with a low conductivity,
the electron--phonon collisions, that change the charge
carrier velocity in the layers-plane, make a negative
contribution into the point contact spectrum in a
longitudinal field. The situation is reversed as the field
$\bs{H}$ turned through a right angle, and the negative
contribution is now made by the scattering processes that
change the electron velocity at a right angle to the
layers. The intensity and sign of the point contact
spectrum are determined by the type of phonons, being
excited, as well as by the shape of the contact. When the
main role in the passage of current through the contact is
played by electrons moving parallel to the layers, the
point contact spectrum is independent of $\bs{H}$ in a strong
magnetic field.

\section*{REFERENCES}
\begin{list}{}{\topsep  0pt \parsep  0pt \itemsep  0pt \partopsep 0pt%
%обязательное включение для ссылок
\settowidth{\labelwidth}{99.}%
\setlength{\labelsep}{1em}%
\setlength{\leftmargin}{\labelwidth}%
\addtolength{\leftmargin}{\labelsep}}
\scriptsize \raggedright \baselineskip 1pt

\item[1.] Kapitza, P. (1928) {\it Proc. Roy. Soc.}, {\bf A129}, 358

\item[2.] Schubnikov, L.V. and de Haas, W.J. (1930) {\it Leiden
Commun.}, {\bf 19}, 207f

\item[3.] de Haas, W.J., Blom, J.W. and Schubnikov, L.V. (1930)
{\it Physica}, {\bf 2}, 907

\item[4.] Landau, L.D. (1939) {\it Proc. Roy. Soc.}, {\bf 170}, 341

\item[5.] Kartsovnic, V.M., Laukhin, V.N., Nizhankovsky, V.I. and
Ignatyev, A.A. (1988) {\it Pis'ma v ZhETF}, {\bf 47}, 302 ({\it JETP Lett},
{\bf 47}, 363 (1988))

\item[6.] Kartsovnic, V.M., Kononovich, P.A., Laukhin, V.N. and
Shchegolev, I.F. (1988) {\it Pis'ma v ZhETF}, {\bf 48}, 498 (JETP
Lett., 48, 541 (1988))

\item[7.] Parker, I.D., Pigram, D.D., Friend, R.H., Kurmo, M. and
Day, P. (1988) {\it Synth. Met.}, {\bf 27}, A387

\item[8.] Oshima, K., Mori, T., Inokuchi, H., Urayama, H.,
Yamochi, H. and Sato, C. (1988) {\it Phys. Rev.}, {\bf B38}, 938

\item[9.] Toyota, N., Sasaki, T., Murata, K., Honda, Y., Tokumoto,
M., Bando, H., Kinoshita, N., Anzai, H., Ishiguro, T. and
Muto, Y. (1988) {\it J. Phys. Soc. Jpn.}, {\bf 57}, 2616

\item[10.] Kang, W., Montambaux, G., Cooper, J.R., Jerome, D.,
Batail, P. and Lenoir, C.  (1989) {\it Phys. Rev. Lett.}, {\bf 62},
2559

\item[11.] Kartsovnic, V.M., Kononovich, P.A., Laukhin, V.N.,
Pesotskii, S.I. and Shchegolev, I.F. (1990) {\it Zh. Eksp. Teor.
Fiz.}, {\bf 97}, 1305 (in Russian)

\item[12.] Shchegolev, I.F., Kononovich, P.A., Kartsovnic, V.M.,
Laukhin, V.N., Pesotskii, S.I., Hilti, B. and Mayer, C.W.
(1990) {\it Synth. Met.}, {\bf 39}, 357

\item[13.] Tokumoto, M., Swanson, A.O., Brooks, J.S., Agosta,
C.C., Hannahs, S.T., Kinoshita, N., Anzai, H. and
J.R.Anderson, (1990) {\it J. Phys. Soc. Jpn.}, {\bf 59}, 2324

\item[14.] Yagi, R., Iye, Y., Osada, T. and Kagoshima, S. (1990)
{\it J. Phys. Soc. Jpn.}, {\bf 59}, 3069

\item[15.] Onsager, L. (1952) {\it Phil. Mag}, {\bf 43}, 1006

\item[16.] Lifshitz, I.M. and Kosevich, A.H. (1955) {\it Zh. Eksp.
Teor. Fiz.}, {\bf 29}, 743 ({\it Sov. Phys.  JETP}, {\bf 2}, 646 (1956))

\item[17.] Azbel', M.Ya. (1960) {\it Zh. Eksp. Teor. Fiz.}, {\bf 39}, 400
({\it Sov. Phys. JETP}, {\bf 12}, 283 (1961))

\item[18.] Onsager, L. (1931) {\it Phys. Rev.}, {\bf 37}, 405

\item[19.] Lifshitz, I.M. and Peschansky, V.G. (1958) {\it Zh. Eksp.
Teor. Fiz.}, {\bf 35}, 1251 ({\it Sov.  Phys. JETP}, {\bf 8}, 875 (1958)).

\item[20.] Ziman, J. (1959) {\it Phil. Mag.}, {\bf 3}, 1117

\item[21.] Stachowiak, H. (1964) {\it Acta Phys. Pol.}, {\bf 26}, 217

\item[22.] Dreizin, Yu.A. and Dykhne, A.M. (1971) {\it Pis'ma v ZhETF},
{\bf 14}, 101 (in Russian)

\item[23.] Peschansky, V.G., Roldan Lopez, J.A. and Toji Gnado Jao
(1991) {\it Journ. de Physik I (France)}, {\bf 1}, 1469

\item[24.] Schoenberg, D. (19??) {\it Magnetic Oscillations} ???

\item[25.] Privorotsky, I.A. (1967) {\it Zh. Eksp. Teor. Fiz.}, {\bf 52}, 1755
(in Russian)

\item[26.] Reuter, E.H.T. and Sondheimer, B.H. (1948) {\it Proc. Roy.
Soc.}, {\bf 195}, 336

\item[27.] Peschansky, V.G., Savel'eva, S.N. and Kheir Bek, H.
(1992) {\it Fiz. Tverd. Tela}, {\bf 34}, 1630 ({\it Sov. Phys. Solid State},
{\bf 34}, 871 (1992))

\item[28.] Peschansky, V.G., Kheir Bek, H. and Savel'eva, S.N.
(1992) {\it Fiz. Nizk. Temp.}, {\bf 18}, 1012 ({\it Sov. J. Low Temp. Phys.},
{\bf 18}, 711 (1992))

\item[29.] Hartmann, L.E. and Luttinger, L.M. (1966) {\it Phys. Rev.},
{\bf 151}, 430

\item[30.] Azbel', M.Ja. (1961) {\it Zh. Eksp. Teor. Fiz.}, {\bf 39}, 400
({\it Sov. Phys. JETP}, {\bf 12}, 283 (1961))

\item[31.] Lur'e, M.A., Peschansky, V.G. and Jiasemides, K. (1984)
{\it J. Low Temp. Phys.}, {\bf 36}, 277

\item[32.] Peschansky, V.G., Dassanaeke, M.S. and Tsybulina, B.V.
(1983) {\it Fiz. Nizk. Temp.}, {\bf 11}, 297 ({\it Sov. Low. Temp. Phys.},
{\bf 11}, 162 (1985))

\item[33.] Pippard, A.B. (1957) {\it Phil. Mag.}, {\bf 2}, 1147

\item[34.] Gurevich, V.L. (1959) {\it Zh. Eksp. Teor. Fiz.}, {\bf 37}, 71 (in
Russian)

\item[35.] Kaner, E.A., Peschansky, V.G. and Privorotsky, I.A.
(1961) {\it Zh. Eksp. Teor. Fiz.}, {\bf 40}, 214 ({\it Sov. Phys. JETP}, {\bf 13},
147 (1961))

\item[36.] Gurevich, V.L., Skobov, V.G. and Firsov, Yu.D. (1961)
{\it Zh. Eksp. Teor. Fiz.}, {\bf 40}, 786 ({\it Sov. Phys. JETP}, {\bf 13}, 552
(1961))

\item[37.] Silin, V.P. (1960) {\it Zh. Eksp. Teor. Fiz.}, {\bf 38}, 977 ({\it Sov.
Phys. JETP}, {\bf 11}, 775 (1960))

\item[38.] Kontorovich, V.M. (1963) {\it Zh. Eksp. Teor. Fiz.}, {\bf 45}, 1633
({\it Sov. Phys. JETP}, {\bf 18}, 1333 (1963))

\item[39.] Andreev, A.F. and Pushkarov, D.I. (1985) {\it Zh. Eksp.
Teor. Fiz.}, {\bf 89}, 1883 ({\it Sov. Phys.  JETP}, {\bf 62}, 1087 (1985))

\item[40.] Akhiezer, A.I. (1938) {\it Zh. Eksp. Teor. Fiz.}, {\bf 8}, 1338

\item[41.] Kirichenko, O.V. and Peschansky, V.G. (1994) {\it Journal de
Physique}, {\bf 4}, 823

\item[42.] Gorhfel'd, V.M., Kirichenko, O.V. and Peschansky, V.G.
(1995) {\it Zh. Eksp. Teor.  Fiz.}, {\bf 108}, 2147 ({\it Sov. Phys. JETP},
{\bf 81}, 1171 (1995))

\item[43.] Kirichenko, O.V. and Peschansky, V.G. (1994) {\it Fiz. Nizk.
Temp.}, {\bf 20}, 574 (Low.  Temp. Phys., 20, 453 (1994))

\item[44.] Landau, L.D. and Lifshits, B.M. (1986) {\it Theory of
Elasticity}, Pergamon, Oxford

\item[45.] Kirichenko, O.V. and Peschansky, V.G. (1996) {\it Pis'ma v
ZhETF}, {\bf 64}, 845 (in Russian)

\item[46.] Gorhfel'd, V.M., Kirichenko, O.V. and Peschansky, V.G.
(1993) {\it Fiz. Nizk. Temp.}, {\bf 19}, 456 ({\it Low. Temp. Phys.}, {\bf 19}, 321
(1993))

\item[47.] Galbova, O., Ivanovski, G., Kirichenko, O.V. and
Peschansky, V.G. (1997) {\it Fiz.  Nizk. Temp.}, {\bf 23}, 173 ({\it Low.
Temp. Phys.}, {\bf 23}, 127 (1997))

\item[48.] Landau, L.D. (1956) {\it Zh. Eksp. Teor. Fiz.}, {\bf 30}, 1058
({\it Sov. Phys. JETP}, {\bf 3}, 920 (1956))

\item[49.] Silin, V.P. (1957) {\it Zh. Eksp. Teor. Fiz.}, {\bf 33}, 495 ({\it Sov.
Phys. JETP}, {\bf 6}, 387 (1957))

\item[50.] Silin, V.P. (1967) {\it Usp. Fiz. Nauk}, {\bf 93}, 185

\item[51.] Bagaev, V.N., Okulov, V.I. and Pamyatnikh, E.A. (1978)
{\it Pis'ma Zh. Eksp. Teor.  Fiz.}, {\bf 27}, 156 ({\it JETP Letters}, {\bf 27},
144 (1978))

\item[52.] Peschansky, V.G., Espeho, G. and Tesgera Bedassa, D.
(1995) {\it Fiz. Nizk. Temp.}, {\bf 21}, 971 ({\it Low Temp. Phys.}, {\bf 21}, 748
(1995))

\item[53.] Galbova, O., Ivanovski, G., Kirichenko, O.V. and
Peschansky, V.G. (1995) {\it Phys.  Low-Dim. Struct.}, {\it 10/11}, 295

\item[54.] Galbova, O., Ivanovski, G., Kirichenko, O.V. and
Peschansky, V.G. (1996) {\it Fiz.  Nizk. Temp.}, {\bf 22}, 425 ({\it Low
Temp. Phys.}, {\bf 22}, 331 (1996))

\item[55.] Sharvin, Yu.V. (1965) {\it Zh. Eksp. Teor. Fiz.}, {\bf 48}, 984
({\it Sov. Phys. JETP}, {\bf 21}, 655 (1965))

\item[56.] Sharvin, Yu.V. and Fisher, L.M. (1965) {\it Pis'ma v ZhETF},
{\bf 1}, 54 ({\it JETP Lett.}, {\bf 1}, 152 (1965))

\item[57.] Hawkins, F.M. and Pippard, A.B. (1965) {\it Proc. Cambridge
Philos. Soc.}, {\bf 61}, 433

\item[58.] Tsoi, V.S. (1974) {\it Pis'ma v ZhETF}, {\bf 19}, 114 ({\it JETP Lett.},
{\bf 19}, 70 (1974))

\item[59.] Goncalves da Silva, C.E.T. (1974) {\it J. Low. Temp. Phys.},
{\bf 16}, 337

\item[60.] Korzh, S.A. (1975) {\it Zh. Eksp. Teor. Fiz.}, {\bf 68}, 144 ({\it Sov.
Phys. JETP}, {\bf 41}, 70 (1975))

\item[61.] Kolesnichenko, Yu.A., Bedassa, T. and Grishaev, V.I.
(1995) {\it Fiz. Nizk. Temp.}, {\bf 21}, 1049 ({\it Low Temp. Phys.}, {\bf 21}, 806
(1995))

\item[62.] Baskin, E.M. and Entin, M.V. (1968) {\it Zh. Eksp. Teor.
Fiz.}, {\bf 49}, 460 (in Russian)

\item[63.] Fal'kovsky, L.A. (1970) {\it Zh. Eksp. Teor. Fiz.}, {\bf 58}, 1830
({\it Sov. Phys. JETP}, {\bf 31}, 981 (1970))

\item[64.] Okulov, V.I. and Ustinov, V.V. (1979) {\it Fiz. Nizk. Temp.},
{\bf 5}, 213 ({\it Sov. J. Low Temp.  Phys.}, {\bf 5}, 101 (1979))

\item[65.] Azbel', M.Ya. and Peschansky, V.G., (1965) {\it Zh. Eksp.
Teor. Fiz.}, {\bf 49}, 52; (1968) {\it Zh. Eksp. Teor. Fiz.}, {\bf 55}, 1980
({\it Sov. Phys. JETP}, {\bf 22}, 399 (1966); {\it Sov. Phys. JETP}, {\bf 22}, 1045
(1969))

\item[66.] Kolesnichenko, Yu.A. and Kulik, I.O. (1992) {\it Fiz. Nizk.
Temp.}, {\bf 18}, 1005 ({\it Sov. J.  Low Temp. Phys.}, {\bf 18}, 706 (1992))

\item[67.] Kolesnichenko, Yu.A. (1992) {\it Fiz. Nizk. Temp.}, {\bf 18}, 1059
({\it Sov. J. Low Temp. Phys.}, {\bf 18}, 741 (1992))

\item[68.] Kulik, I.O., Omel'anchuk, A.N. and Shekhter, R.I.
(1977) {\it Fiz. Nizk. Temp.}, {\bf 3}, 1543 ({\it Sov. J. Low Temp. Phys.},
{\bf 3}, 740 (1977))

\item[69.] Lifshits, I.M. Azbel', M.Ya. and Kaganov, M.I. (1972)
{\it Electron Theory of Metals}, Moscow: Nauka (New York:
Consultant Bureau (1973))

\item[70.] Kulik, I.O., Shekhter, R.I. and Shkorbatov, A.G. (1981)
{\it Zh. Eksp. Teor. Fiz.}, {\bf 81}, 2126 ({\it Sov. Phys. JETP}, {\bf 54}, 1130
(1981))

\item[71.] Bogachek, E.N., Kulik, I.O. and Shekhter, R.I. (1987)
{\it Zh. Eksp. Teor. Fiz.}, {\bf 92}, 730 ({\it Sov. Phys. JETP}, {\bf 65}, 411
(1987))

\item[72.] Kolesnichenko, Yu.A., Tuluzov, I.G. and Khotkevich,
A.V. (1993) {\it Fiz. Nizk.Temp.}, {\bf 19}, 402 ({\it Low Temp. Phys.}, {\bf 19},
282 (1993))

\item[73.] Kolesnichenko, Yu.A., Tuluzov, I.G. and Khotkevich,
A.V. (1993) {\it Fiz. Nizk.Temp.}, {\bf 19}, 901 ({\it Low Temp. Phys.}, {\bf 19},
642 (1993))

\item[74.] Yanson, I.K. (1974) {\it Zh. Eksp. Teor. Fiz.}, {\bf 66}, 1035
({\it Sov. Phys. JETP}, {\bf 39}, 506 (1974))

\item[75.] Yanson, I.K. and Khotkevich, A.V. (1986) {\it Atlas of Point
Contact Spectra of Electron--Phonon Interactions in Metals},
Kiev: Naukova dumka (Kluwer Academic Publishers (1995))

\item[76.] Novack, A., Weger, M., Schweitzer, D. and Keller, H.J.
(1986) {\it Solid State Commun.}, {\bf 60}, 199

\item[77.] Novak, A., Poppe, U., Weger, M. {\it et al.} (1987) {\it Zh.
Phys.}, {\bf B68}, 41

\item[78.] Kamarchuk, G.V., Pokhodnya, K.I., Khotkevich, A.V. and
Yanson, I.K. (1990) {\it Fiz.  Nizk. Temp.}, {\bf 16}, 711 ({\it Sov. J. Low
Temp. Phys.}, {\bf 16}, 419 (1990))

\item[79.] Kamarchuk, G.V., Khotkevich, A.V., Kozlov, M.E. and
Pokhodnya, K.I. (1992) {\it Fiz. Nizk. Temp.}, {\bf 18}, 967 ({\it Sov. J.
Low Temp. Phys.}, {\bf 18}, 679 (1992))

\item[80.] Kamarchuk, G.V., Khotkevich, A.V., Kolesnichenko,
Yu.A., Pokhodnya, K.I. and Tuluzov, I.G. (1994) {\it J. Phys.:
Condens. Matter}, {\bf 6}, 3559

\item[81.] Kulik, I.O. (1985) {\it Pis'ma v ZhETF}, {\bf 41}, 302 ({\it JETP Lett.},
{\bf 41}, 370 (1985))

\item[82.] Syrkin, E.S. and Feodos'ev, S.B. (1991) {\it Fiz. Nizk.
Temp.}, {\bf 17}, 1055 ({\it Sov. J. Low Temp. Phys.}, {\bf 19}, 549 (1991))
\end{list}

\cleardoublepage
\thispagestyle{myfirst}
\section*{INDEX}

\noindent{}Acoustic transparency

\noindent{}Acousto-electronic tensors

\noindent{}Anomalous skin effect

\noindent{}Boundary condition

\noindent{}Collision integral

\noindent{}Deformation potential

\noindent{}Distribution function of electrons

\noindent{}Electron focusing

\noindent{}Electron--phonon interaction

\noindent{}Electrical neutrality

\noindent{}Fermi liquid-interaction

\noindent{}Fermi surface

\noindent{}Kinetic equation

\noindent{}Magnetoresistance

\noindent{}Magnetoacoustic resonance

\noindent{}Maxwell's equation

\noindent{}Normal skin effect

\noindent{}Open electron orbit

\noindent{}Penetration depth

\noindent{}Point contacts

\noindent{}Point contact spectroscopy

\noindent{}Quasi-two-dimensional energy spectrum

\noindent{}Rate of sound attenuation

\noindent{}Scattering indicatrix

\noindent{}Specular parameter

\noindent{}Weakly damping waves

\end{document}